\documentclass[a4paper,iop,twocolumn,numberedappendix,appendixfloats]{emulateapj}
\usepackage[breaklinks,colorlinks=true,linkcolor=blue,anchorcolor=blue,citecolor=blue,urlcolor=blue,draft=false]{hyperref}
\usepackage{color}
\usepackage{graphicx}
\usepackage{amsmath}
\usepackage{amssymb}
\usepackage{mathrsfs}
\usepackage{placeins}
\usepackage{times}
\usepackage{xcolor}
\usepackage{xspace}
\usepackage{datetime}
\usepackage{version}
\usepackage{bm}

\hypersetup{pdfauthor={I-Non Chiu},pdftitle={Triaxial Modeling of Galaxy Clusters}}
\lefthead{Chiu et. al.}
\righthead{Triaxial Modeling of Galaxy Clusters} 

\newcommand{\LCDM}{\ensuremath{\Lambda\textrm{CDM}}}
\newcommand{\OmegaM}{\ensuremath{\Omega_{\mathrm{m}}}}

\newcommand{\Hnow}{\ensuremath{H_{0}}}
\newcommand{\seight}{\ensuremath{\sigma_{8}}}
\newcommand{\Msun}{\ensuremath{\mathrm{M}_{\odot}\,h^{-1}}}

\newcommand{\Rtwooo}{\ensuremath{R_{200\mathrm{c}}}}
\newcommand{\Mtwooo}{\ensuremath{M_{200\mathrm{c}}}}
\newcommand{\Ctwooo}{\ensuremath{c_{200\mathrm{c}}}}
\newcommand{\rhocrit}{\ensuremath{\rho_{\mathrm{c}}}}
\newcommand{\redshift}{\ensuremath{z}}
\newcommand{\dif}{\ensuremath{\mathrm{d}}}

\newcommand{\HST}{\emph{HST}}
\newcommand{\PLANCK}{\emph{Planck}}
\newcommand{\WMAP}{\emph{WMAP}}

\newcommand{\CLASH}{\textrm{CLASH}}

\newcommand{\zd}{\ensuremath{z_{\mathrm{d}}}}

\newcommand{\shearone}{\ensuremath{\gamma_{1}}}
\newcommand{\sheartwo}{\ensuremath{\gamma_{2}}}
\newcommand{\shear}{\ensuremath{\gamma}}
\newcommand{\sigmam}{\ensuremath{\Sigma}}
\newcommand{\sigmacrit}{\ensuremath{\Sigma_{\mathrm{c}}}}
\newcommand{\dl}{\ensuremath{D_{\mathrm{l}}}}
\newcommand{\ds}{\ensuremath{D_{\mathrm{s}}}}
\newcommand{\dls}{\ensuremath{D_{\mathrm{ls}}}}

\newcommand{\percent}{\ensuremath{\%}}

\newcommand{\rhos}{\ensuremath{\rho_{\mathrm{s}}}}
\newcommand{\Rs}{\ensuremath{R_{\mathrm{s}}}}
\newcommand{\qa}{\ensuremath{q_{\mathrm{a}}}}
\newcommand{\qb}{\ensuremath{q_{\mathrm{b}}}}
\newcommand{\fgeo}{\ensuremath{f_{\mathrm{geo}}}}
\newcommand{\triaxiality}{\ensuremath{\mathcal{T}}}

\newcommand{\Msl}{\ensuremath{M_{\mathrm{SL}}}}

\newcommand{\sphericalmodel}{\textsf{Spherical}}
\newcommand{\triaxialfmodel}{\textsf{Triaxial}}
\newcommand{\triaxialbmodel}{\textsf{Triaxial+B15}}

\newcommand{\appropto}{\mathrel{\vcenter{
  \offinterlineskip\halign{\hfil$##$\cr
    \propto\cr\noalign{\kern2pt}\sim\cr\noalign{\kern-2pt}}}}}

\def\le{\leqslant}

\definecolor{inon}{rgb}{1.00,0.27,0.00}

\newcommand{\numAconcenSph}{\ensuremath{ 4.51 \pm 0.14 }} 
\newcommand{\numBconcenSph}{\ensuremath{ -0.47 \pm 0.07 }} 
\newcommand{\numDconcenSph}{\ensuremath{ 0.05 }} 
\newcommand{\numAconcenTri}{\ensuremath{ 4.82 \pm 0.30 }} 
\newcommand{\numBconcenTri}{\ensuremath{ -0.36 \pm 0.11 }} 
 
\newcommand{\numAconcenBon}{\ensuremath{ 4.73 \pm 0.28 }} 
\newcommand{\numBconcenBon}{\ensuremath{ -0.36 \pm 0.10 }} 
\newcommand{\numDconcenBon}{\ensuremath{ 0.04 }}

\newcommand{\numAqaTri}{\ensuremath{ 0.52 \pm 0.04 }} 
\newcommand{\numBqaTri}{\ensuremath{ -0.36 \pm 0.13 }} 
\newcommand{\numDqaTri}{\ensuremath{ 0.05 }} 
\newcommand{\numAqaBon}{\ensuremath{ 0.44 \pm 0.02 }} 
\newcommand{\numBqaBon}{\ensuremath{ -0.14 \pm 0.07 }} 
\newcommand{\numDqaBon}{\ensuremath{  <~0.06~(2\sigma~\mathrm{upper~bound}) }}

\newcommand{\numAfgeoTri}{\ensuremath{ 0.93 \pm 0.07 }} 
\newcommand{\numBfgeoTri}{\ensuremath{ -0.14 \pm 0.15 }} 
\newcommand{\numDfgeoTri}{\ensuremath{  <~0.34~(2\sigma~\mathrm{upper~bound}) }} 
\newcommand{\numAfgeoBon}{\ensuremath{ 0.96 \pm 0.07 }} 
\newcommand{\numBfgeoBon}{\ensuremath{ -0.21 \pm 0.12 }} 
\newcommand{\numDfgeoBon}{\ensuremath{  <~0.26~(2\sigma~\mathrm{upper~bound}) }}

\newcommand{\numBtriaxialityTri}{\ensuremath{ 0.07 \pm 0.19 }} 
 
\newcommand{\numAtriaxialityBon}{\ensuremath{ 0.79 \pm 0.03 }} 
\newcommand{\numBtriaxialityBon}{\ensuremath{ 0.06 \pm 0.07 }} 
\newcommand{\numDtriaxialityBon}{\ensuremath{  <~0.13~(2\sigma~\mathrm{upper~bound}) }} 

\begin{document}

%
%

\title{
CLUMP-3D: Three-dimensional Shape and Structure of 20 CLASH Galaxy 
Clusters from Combined Weak and Strong Lensing
}

%
%

\author{I-Non Chiu\altaffilmark{1}}
\author{Keiichi Umetsu\altaffilmark{1}}
\author{Mauro Sereno\altaffilmark{2,3}}
\author{Stefano Ettori\altaffilmark{2,4}}
\author{Massimo Meneghetti\altaffilmark{5}}
\author{Julian Merten\altaffilmark{6}}
\author{Jack Sayers\altaffilmark{7}}
\author{Adi Zitrin\altaffilmark{8}}

\altaffiltext{1}{Institute of Astronomy and Astrophysics, Academia Sinica, P.O. Box 23-141, Taipei 10617, Taiwan}
\altaffiltext{2}{INAF - Osservatorio Astronomico di Bologna, via Piero Gobetti 93/3, I-40129 Bologna, Italy}
\altaffiltext{3}{INAF - Osservatorio di Astrofisica e Scienza dello Spazio di Bologna, via Piero Gobetti 93/3, I-40129 Bologna, Italy}
\altaffiltext{4}{INFN, Sezione di Bologna, viale Berti Pichat 6/2, I-40127 Bologna, Italia}
\altaffiltext{5}{INAF - Osservatorio Astronomico di Bologna, via Ranzani 1, I-40127 Bologna, Italy}
\altaffiltext{6}{Oxford University, Keble Road, Oxford OX1 3RH, United Kingdom}
\altaffiltext{7}{Division of Physics, Math, and Astronomy, California Institute of Technology, Pasadena, CA 91125}
\altaffiltext{8}{Physics Department, Ben-Gurion University of the Negev,
P.O. Box 653, Be'er-Sheva 8410501, Israel}

%
%

\begin{abstract}

We perform a three-dimensional triaxial analysis of 16 X-ray regular and 4 high-magnification galaxy clusters selected from the \CLASH\ survey by combining two-dimensional weak-lensing and central strong-lensing constraints.  
In a Bayesian framework, we constrain the intrinsic structure and geometry of each individual cluster assuming a triaxial Navarro--Frenk--White halo with arbitrary orientations, characterized by the mass \Mtwooo, halo concentration \Ctwooo, and triaxial axis ratios ($\qa \le \qb$), and investigate scaling relations between these halo structural parameters.  
From triaxial modeling of the X-ray-selected subsample, we find that the halo concentration decreases with increasing cluster mass, with a mean concentration of $\Ctwooo = 4.82\pm0.30$ at the pivot mass $\Mtwooo=10^{15}\Msun$. This is consistent with the result from spherical modeling, $\Ctwooo=4.51\pm 0.14$. 
Independently of the priors, the minor-to-major axis ratio \qa\ of our full sample exhibits a clear deviation from the spherical configuration ($\qa=\numAqaTri$ at $10^{15}\Msun$ with uniform priors), with a weak dependence on the cluster mass. 
Combining all 20 clusters, we obtain a joint ensemble constraint on the minor-to-major axis ratio of $\qa=0.652^{+0.162}_{-0.078}$ and a lower bound on the intermediate-to-major axis ratio of $\qb>0.63$ at the $2\sigma$ level from an analysis with uniform priors. 
Assuming priors on the axis ratios derived from numerical simulations, we constrain the degree of triaxiality for the full sample to be $\triaxiality=\numAtriaxialityBon$ at $10^{15}\Msun$, indicating a preference for a prolate geometry of cluster halos. 
We find no statistical evidence for an orientation bias ($\fgeo=\numAfgeoTri$), which is insensitive to the priors and in agreement with the theoretical expectation for the \CLASH\ clusters.

\end{abstract}

%
%

\keywords{cosmology: observations --- dark matter --- galaxies:
clusters: general --- gravitational lensing: weak --- gravitational
lensing: strong}

%
%

\section{Introduction}
\label{sec:introduction}

Galaxy clusters are gravitationally dominated by dark matter and serve as a wealth of ideal laboratories to study structure formation in the universe. 
In particular, an accurate mass estimation of galaxy clusters is crucial not only for utilizing them as cosmological probes \citep{planck2015clstrcnts,mantz15,bocquet15,dehaan16} but also for understanding the root cause of various astrophysical processes in
massive halos, such as environmental quenching of galaxies \citep{dressler80}.  
Conventionally, the total cluster mass is determined from projected measurements assuming spherical symmetry. 
In this context, $N$-body simulations in the standard $\Lambda$ cold dark matter (\LCDM) model established a nearly self-similar form for the spherically averaged density profile $\rho(r)$ of dark-matter halos \citep{navarro96}, which can be characterized by two parameters, namely the characteristic density and radius of halos.
This two-parameter model gives a satisfactory success in terms of statistically quantifying the ensemble mass of observed clusters over a  sizable sample \citep[e.g.,][]{umetsu11b,newman13,umetsu14,vonderlinden14a,hoekstra15,umetsu16,okabe16}. 

However, cluster halos are predicted to be non-spherical, with a preference for prolate shapes according to $N$-body simulations in the \LCDM\ model \citep{frenk1988,dubinski1991,warren92,jing02}.
Besides, the shape of halos is predicted to depend on the redshift, halo mass, and cluster-centric radius  \citep{bailin05,hopkins05,allgood06,bett07,bonamigo15}, as well as on baryonic effects \citep{flores05}, large-scale environments \citep{kasun05}, and the background cosmology \citep{allgood06,despali14}. 
Therefore, cluster mass estimates assuming spherical symmetry cause a substantial scatter around their true mass \citep[e.g.,][]{battaglia11b}. 
Importantly, an inappropriate assumption about the cluster shape and
orientation could significantly bias individual mass measurements \citep[e.g.,][]{oguri05}. 
There have been initial attempts to compare the observed shapes of galaxy clusters with those predicted by cosmological numerical simulations \citep[e.g.,][]{oguri10b}, opening up a new avenue of testing models of structure formation.
It is thus important to perform a statistical analysis of the shape of clusters by extending cluster mass determinations beyond spherical modeling.

Compared to the theoretical efforts to characterize the shape of galaxy clusters in numerical simulations, significantly less progress has been made on the observational side \citep{defilippis05,sereno06,corless08b,morandi12,limousin13,umetsu15,sereno17}.
Detailed observational work thus far was subject to case studies instead of a statistical interpretation from a large cluster sample, because of difficulty in acquiring data sets that are needed to achieve the required precision.
Moreover, the shape of the total mass distribution in clusters was often inferred indirectly from observations of the intracluster medium (ICM) assuming hydrostatic equilibrium, which however would be violated in the presence of, for example, turbulent and bulk motions of hot gas \citep{lau09,molnar10,chiu12}.
Astrophysical processes, such as radiative cooling of ICM and entropy
injection from active galactic nuclei, as well as the cluster dynamical
state \citep{cialone17}, further complicate the
interpretation of cluster shapes inferred from X-ray or the
Sunyaev--Zel'dovitch Effect \citep[SZE hereafter;][]{sunyaev70,sunyaev72} observations.   
Thus, investigating the shape of galaxy clusters is observationally challenging.

Gravitational lensing provides a direct access to the underlying mass distribution of galaxy clusters without requiring any assumptions about their dynamical or physical state.
There have been many successful attempts to determine the cluster mass by weak lensing \citep{okabe10,vonderlinden14a,hoekstra15,melchior17,medezinski17,schrabback18}, strong lensing \citep{broadhurst05,richard10,zitrin15,grillo15} and the combination of both \citep{bradac06,oguri12,umetsu13,umetsu16}.
With recent progress in controlling systematics in weak lensing \citep[e.g., intensive calibration against simulations;][]{bridle10,kitching12,mandelbaum15}, together with advances in instrumentation and observing techniques, we are in a great position to utilize gravitational lensing with high-quality lensing data.

In this work, we aim to use both weak and strong lensing to constrain the three-dimensional (3D) structure and shape of clusters targeted by the CLUster Multi-Probes in Three Dimensions (CLUMP-3D) program \citep{sereno17}. 
Our sample consists of 20 high-mass clusters that were selected by the Cluster Lensing And Supernova survey with Hubble  \citep[CLASH, hereafter;][]{postman12}.  
Importantly, these 20 clusters were all deeply followed up from the ground and from space in different wavelengths \citep{donahue14, umetsu14, rosati14, czakon15}, with the goal of precisely characterizing the cluster mass distribution.
Besides, these clusters have been intensively studied in previous work in the context of galaxy evolution \citep{annunziatella14,demaio15,gupta16}, characterization of strongly lensed arcs \citep{zitrin15}, wide-field weak-lensing analysis \citep{umetsu12,umetsu14}, and exploration of the high-redshift universe \citep{zheng12,coe13,balestra13,monna14,mcleod16}.

In the first paper of the CLUMP-3D program, \citet{sereno17} carried out a full triaxial analysis of MACS~J1206.2$-$0847 using multi-probe data sets from weak-lensing, strong-lensing, X-ray, and SZE observations, demonstrating the power of multi-probe cluster analysis.
In our companion paper, \citet{umetsu18} present direct
reconstructions of the two-dimensional (2D) matter distribution in the
20 \CLASH\ clusters from a joint analysis of 2D shear and azimuthally
averaged magnification measurements. 
This work is the third paper of the series, where we focus on
characterizing the 3D mass distribution of the 20 \CLASH\ clusters by
combining weak and strong lensing.  
A multi-probe triaxial analysis of 16 X-ray-selected \CLASH\ clusters
using weak-lensing, strong-lensing, X-ray, and SZE data sets is
presented in another companion paper \citep{sereno18}.  
We note that even though a joint analysis of multi-probe data sets can
formally achieve a better precision, studies using gravitational lensing
alone have the advantage of being free from assumptions about baryonic
components in clusters. Therefore, both approaches are required and
complementary to each other.

This paper is organized as follows.
We will briefly introduce the basics of gravitational lensing in Section~\ref{sec:basics}.
We then describe the cluster sample and the lensing data products in Section~\ref{sec:sample_and_data}.
In Section~\ref{sec:methodology} we outline our methodology for triaxial modeling.
We discuss our results in Section~\ref{sec:results_and_discussion}, followed by the conclusions made in Section~\ref{sec:conclusions}. 
Throughout this work, we assume a flat \LCDM\ cosmological model with $\OmegaM=0.27$, $\Hnow = h\times100$\,km\,s\,$^{-1}$\.Mpc$^{-1}$ with $h=0.7$, and $\seight = 0.8$.
We define an ellipsoidal overdensity radius $R_\Delta$ \citep[e.g.,][]{corless09b,umetsu15} such that the mean interior density contained within an ellipsoidal volume of semimajor axis $R_\Delta$ is $\Delta$ times the critical density of the universe $\rhocrit(z)$ at the cluster redshift $z$.
We use $\Delta=200$ to define the halo mass, $M_\mathrm{200c}$\footnote{See Equation~(\ref{eq:def_mass}).}.
All quoted errors are $68\percent$ confidence limits (i.e., $1\sigma$) unless otherwise stated.
We use the AB magnitude system.
The notation $\mathcal{U}(x,y)$ stands for a uniform distribution between $x$ and $y$.

%
%

\section{Theory of Gravitational Lensing}
\label{sec:basics}

In this section, we briefly review the basics of gravitational lensing
with emphasis on cluster lensing. In this case, we can approximate the
lensing cluster of interest at redshift \zd\ as a single thin lens
embedded in a homogeneous universe where background sources at redshift
$\redshift>\zd$ are all lensed.  
We refer the readers to \cite{bartelmann01}, \cite{umetsu10} and \cite{hoekstra13} for a more complete overview of gravitational lensing.

To the first order, the deformation of observed background images due to gravitational lensing can be described by the lensing Jacobian matrix (see Equation~(\ref{eq:jacobian})), which is characterized by the convergence $\kappa$ and the shear
$\shear \equiv \shearone + i\sheartwo$ at the position $\bm{\vartheta}$ on the lens plane,  
\begin{equation}
\label{eq:jacobian}
J(\bm{\vartheta}) = 
\begin{pmatrix}
1 - \kappa - \shearone & -\sheartwo \\
-\sheartwo & 1 - \kappa + \shearone
\end{pmatrix},
\end{equation}
where $\kappa$, $\shearone$, and $\sheartwo$ are written as linear combinations of second derivatives of the lensing potential. 
The convergence $\kappa(\bm{\vartheta})$ is the surface mass density normalized by the critical surface mass density for lensing \sigmacrit,
\begin{equation}
\label{eq:kappa_def}
\kappa(\bm{\vartheta}) = \frac{\sigmam(\bm{\vartheta})}{\sigmacrit},
\end{equation}
where $\sigmam(\bm{\vartheta})$ is the surface mass density of the cluster projected along the line of sight, and 
\begin{equation}
\label{eq:crit_def}
 \sigmacrit = \frac{c^2}{4\pi G} \frac{\ds}{\dl\dls}
\end{equation}
with $G$ the Newton's constant,
and \dl, \ds\ and \dls\ the angular diameter distances between the observer-to-cluster, observer-to-source, and the cluster-to-source pairs, respectively. 
The complex shear \shear\ is related to the convergence $\kappa$ by
\begin{equation}
\label{eq:shear_field}
\shear(\bm{\vartheta}) = \int\dif^2\bm{\vartheta^{\prime}}D(\bm{\vartheta} - \bm{\vartheta^{\prime}})\kappa(\bm{\vartheta^{\prime}}),
\end{equation}
where the convolution kernel is defined as $D(\bm{y})\equiv\left(y_2^2 - y_1^2 - 2iy_1y_2\right)/(\pi|\bm{y}|^4)$.

In general, the observable quantity for weak lensing is not the gravitational shear \shear\, but the reduced shear $g$ in the subcritical regime,
\begin{equation}
\label{eq:reduced_shear}
g = \frac{\shear}{1 - \kappa}.
\end{equation}
The reduced shear $g$ remains invariant under the global transformation, 
$\kappa\rightarrow\lambda\kappa+1-\lambda$ and
$\shear\rightarrow\lambda\shear$, for any $\lambda \neq 0$.
This is referred to as the mass-sheet degeneracy \citep{bartelmann01},
which can be broken, for example, by including the lensing magnification
effect.

The lensing magnification is characterized by the inverse determinant of the Jacobian matrix,
\begin{equation}
\label{eq:mu_def}
\mu = \frac{1}{\left(1 - \kappa\right)^2 - |\shear|^2}.
\end{equation}
The magnification factor transforms differently as
$\mu\rightarrow\lambda^{-2}\mu$, which can be used to break the mass-sheet degeneracy.
In the subcritical regime where $\mu > 0$  and $|g|<1$, the magnification introduces two
competing effects: the reduction (increase) of observed area on the source plane
given a solid angle, and the amplificatin (deamplification) of flux of background sources. 
As a net result, the surface number density of a ``flux-limited''
background sample is altered due to the presence of lensing
magnification depending on the intrinsic slope of the background
luminosity function. This effect is known as magnification bias
\citep{broadhurst95,taylor98}.

The effect of magnification bias can be measured by comparing cumulative
number counts of flux-limited background galaxies with and without
gravitational lensing as 
\begin{equation}
\label{eq:magnification_bias}
\mu^{2.5s - 1} = \frac{n(<m)}{n_{0}(<m)},
\end{equation}
where $n(<m)$ and $n_{0}(<m)$ represent the lensed and unlensed surface
number densities of background galaxies brighter than the apparent magnitude
$m$, respectively, and $s\equiv \dif\log n_{0}(<m)/\dif{m}$ is the
logarithmic slope of the cumulative magnitude distribution.
It has been shown that, with a sizable sample of galaxy clusters, this effect can
be solely used to calibrate the cluster mass proxies \citep[e.g.,][]{hildebrandt09,ford12,chiu16b,tudorica17}.
By combining complementary observables of shear and magnification, one
can break the mass-sheet degeneracy \citep{broadhurst05b,umetsu08}.   
In this work, we combine both observables to derive an unbiased
convergence map for each individual cluster (see
Section~\ref{sec:weak_lensing_data}).

In the regime of strong lensing, detailed modeling with many sets of
multiple images with known redshifts allows to determine the location of
critical curves, which then returns robust estimates of the Einstein
mass, that is, the projected mass enclosed by the critical area
$A_\mathrm{c}$ of an effective Einstein radius
$\theta_{\mathrm{Ein}}=\sqrt{A_\mathrm{c}/\pi}$ \citep{zitrin15}, 
\begin{equation}
\label{eq:strong_intro}
\Msl(<r) = \sigmacrit{\dl}^2\int_{|\bm{\vartheta}| \leq r }
\kappa(\mathbf{\bm{\vartheta}}) \dif^2\mathbf{\bm{\vartheta}}. 
\end{equation}
In this work, we use strong-lensing constraints in the form of the
enclosed projected mass profile around the effective Einstein
radius. These constraints were obtained by
\citet{umetsu16} using detailed lens models constructed by
\citet{zitrin15} from a combined strong and weak lensing analysis of
{\em Hubble Space Telescope} (\HST) observations.
We give further details in Section~\ref{sec:strong_lensing_data}.

%
%

\section{Cluster Sample and Data}
\label{sec:sample_and_data}

We first describe the cluster sample in Section~\ref{sec:sample}. 
The data products of weak and strong lensing are
presented in Section~\ref{sec:weak_lensing_data} and
Section~\ref{sec:strong_lensing_data}, respectively. 

\subsection{Cluster Sample}
\label{sec:sample}

In this work, we study a sample of 16 X-ray regular and 4
high-magnification galaxy clusters targeted by the CLUMP-3D program
\citep{sereno17,sereno18,umetsu18}.
Our sample stems from the \CLASH\ wide-field weak-lensing analysis of
\citet{umetsu14}, and comprises two subsmples, both taken from the
CLASH survey \citep{postman12}
targeting 25 high-mass clusters.
Here, 20 clusters in the first CLASH subsample were selected to have
X-ray temperatures greater than 5\,keV and to have regular X-ray
morphology. Numerical simulations suggest that this subsample is
largely composed of relaxed clusters and free of orientation
bias \citep{meneghetti14}.
The second subset of 5 clusters were selected for their high
magnification properties. These clusters turn out to be dynamically
disturbed, merging systems \citep{zitrin13,medezinski13,balestra16,jauzac17}.
Accordingly, modeling with a single-halo component may not be
adequate to describe the high-magnification subsample \citep{medezinski13},
in contrast to the X-ray-selected subsample that can be well
described by a single Navarro--Frenk--White \citep[][hereafter
NFW]{navarro1997} profile out to large cluster radii 
\citep{umetsu16,umetsu17}.
For the sake of homogeneity, however, we analyze all clusters in the
full sample in a consistent manner. We will also split
the sample into several subsamples and statistically characterize each
of them (see Section~\ref{sec:fitting_strategy}).

This sample spans a factor of $\approx 5$ in mass \citep[$4\times
10^{14}\Msun < \Mtwooo < 20\times10^{14}\Msun$;][]{umetsu16} and a
redshift range of $0.18 < \redshift < 0.69$.
Following \citet{umetsu14}, we adopt the location of the
brightest cluster galaxy (BCG) as the center for each cluster. 
As discussed in \cite{umetsu14}, the rms of positional offsets between
the BCGs and X-ray peaks for the full sample is
$\approx30$\,kpc\,$h^{-1}$, and it reduces to
$\lesssim10$\,kp\,$h^{-1}$ for the X-ray-selected subsample.
Therefore, the effect of miscentering is not expected to be significant
in this work \citep{johnston07,umetsu11,umetsu16}.
We tabulate the basic information of our 20 clusters in
Table~\ref{tab:basics}.

We note that this sample has been intensively studied in previous
\CLASH\ work, especially by \citet[][hereafter U14]{umetsu14} and
\citet[][hereafter U16]{umetsu16}, who performed recnstructions of the 
azimuthally averaged surface mass density profile from weak and
weak+strong lensing data, respectively. Accordingly, both U14 and U16
focused on spherical mass estimates of these clusters.
In this work, we analyze {\em HST} Einstein-mass constraints in
combination with 2D weak-lensing mass maps of \citet{umetsu18}
reconstructed from a joint analysis of 2D shear and
azimuthally averaged magnification constraints.
We extend the analyses of U14 and U16 to investigate the 3D structure
and shape of the 20 clusters using combined strong and weak
lensing data sets.

\begin{table}
\centering
\caption{Basic information of the cluster sample}
\label{tab:basics}
\begin{tabular}{lccc}
\hline\hline
Name & Redshift & $\alpha_{\mathrm{BCG}}$ & $\delta_{\mathrm{BCG}}$ \\
\hline\hline
    Abell~383    & 0.187  & 02:48:03.40 & $-$03:31:44.9  \\[3pt] 
    Abell~209    & 0.206  & 01:31:52.54 & $-$13:36:40.4  \\[3pt] 
    Abell~2261    & 0.224  & 17:22:27.18 & $+$32:07:57.3  \\[3pt] 
    RX~J2129$+$0005    & 0.234  & 21:29:39.96 & $+$00:05:21.2  \\[3pt] 
    Abell~611    & 0.288  & 08:00:56.82 & $+$36:03:23.6  \\[3pt] 
    MS2137$-$2353    & 0.313  & 21:40:15.17 & $-$23:39:40.2  \\[3pt] 
    RX~J2248$-$4431    & 0.348  & 22:48:43.96 & $-$44:31:51.3  \\[3pt] 
    MACS~J1115$+$0129    & 0.352  & 11:15:51.90 & $+$01:29:55.1  \\[3pt] 
    MACS~J1931$-$2635    & 0.352  & 19:31:49.62 & $-$26:34:32.9  \\[3pt] 
    RX~J1532$+$3021    & 0.363  & 15:32:53.78 & $+$30:20:59.4  \\[3pt] 
    MACS~J1720$+$3536    & 0.391  & 17:20:16.78 & $+$35:36:26.5  \\[3pt] 
    MACS~J0429$-$0253    & 0.399  & 04:29:36.05 & $-$02:53:06.1  \\[3pt] 
    MACS~J1206$-$0847    & 0.440  & 12:06:12.15 & $-$08:48:03.4  \\[3pt] 
    MACS~J0329$-$0211    & 0.450  & 03:29:41.56 & $-$02:11:46.1  \\[3pt] 
    RX~J1347$-$1145    & 0.451  & 13:47:31.05 & $-$11:45:12.6  \\[3pt] 
    MACS~J0744$+$3927    & 0.686  & 07:44:52.82 & $+$39:27:26.9  \\[3pt] 
    MACS~J0416$-$2403    & 0.396  & 04:16:08.38 & $-$24:04:20.8  \\[3pt] 
    MACS~J1149$+$2223    & 0.544  & 11:49:35.69 & $+$22:23:54.6  \\[3pt] 
    MACS~J0717$+$3745    & 0.548  & 07:17:32.63 & $+$37:44:59.7  \\[3pt] 
    MACS~J0647$+$7015    & 0.584  & 06:47:50.27 & $+$70:14:55.0  \\[3pt] 

\hline\hline
\end{tabular}
\tablecomments{
The right ascension $\alpha_{\mathrm{BCG}}$ and declination
 $\delta_{\mathrm{BCG}}$ of the BCG position are adopted as the cluster
 center.
The first 16 clusters are taken from the \CLASH\ X-ray-selected subsample,
 while the other 4 clusters are from the \CLASH\ high-magnification subsample.
}
\end{table} 

\subsection{Weak-lensing Data}
\label{sec:weak_lensing_data}

In this section, we briefly summarize the weak-lensing data products
used in this study, and refer the reader to our companion paper \citep{umetsu18} for full details.
Our weak-lensing analysis is based on deep 
multi-band imaging taken primarily with Suprime-Cam \citep{miyazaki12}
on the Subaru Telescope \citep[typically, 5 Suprime-Cam bands; Table 1 of][]{umetsu14},
as obtained by the \CLASH\ collaboration \citep{umetsu14}.  
For our southernmost cluster (RX~J2248$-$4431),
we used data taken with the Wide-Field Imager at the ESO 2.2\,m MPG/ESO
telescope at La Silla \citep{gruen13}. 
General data products from the \CLASH\ survey, including the reduced
Subaru/Suprime-Cam data, weight maps, and photometric catalogs, are
available at the Mikulski Archive for Space Telescopes 
(MAST)\footnote{\href{https://archive.stsci.edu/prepds/clash/}{https://archive.stsci.edu/prepds/clash/}}.
Details of the image reduction, photometry, background galaxy
selection, and the creation of weak-lensing shear catalogs are presented 
in \citet{umetsu14}.

In our companion paper, \cite{umetsu18} have presented a 2D
weak-lensing analysis for the 20 \CLASH\ clusters using the
background-selected shear catalogs and azimuthaly averaged magnification
profiles, both published in \citet{umetsu14}.
In this study, we use pixelized 2D surface mass density maps 
obtained in \citet{umetsu18} as our weak-lensing constraints.
For each cluster, the mass map is pixelized on a regular 
grid of $48\times 48$ pixels covering the central $24\arcmin\times
24\arcmin$ region.
\citet{umetsu18} accounted for various sources of errors associated
with their weak-lensing shear and magnification measurements (see their
Section 3), including the covariance due to uncorrelated large-scale
structures projected along the line of sight. All these errors are
encoded in the covariance matrix used in our analysis.

As summarized in Section 5.1 of \cite{umetsu18},
we quantified major sources of systematic errors 
in the \CLASH\ weak-lensing analysis. In particular, we consider the
following systematic effects:
(1) dilution of the lensing signal caused by residual contamination from
cluster members ($2.4\percent\pm 0.7\percent$),
(2) photometric-redshift bias in estimates of the mean lensing depth
($0.27\percent$), and (3) uncertainty in the shear calibration factor
($5\percent$). These errors add to $5.6\percent$ in 
quadrature. This corresponds to the mass calibration uncertainty of
$5.6\percent/\Gamma \simeq 7\percent$ with $\Gamma\simeq 0.75$ being the typical value of 
the logarithmic derivative of the lensing signal with respect to
cluster mass \citep{melchior17}.

On the other hand, by performing a shear--magnification consistency
test, \citet{umetsu14} estimated a systematic uncertainty in the \CLASH\
mass calibration to be $8\percent$. In the CLUMP-3D program,
we conservatively use this value as the systematic uncertainty in the
ensemble mass calibration.

\subsection{Strong-lensing Data}
\label{sec:strong_lensing_data}

\citet{zitrin15} obtained detailed lens models for the \CLASH\ sample
using two different parameterizations, one assuming that light traces mass
for both DM and galaxy components, and the other using an analytical 
elliptical NFW form for the DM-halo components.
Here we include {\em HST} lensing constrints of \citet{zitrin15} to improve
modeling of cluster cores, which are unresolved by the wide-field
weak-lensing observations.
Full details of data acquisition, reduction, and analysis of {\em HST}
lensing data are fully given in \cite{zitrin15} and U16,
to which we refer the reader for more details.

Here we give a brief summary of our {\em HST} lensing data.
Specifically,
for each cluster except RX~J1532.9$+$3021 for
which no secure identification of multiple images has been
made \citep{zitrin15},
we use enclosed projected mass
constraints $\Msl(<r)$ for a set of four fixed integration radii,
$r=10\arcsec, 20\arcsec, 30\arcsec$, and $40\arcsec$. These constraints
are presented in Table~1 of \citet{umetsu16}.
The measurement errors $\sigma_{\Msl(<r)}$ include
systematic as well as statistical uncertainties, by accounting for
modeling discrepancies between the two modeling methods of \citet{zitrin15}.
The integrated signal-to-noise ratio of the enclosed mass constraints is
on average $\approx12$, comparable to that of the weak lensing constraints
\citet{umetsu14}.

%
%

%
\begin{table*}
\caption{Prior Distributions}
\label{tab:priors}
\centering
\resizebox{1.0\textwidth}{!}{%
\begin{tabular}{lccccccc}
\hline\hline
Modeling &$\log\left(\Mtwooo\right)$ &$\log\left(\Ctwooo\right)$ &\qa &\qb &$\cos\theta$ &$\phi$ &$\psi$ \\
\hline\hline
\sphericalmodel\ &$\mathcal{U}(14,16)$  &$\mathcal{U}(-1,1)$  & --  & --
		 & --  & --  & --  \\
\triaxialfmodel\    &$\mathcal{U}(14,16)$  &$\mathcal{U}(-1,1)$  &Equation~(\ref{eq:shape_prior_1})  &Equation~(\ref{eq:shape_prior_2})  &$\mathcal{U}(0,1)$  &$\mathcal{U}(-\pi/2,\pi/2)$  &$\mathcal{U}(-\pi/2,\pi/2)$  \\
\triaxialbmodel        &$\mathcal{U}(14,16)$  &$\mathcal{U}(-1,1)$  &\cite{bonamigo15}  &\cite{bonamigo15}  &$\mathcal{U}(0,1)$  &$\mathcal{U}(-\pi/2,\pi/2)$  &$\mathcal{U}(-\pi/2,\pi/2)$  \\
\hline\hline
\end{tabular}
}
\tablecomments{
Uniform priors of $\mathcal{U}(14,16)$ and $\mathcal{U}(-1,1)$ are used
 for $\log\left(\Mtwooo\right)$ and $\log\left(\Ctwooo\right)$,
 respectively. Masses are expressed in the units of \Msun. All
 parameters other than $\Mtwooo$ and $\Ctwooo$ are fixed in \sphericalmodel\ modeling.
In \triaxialfmodel\ modeling, we use uniform priors on the shape
 (\qa\ and \qb) and orientation ($\cos\theta$, $\phi$, and $\psi$) parameters. 
In \triaxialbmodel\ modeling we assume informative shape priors taken
 from cosmological numerical simulations of \cite{bonamigo15}, while
 keeping the uniform priors for the orientation parameters.  
}
\end{table*}

\section{Methodology}
\label{sec:methodology}

In this section, we describe our methodology for triaxial modeling of
galaxy clusters. We first describe the formalism for halo
modeling in Section~\ref{sec:modeling}, and outline Bayesian methods in
Section~\ref{sec:likelihood_construction}.    
In Section~\ref{sec:fitting_strategy}, we perform Bayesian inference for
individual and ensemble clusters using the combined weak and strong
lensing data sets. 
In Section~\ref{sec:sr_fitting}, we examine scaling relations with halo
mass for our clusters.

\subsection{Halo Modeling}
\label{sec:modeling}

In this section, we describe triaxial halo modeling based on the 2D
weak-lensing and central \HST\ lensing data sets.
To this end, we closely follow the forward-modeling approach of
\cite{umetsu15} and \cite{sereno17}.
Specifically, we forward-model the projected cluster lensing
observations by projecting a triaxial NFW halo \citep{corless09b} along
the line of sight.

The density profile of the triaxial NFW model is written as a function
of the ellipsoidal radius $R$ as
\begin{equation}
\label{eq:triaxial_nfw}
\rho(R) = \frac{\rhos}{\left(R/\Rs\right)\left(1 + R/\Rs\right)^2},
\end{equation}
where \rhos\ is the characteristic densty, and \Rs\ is the ellipsoidal
scale radius measured along the major axis of the halo ellipsoid.
The ellipsoidal radius $R$ is related to the principal coordinates
($X,Y,Z$) centered on the cluster as
\begin{equation}
\label{eq:ellipsoid}
R^2 = \frac{X^2}{\qa^2} + \frac{Y^2}{\qb^2} + Z^2,
\end{equation}
with \qa\ the minor-to-major axis ratio  and \qb\ the
intermediate-to-major axis ratio.
By definition, we have $0< \qa \leq \qb \leq
1$. Equation~(\ref{eq:triaxial_nfw}) reduces to the spherical NFW model
if $\qa = \qb = 1$. 

The \Mtwooo\ mass and the \Rtwooo\ radius for a cluster at redshift \zd\
are related to each other by
\begin{equation}
\label{eq:def_mass}
\Mtwooo = \frac{4\pi}{3}200\rhocrit(\zd)\qa\qb\Rtwooo^3.
\end{equation}
On the other hand, \Mtwooo\ can be expressed as
\begin{equation}
\label{eq:def_mass_int}
\Mtwooo = 4\pi\qa\qb \int_{0}^{\Rtwooo}\! \rho(R) R^2 \dif R.
\end{equation}
We define the concentration parameter \Ctwooo\ as the ratio of the
cluster radius to the scale radius along the major axis,
\begin{equation}
\label{eq:def_concen}
\Ctwooo \equiv \frac{\Rtwooo}{\Rs}.
\end{equation}
Combining Equations~(\ref{eq:def_mass}), (\ref{eq:def_mass_int}), and
(\ref{eq:def_concen}), one can express \rhos\ as 
\begin{equation}
\label{eq:def_rhos}
\rhos = \frac{200\rhocrit(\zd)}{3} \frac{{\Ctwooo}^3}{\ln\left(1 + \Ctwooo\right) - \Ctwooo/(1 + \Ctwooo)}.
\end{equation}
We specify the radial density profile of the triaxial NFW model
(see Equation~(\ref{eq:triaxial_nfw})) with $(\Mtwooo, \Ctwooo)$, instead of
$(\rho_\mathrm{s}, r_\mathrm{s})$.

A triaxial halo is projected onto the lens plane as elliptical
isodensity contours \citep{stark77}, which can be specified by the
intrinsic axis ratios ($q_a,q_b$) and 
orientation angles ($\theta,\phi,\psi$) defined with respect to the line 
of sight of the observer. Following \citet{umetsu15} and
\citet{sereno17b}, we adopt the $z$-$x$-$z$ convention of Euler angles \citep{stark77}.
The angle $\theta$ describes the inclination of the major ($Z$) axis with
respect to the line of sight.

After a coordinate transformation of the first two Euler angles, 
elliptical isodensity contours of the projected ellipsoid can be
described as a function of the elliptical radius $\zeta$ defined in
terms of the observer's sky coordinates ($\mathcal{X},\mathcal{Y}$) as
\begin{equation}
 \label{eq:elliptical}
  \begin{aligned}
\zeta^2 &= \frac{1}{f} \left( j \mathcal{X}^2 + 2k\mathcal{X}\mathcal{Y} + l\mathcal{Y}^2\right),\\
j &= \cos^2\theta\left(\frac{\cos^2\phi}{\qa^2} + \frac{\sin^2\phi}{\qb^2} \right) + \frac{\sin^2\theta}{\qa^2\qb^2},\\
k &= \sin\phi\cos\phi\cos\theta\left(\frac{1}{\qa^2} - \frac{1}{\qb^2}\right),\\
l &= \left(\frac{\sin^2\phi}{\qa^2} + \frac{\cos^2\phi}{\qb^2}\right),\\
f &= \sin^2\theta \left( \frac{\sin^2\phi}{\qa^2} + \frac{\cos^2\phi}{\qb^2}\right) + \cos^2\theta.
  \end{aligned}
\end{equation}
The third Euler angle $\psi$ represents the rotational degree of freedom
in the sky plane to specify the observer's coordinate system.

To sum up, our triaxial NFW model is specified by seven parameters,
namely, halo mass and concentration ($\Mtwooo,\Ctwooo$),
intrinsic axis ratios ($\qa,\qb$) characterizing the intrinsic halo shape,
and three Euler angles ($\theta, \phi, \psi$) describing the halo
orientation with respect to the line of sight.
In this way, for a given set of the parameters,
we can project a triaxial NFW halo onto the lens plane and
compute the surface mass density at each position.

In this work, we pay a special attention to two geometric quantities
that characterize the intrinsic shape and orientation of clusters.
The first quantity is a geometrical factor \fgeo\ that describes the
degree of elongation of the cluster mass distribution along the line of sight.
Specifically, a cluster halo is elongated along the line of sight if
$f_\mathrm{geo}>1$, while it is elongated in the plane of the sky if
$f_\mathrm{geo}<1$. 
We stress that the case of $\fgeo=1$ does not necessarily correspond to a
spherical halo configuration, but indicates that the halo sizescale
along the line of sight is eqaul to that in the plane of the sky.
Following \citet{sereno10} and \citet{umetsu15}, we define \fgeo\ by
\begin{equation}
\label{eq:fgeo_def}
\begin{aligned}
\fgeo &\equiv \frac{ L_{\parallel} }{ \xi_{\mathrm{s}}\sqrt{q_{\bot}} },\\
L_{\parallel} &= \xi_{\mathrm{s}}\left(\frac{q_{\bot}}{\qa\qb}\right)^{\frac{1}{2}} f^{-\frac{3}{4}},\\
q_{\bot} &= \left(\frac{j + l - \sqrt{(j-l)^2 + 4k^2}}{j + l + \sqrt{(j-l)^2 + 4k^2}}\right)^{\frac{1}{2}},
\end{aligned}
\end{equation}
where
$q_{\bot}$ is the minor-to-major axis ratio of the projected ellipsoid,
$L_\parallel$ represents the line of sight half length of the ellipsoid
of ellipsoidal radius $R=R_\mathrm{s}$,
$\xi_{\mathrm{s}}$ is the projected scale radius (semi-major axis) in the sky plane.
That is, for a given ellipsoid, \fgeo\ is the ratio of the line-of-sight
half length $L_{\parallel}$ to the geometric mean of the semi-major and
semi-minor axes of the projected isodensity contour.

The second quantity of interest is the degree of triaxiality, \triaxiality.
Following the definition in \cite{umetsu15}, the triaxiality is defined by
\begin{equation}
\label{eq:triaxiality_def}
\triaxiality \equiv \frac{1 - {\qb}^2}{1 - {\qa}^2} \, .
\end{equation}
By construction, $0 \leq \triaxiality \leq 1$.
The degree of triaxiality \triaxiality\ approaches unity, or $\qa=\qb$
(zero, or $q_b=1$) if the halo shape is maximally prolate (oblate).

\subsection{Bayesian Inference}
\label{sec:likelihood_construction}

In what follows, we describe the Bayesian inference formalism that we
used to explore parameter space given the combined weak and strong
lensing data sets.
The joint posterior probability distribution of model parameters
$\mathbf{p}$ given data $\mathcal{D}$ is written as
\begin{equation}
\label{eq:likelihood}
P(\mathbf{p}|\mathcal{D}) \propto  
\mathcal{L}(\mathcal{D}|\mathbf{p}) \mathcal{P(\mathbf{p})},
\end{equation}
where $\mathcal{P(\mathbf{p})}$ denotes the prior distribution of
$\mathbf{p}$. In this work, our model includes up to seven parameters,
namely, $(\Mtwooo, \Ctwooo, \qa, \qb, \theta, \phi, \psi)$, depending on
modeling approaches (see below).

The likelihood $\mathcal{L}(\mathcal{D}|\mathbf{p})$ describes the
probability of observing data $\mathcal{D}$ given the model
$\mathbf{p}$. Here we explicitly express the data as 
\begin{equation}
\mathcal{D} = \left\lbrace\mathcal{D}_{\mathrm{SL}}, \mathcal{D}_{\mathrm{WL}} \right\rbrace,
\end{equation}
with $\mathcal{D}_{\mathrm{SL}} = \left\lbrace \Msl\left(<r\right) | r = 10\arcsec, \cdots, 40\arcsec\right\rbrace$ 
the data vector containing a set of encloesd projected mass constraints,
and $\mathcal{D}_{\mathrm{WL}}$ the concatenated data vector containing
pixelized values of the weak-lensing mass map.
For each cluster, we evaluate the log-likelihood
\begin{equation}
\label{eq:likelihood2}
\ln\mathcal{L}(\mathcal{D}|\mathbf{p}) =
\ln\mathcal{L}_{\mathrm{SL} }(\mathcal{D}_{\mathrm{SL} }|\mathbf{p}) + 
\ln\mathcal{L}_{\mathrm{WL}}(\mathcal{D}_{\mathrm{WL}}|\mathbf{p}),
\end{equation}
where
\begin{equation}
 \label{eq:likelihood3}
 \begin{aligned}
\ln\mathcal{L}_{\mathrm{SL} }(\mathcal{D}_{\mathrm{SL}}|\mathbf{p}) &= 
-\frac{1}{2}\sum_{i}\left(\frac{\Msl(<r_{i}) -  \mathcal{M}_{\mathrm{2D}}(<r_{i}) }{\sigma_{\Msl(<r_{i})}}\right)^2,\\
\ln\mathcal{L}_{\mathrm{WL}}(\mathcal{D}_{\mathrm{WL}}|\mathbf{p}) &= 
-\frac{1}{2}\left(\mathcal{D}_{\mathrm{WL}} - \mathcal{M}\right)^\mathrm{T}
\cdot\mathfrak{C}^{-1}\cdot
\left(\mathcal{D}_{\mathrm{WL}} - \mathcal{M}\right),
 \end{aligned}
\end{equation}
where the index $i$ runs over the four strong-lensing constraints,
$\sigma_{\Msl(<r_{i})}$ is the uncertainty in the enclosed projected
mass estimate $\Msl(<r_i)$,
$\mathcal{M}_{\mathrm{2D}}(<r_{i})$ is the model prediction,
and
$\mathfrak{C}$ represents the error covariance matrix for the
weak-lensing data.
In this work, we fit weak-lensing data across the entire
$24\arcmin\times 24\arcmin$ region centered on the cluster. 
We checked that restricting the fitting range to the central
$4\mathrm{Mpc}/h\times4\mathrm{Mpc}/h$ region (side length corrsponding
approximately to twice the virial radius) 
does not significantly
change the results, indicating that our analysis is not sensitive to the
2-halo term. The enclosed projected mass measurements from the {\em HST}
lensing analysis impose a set of integrated constraints on the inner
density profile. We note that, by doing this, no assumption is made of
azimuthal symmetry or isotropy of the underlying mass distribution.

We use the python implementation of the affine-invariant ensemble Markov
Chain Monte Carlo (MCMC) sampler, \texttt{emcee} \citep{foreman13}, to
explore parameter space.  
We consider the following three different modeling approaches:
(1) spherical modeling with uniform priors on $\log\Mtwooo$ and $\log\Ctwooo$,
(2) triaxial modeling with uniform priors on all parameters, and 
(3) triaxial modeling incorporating informative shape priors from
cosmological $N$-body simulations \citep[][hereafter B15]{bonamigo15}.
For simplicity, we refer to these three approaches as
\sphericalmodel,
\textsf{Triaxial},
and \triaxialbmodel\ modeling, respectively.
Here we briefly describe each case.
\begin{itemize}
\item \sphericalmodel\ modeling: We float only two parameters
       $(\Mtwooo,\Ctwooo)$ and fix the remaining parameters
      ($\qa\  = \qb = 1$ and $\theta = \phi = \psi = 0$).

\item \triaxialfmodel\ modeling: We use  uniform priors on
       $\log\Mtwooo$, $\log\Ctwooo$, intrinsic shapes
       ($\qa, \qb$), and orientation angles
       $(\cos\theta, \phi, \psi)$. 
       We assume the following form of the prior probability
       distribution for the intrinsic axis ratios,
\begin{equation} 
\label{eq:shape_prior_0}
\mathcal{P}(\qa, \qb) = \mathcal{P}(\qb|\qa) \times \mathcal{P}(\qa),
\end{equation}
where
\begin{equation}
\label{eq:shape_prior_1}
\mathcal{P}(\qa) = \begin{cases}
1/\left( 1 - q_{\mathrm{min}} \right) & \mathrm{if}~q_{\mathrm{min}} < \qa < 1\cr
0,  &\mathrm{otherwise} \cr
\end{cases}
\end{equation}
and 
\begin{equation}
\label{eq:shape_prior_2}
\mathcal{P}(\qb|\qa) = \begin{cases}
1/\left( 1 - \qa \right) & \mathrm{if}~\qa \leq \qb < 1\cr
0,  &\mathrm{otherwise} \, , \cr
\end{cases}
\end{equation}
and $q_{\mathrm{min}} = 0.1$ is the lower bound of the minor-to-major axis ratio \citep{sereno11}.

\item \triaxialbmodel\ modeling: We adopt informative shape priors from
      $N$-body simulations of B15, who characterized the distribution of 
      intrinsic axis ratios of $N$-body CDM halos as function of the
      halo peak height and redshift. We self-consistently
      update the shape prior for a given set of $\Mtwooo, \Ctwooo$, and
      redshift.
\end{itemize}
Table~\ref{tab:priors} summarizes the prior distributions assumed in
this study.

\begin{table*}
\centering
\caption{Marginalized Posterior Concstraints on Cluster Model Parameters}
\label{tab:results}
\resizebox{1.0\textwidth}{!}{
\begin{tabular}{lcccccccccc}
\hline\hline
& \multicolumn{2}{c}{\sphericalmodel\ Modeling}
& \multicolumn{4}{c}{\triaxialfmodel\ Modeling}
& \multicolumn{4}{c}{\triaxialbmodel\ Modeling}\\[6pt]
Name 
& $\frac{\Mtwooo}{10^{15}\Msun}$
& $\Ctwooo$
& $\frac{\Mtwooo}{10^{15}\Msun}$
& $\Ctwooo$
& \qa
& \qb
& $\frac{\Mtwooo}{10^{15}\Msun}$
& $\Ctwooo$
& \qa
& \qb \\[6pt]
\hline
\multicolumn{1}{l}{Individual constraints} &\multicolumn{10}{c}{}\\[6pt]
\hline
 		Abell~383  &  $0.385^{+0.149}_{-0.094}$  &  $6.7^{+1.5}_{-1.6}$  &  $0.40^{+0.16}_{-0.11}$  &  $6.9^{+1.6}_{-1.8}$  &  $0.82^{+0.18}_{-0.26}$  & $ >0.446 $ &  $0.41^{+0.16}_{-0.12}$  &  $6.5^{+1.9}_{-1.7}$  &  $0.474^{+0.077}_{-0.098}$  &  $0.60^{+0.14}_{-0.11}$ \\ [3pt]
 		Abell~209  &  $1.21^{+0.27}_{-0.21}$  &  $2.76^{+0.44}_{-0.45}$  &  $1.30^{+0.43}_{-0.30}$  &  $2.86^{+0.69}_{-0.71}$  &  $0.51^{+0.11}_{-0.25}$  &  $0.662^{+0.337}_{-0.029}$  &  $1.32^{+0.42}_{-0.22}$  &  $2.78^{+0.72}_{-0.49}$  &  $0.424^{+0.071}_{-0.075}$  &  $0.523^{+0.141}_{-0.076}$ \\ [3pt]
 		Abell~2261  &  $1.61^{+0.24}_{-0.22}$  &  $3.87^{+0.50}_{-0.47}$  &  $1.66\pm 0.28$  &  $3.86^{+0.70}_{-0.62}$  &  $0.67^{+0.22}_{-0.17}$  & $ >0.491 $ &  $1.53^{+0.41}_{-0.26}$  &  $3.75^{+0.80}_{-0.85}$  &  $0.427^{+0.105}_{-0.065}$  &  $0.63^{+0.12}_{-0.13}$ \\ [3pt]
 		RX~J2129$+$0005  &  $0.46^{+0.13}_{-0.11}$  &  $4.9^{+1.2}_{-1.0}$  &  $0.46^{+0.17}_{-0.12}$  &  $5.0\pm 1.3$  &  $0.75^{+0.25}_{-0.16}$  & $ >0.519 $ &  $0.43^{+0.15}_{-0.10}$  &  $5.1^{+1.2}_{-1.5}$  &  $0.509^{+0.075}_{-0.101}$  &  $0.632^{+0.145}_{-0.096}$ \\ [3pt]
 		Abell~611  &  $0.90^{+0.25}_{-0.19}$  &  $4.36^{+0.96}_{-0.73}$  &  $1.00^{+0.23}_{-0.30}$  &  $4.9^{+1.0}_{-1.2}$  &  $0.48^{+0.19}_{-0.24}$  & $ >0.349 $ &  $0.94^{+0.27}_{-0.24}$  &  $4.50^{+1.29}_{-0.86}$  &  $0.431^{+0.071}_{-0.083}$  &  $0.56^{+0.13}_{-0.11}$ \\ [3pt]
 		MS2137$-$2353  &  $0.53^{+0.18}_{-0.17}$  &  $4.5^{+1.7}_{-1.1}$  &  $0.53^{+0.22}_{-0.16}$  &  $5.2^{+1.3}_{-1.8}$  &  $0.72^{+0.28}_{-0.14}$  & $ >0.487 $ &  $0.51^{+0.20}_{-0.16}$  &  $5.2^{+1.4}_{-2.0}$  &  $0.474^{+0.093}_{-0.090}$  &  $0.64^{+0.10}_{-0.13}$ \\ [3pt]
 		RX~J2248$-$4431  &  $1.02^{+0.27}_{-0.30}$  &  $4.14^{+1.25}_{-0.90}$  &  $1.02^{+0.41}_{-0.26}$  &  $3.78^{+2.02}_{-0.67}$  &  $0.46^{+0.24}_{-0.23}$  & $ >0.361 $ &  $1.00^{+0.37}_{-0.27}$  &  $4.5^{+1.5}_{-1.3}$  &  $0.384^{+0.113}_{-0.045}$  &  $0.592^{+0.078}_{-0.156}$ \\ [3pt]
 		MACS~J1115$+$0129  &  $1.26^{+0.24}_{-0.26}$  &  $2.89^{+0.53}_{-0.58}$  &  $1.21^{+0.41}_{-0.25}$  &  $2.90^{+0.99}_{-0.65}$  &  $0.45^{+0.17}_{-0.22}$  & $ >0.364 $ &  $1.28^{+0.32}_{-0.29}$  &  $3.14^{+0.84}_{-0.72}$  &  $0.413^{+0.078}_{-0.072}$  &  $0.541^{+0.134}_{-0.095}$ \\ [3pt]
 		MACS~J1931$-$2635  &  $0.60^{+0.18}_{-0.12}$  &  $7.3^{+1.6}_{-1.5}$  &  $0.65^{+0.19}_{-0.16}$  &  $7.6^{+1.6}_{-1.7}$  &  $0.906^{+0.072}_{-0.426}$  & $ >0.399 $ &  $0.63^{+0.23}_{-0.15}$  &  $7.8^{+1.7}_{-1.6}$  &  $0.449^{+0.081}_{-0.087}$  &  $0.625^{+0.095}_{-0.149}$ \\ [3pt]
 		RX~J1532$+$3021  &  $0.479^{+0.110}_{-0.092}$  &  $6.7^{+2.4}_{-1.2}$  &  $0.472^{+0.121}_{-0.089}$  &  $6.9^{+1.9}_{-1.6}$  &  $0.936^{+0.061}_{-0.299}$  & $ >0.480 $ &  $0.479^{+0.087}_{-0.142}$  &  $5.91^{+3.01}_{-0.76}$  &  $0.464^{+0.093}_{-0.084}$  &  $0.593^{+0.158}_{-0.096}$ \\ [3pt]
 		MACS~J1720$+$3536  &  $0.79^{+0.14}_{-0.18}$  &  $4.88^{+1.12}_{-0.71}$  &  $0.72^{+0.25}_{-0.13}$  &  $4.96^{+1.42}_{-0.99}$  &  $0.57^{+0.32}_{-0.16}$  & $ >0.471 $ &  $0.80^{+0.20}_{-0.19}$  &  $5.4^{+1.0}_{-1.4}$  &  $0.446^{+0.085}_{-0.073}$  &  $0.556^{+0.161}_{-0.065}$ \\ [3pt]
 		MACS~J0429$-$0253  &  $0.60^{+0.13}_{-0.12}$  &  $5.6^{+1.1}_{-1.0}$  &  $0.562^{+0.182}_{-0.096}$  &  $5.7^{+1.3}_{-1.2}$  &  $0.934^{+0.064}_{-0.289}$  & $ >0.491 $ &  $0.61^{+0.11}_{-0.15}$  &  $5.9^{+1.2}_{-1.5}$  &  $0.464^{+0.094}_{-0.080}$  &  $0.63^{+0.12}_{-0.13}$ \\ [3pt]
 		MACS~J1206$-$0847  &  $1.01^{+0.19}_{-0.20}$  &  $5.21^{+0.84}_{-1.08}$  &  $0.99^{+0.28}_{-0.18}$  &  $5.0^{+1.3}_{-1.1}$  &  $0.63^{+0.36}_{-0.11}$  & $ >0.480 $ &  $0.96^{+0.27}_{-0.19}$  &  $5.1\pm 1.2$  &  $0.426^{+0.095}_{-0.067}$  &  $0.569^{+0.136}_{-0.096}$ \\ [3pt]
 		MACS~J0329$-$0211  &  $0.706^{+0.142}_{-0.095}$  &  $5.70^{+1.20}_{-0.85}$  &  $0.86^{+0.26}_{-0.14}$  &  $5.0^{+1.6}_{-1.1}$  &  $0.34^{+0.14}_{-0.13}$  &  $0.526^{+0.360}_{-0.080}$  &  $0.87^{+0.23}_{-0.13}$  &  $5.6\pm 1.1$  &  $0.396^{+0.067}_{-0.057}$  &  $0.507^{+0.112}_{-0.092}$ \\ [3pt]
 		RX~J1347$-$1145  &  $2.18^{+0.34}_{-0.28}$  &  $4.09^{+0.61}_{-0.50}$  &  $2.64^{+0.64}_{-0.56}$  &  $3.44^{+1.13}_{-0.68}$  &  $0.38^{+0.13}_{-0.16}$  &  $0.499^{+0.346}_{-0.060}$  &  $2.65^{+0.62}_{-0.37}$  &  $3.82^{+0.82}_{-0.68}$  &  $0.372^{+0.065}_{-0.055}$  &  $0.510^{+0.092}_{-0.105}$ \\ [3pt]
 		MACS~J0744$+$3927  &  $1.64^{+0.41}_{-0.33}$  &  $2.68^{+0.65}_{-0.56}$  &  $1.37^{+0.68}_{-0.24}$  &  $4.02^{+0.78}_{-1.44}$  &  $0.25^{+0.13}_{-0.12}$  & $ >0.242 $ &  $1.69^{+0.37}_{-0.41}$  &  $3.63^{+1.11}_{-0.85}$  &  $0.358^{+0.053}_{-0.068}$  &  $0.452^{+0.104}_{-0.092}$ \\ [3pt]
 		MACS~J0416$-$2403  &  $0.87^{+0.21}_{-0.14}$  &  $2.69^{+0.42}_{-0.46}$  &  $0.88^{+0.31}_{-0.14}$  &  $2.67^{+0.58}_{-0.62}$  &  $0.71^{+0.22}_{-0.26}$  & $ >0.445 $ &  $0.84^{+0.32}_{-0.13}$  &  $2.64^{+0.63}_{-0.61}$  &  $0.437^{+0.096}_{-0.075}$  &  $0.60^{+0.13}_{-0.12}$ \\ [3pt]
 		MACS~J1149$+$2223  &  $1.84^{+0.43}_{-0.32}$  &  $1.94^{+0.39}_{-0.42}$  &  $1.88^{+0.67}_{-0.45}$  &  $2.04^{+0.59}_{-0.52}$  &  $0.35^{+0.16}_{-0.15}$  & $ >0.314 $ &  $2.01^{+0.59}_{-0.46}$  &  $2.21^{+0.54}_{-0.56}$  &  $0.372^{+0.071}_{-0.060}$  &  $0.50\pm 0.10$ \\ [3pt]
 		MACS~J0717$+$3745  &  $2.33^{+0.37}_{-0.34}$  &  $1.34^{+0.25}_{-0.16}$  &  $2.36^{+0.88}_{-0.61}$  &  $1.38^{+0.54}_{-0.34}$  &  $0.363^{+0.080}_{-0.176}$  & $ >0.292 $ &  $2.63^{+0.70}_{-0.48}$  &  $1.62^{+0.36}_{-0.31}$  &  $0.351^{+0.064}_{-0.057}$  &  $0.471^{+0.097}_{-0.098}$ \\ [3pt]
 		MACS~J0647$+$7015  &  $1.02^{+0.30}_{-0.21}$  &  $3.49^{+1.01}_{-0.80}$  &  $1.06^{+0.27}_{-0.30}$  &  $3.7^{+1.2}_{-1.0}$  & $ >0.193 $ & $ >0.452 $ &  $0.99^{+0.32}_{-0.25}$  &  $3.53^{+1.62}_{-0.85}$  &  $0.441^{+0.070}_{-0.091}$  &  $0.608^{+0.097}_{-0.132}$ \\ [3pt]
\hline
\multicolumn{1}{l}{Joint ensemble constraints} &\multicolumn{10}{c}{}\\[6pt]
\hline
 		Full  &  $1.089^{+0.050}_{-0.052}$  &  $3.42^{+0.14}_{-0.15}$  &  $1.07^{+0.11}_{-0.13}$  &  $3.26^{+0.71}_{-0.11}$  &  $0.652^{+0.162}_{-0.078}$  & $ >0.632 $ &  $1.027^{+0.111}_{-0.100}$  &  $3.64^{+0.40}_{-0.24}$  &  $0.499^{+0.018}_{-0.056}$  &  $0.636^{+0.078}_{-0.045}$ \\ [3pt]
 		Low-mass  &  $0.721^{+0.052}_{-0.051}$  &  $4.39^{+0.30}_{-0.26}$  &  $0.718^{+0.090}_{-0.092}$  &  $4.28^{+0.71}_{-0.39}$  &  $0.63^{+0.13}_{-0.16}$  & $ >0.613 $ &  $0.659^{+0.128}_{-0.038}$  &  $4.72^{+0.58}_{-0.45}$  &  $0.467^{+0.061}_{-0.030}$  &  $0.666^{+0.074}_{-0.070}$ \\ [3pt]
 		High-mass  &  $1.602^{+0.096}_{-0.095}$  &  $2.73^{+0.14}_{-0.15}$  &  $1.65^{+0.15}_{-0.16}$  &  $2.96^{+0.35}_{-0.20}$  &  $0.481^{+0.089}_{-0.060}$  & $ >0.510 $ &  $1.68^{+0.11}_{-0.13}$  &  $3.07^{+0.26}_{-0.18}$  &  $0.409^{+0.023}_{-0.025}$  &  $0.509^{+0.059}_{-0.021}$ \\ [3pt]
 		X-ray selected  &  $0.962^{+0.049}_{-0.052}$  &  $4.18^{+0.20}_{-0.19}$  &  $0.99\pm 0.11$  &  $3.87^{+0.76}_{-0.11}$  &  $0.541^{+0.188}_{-0.092}$  & $ >0.631 $ &  $1.050^{+0.048}_{-0.144}$  &  $4.42^{+0.41}_{-0.40}$  &  $0.466^{+0.035}_{-0.028}$  &  $0.654^{+0.054}_{-0.061}$ \\ [3pt]
 		High magnification  &  $1.53^{+0.13}_{-0.16}$  &  $2.03^{+0.19}_{-0.16}$  &  $1.66^{+0.14}_{-0.27}$  &  $1.937^{+0.406}_{-0.097}$  & $ >0.323 $ & $ >0.498 $ &  $1.73^{+0.18}_{-0.20}$  &  $2.31^{+0.25}_{-0.23}$  &  $0.408^{+0.026}_{-0.045}$  &  $0.532^{+0.058}_{-0.056}$ \\ [3pt]

\hline\hline
\end{tabular}
}
\tablecomments{
The first column lists the cluster name, followed by marginalized
 posterior constraints on respective parameters from \sphericalmodel,
 \triaxialfmodel\ and \triaxialbmodel\ modeling.
The cluster masses are expressed in the unit of $10^{15}\Msun$.
The first twenty rows show the results of individual modeling, and the
 last five rows show the results from joint ensemble modeling.
We provide $2\sigma$ lower limits on the axis-ratio parameters when they
 are ill-constrained.
}
\end{table*}

\subsection{Modeling Strategy}
\label{sec:fitting_strategy}

In this study, we perform both individual and joint ensemble modeling of
clusters. For the latter case, we simultaneously fit a single density
profile to a (sub)sample of clusters.
Specifically, we consider the following five (sub)samples of clusters:
\begin{itemize}
\item full sample of 20 clusters,
\item low-mass subsample containing the ten lowest $\Mtwooo$ mass
      clusters from U14,
\item high-mass subsample containing the ten highest $\Mtwooo$ mass
      clusters from U14,
\item \CLASH\ X-ray-selected subsample of 16 clusters,
\item \CLASH\ high-magnification-selected subsample of 4 clusters.
\end{itemize} 

For joint ensemble modeling of clusters, we assume that all clusters have the same
mass, concentration, and intrinsic axis ratios, and fit the orientation
angles for each individual cluster.
Specifically, the joint posterior probability distribution for ensemble
modeling is written as
\begin{equation}
\label{eq:jointfit}
P(\mathbf{q}) \propto \mathcal{P}(\mathbf{q})\prod_{i\in\mathrm{sample}} \mathcal{L}(\mathcal{D}_{i}|\mathbf{p}_{i}),
\end{equation}
where $i$ runs over all clusters in the (sub)sample,
$\mathbf{q}$ denotes the vector containing model parameters
$(\Mtwooo, \Ctwooo, \qa, \qb, \{\theta_{i}, \phi_{i},\psi_{i}\}_{\mathrm{i\in\mathrm{sample}}})$,
$\mathcal{P}(\mathbf{q})$ is the prior distribution of $\mathbf{q}$,
$\mathcal{D}_{i}=\{\mathcal{D}_{\mathrm{WL},i}, \mathcal{D}_{\mathrm{SL},i}\}$ is the lensing data of the $i$th clusters,
and
$\mathbf{p}_{i}=(\Mtwooo, \Ctwooo, \qa, \qb, \theta_{i}, \phi_{i}, \psi_{i})$ is the model of the $i$th cluster.
The individual and ensemble modeling approaches are complementary to
 each other. We will present both results in
 Section~\ref{sec:results_and_discussion}.

\begin{table*}
\centering
\caption{
Best-fit Parameters for the Scaling Relations
}
\label{tab:sr}
\resizebox{\textwidth}{!}{
\begin{tabular}{lccccccccc}
\hline\hline
&\multicolumn{3}{c}{Full} &\multicolumn{3}{c}{X-ray selected} &\multicolumn{3}{c}{High magnification} \\
Modeling &$A_{\mathcal{X}}$ &$B_{\mathcal{X}}$ &$D_{\mathcal{X}}$ &$A_{\mathcal{X}}$ &$B_{\mathcal{X}}$ &$D_{\mathcal{X}}$ &$A_{\mathcal{X}}$ &$B_{\mathcal{X}}$ &$D_{\mathcal{X}}$   \\
\hline
\Ctwooo--\Mtwooo\ relation & & & & & & & & & \\ [3pt]
\sphericalmodel\ 
& $  4.03 \pm 0.10  $ & $  -0.65 \pm 0.06   $ & $  0.10 \pm 0.01   $ & $  4.51 \pm 0.14  $ & $  -0.47 \pm 0.07   $ & $  0.05 \pm 0.02   $ & $  2.81 \pm 0.24  $ & $  -0.70 \pm 0.15   $ & $   < 0.10    $  \\ [3pt]
\triaxialfmodel\ 
& $  4.41 \pm 0.25  $ & $  -0.49 \pm 0.11   $ & $  0.08 \pm 0.05   $ & $  4.82 \pm 0.30  $ & $  -0.36 \pm 0.11   $ & $   < 0.12    $ & $  3.00 \pm 0.65  $ & $  -0.47 \pm 0.32   $ & $   < 0.19    $ \\ [3pt]
\triaxialbmodel\      
& $  4.33 \pm 0.24  $ & $  -0.48 \pm 0.09   $ & $  0.09 \pm 0.04   $ & $  4.73 \pm 0.28  $ & $  -0.36 \pm 0.10   $ & $  0.04 \pm 0.04   $ & $  3.00 \pm 0.58  $ & $  -0.46 \pm 0.26   $ & $   < 0.16    $ \\ [3pt]
\hline
\qa--\Mtwooo\ relation & & & & & & & & & \\ [3pt]
\triaxialfmodel\ 
& $  0.52 \pm 0.04  $ & $  -0.36 \pm 0.13   $ & $  0.05 \pm 0.05   $ & $  0.51 \pm 0.05  $ & $  -0.36 \pm 0.15   $ & $  0.05 \pm 0.06   $ & $  0.61 \pm 0.13  $ & $  -0.54 \pm 0.35   $ & $   < 0.13    $ \\ [3pt]
\triaxialbmodel\      
& $  0.44 \pm 0.02  $ & $  -0.14 \pm 0.07   $ & $   < 0.06    $ & $  0.44 \pm 0.02  $ & $  -0.13 \pm 0.08   $ & $   < 0.07    $ & $  0.44 \pm 0.06  $ & $  -0.17 \pm 0.17   $ & $   < 0.07    $ \\ [3pt]
\hline
\fgeo--\Mtwooo\ relation & & & & & & & & & \\ [3pt]
\triaxialfmodel\ 
& $  0.93 \pm 0.07  $ & $  -0.14 \pm 0.15   $ & $   < 0.34    $ & $  0.92 \pm 0.08  $ & $  -0.14 \pm 0.17   $ & $   < 0.35    $ & $  0.97 \pm 0.18  $ & $  -0.16 \pm 0.31   $ & $   < 0.28    $ \\ [3pt]
\triaxialbmodel\      
& $  0.96 \pm 0.07  $ & $  -0.21 \pm 0.12   $ & $   < 0.26    $ & $  0.94 \pm 0.08  $ & $  -0.21 \pm 0.15   $ & $   < 0.29    $ & $  1.00 \pm 0.23  $ & $  -0.24 \pm 0.26   $ & $   < 0.29    $ \\ [3pt]
\hline
\triaxiality--\Mtwooo\ relation & & & & & & & & & \\ [3pt]
\triaxialfmodel\ 
& $  < 0.69  $ & $  0.07 \pm 0.19   $ & $   < 0.16    $ & $  < 0.70  $ & $  0.07 \pm 0.22   $ & $   < 0.17    $ & $  < 0.87  $ & $  0.11 \pm 0.56   $ & $   < 0.20    $ \\ [3pt]
\triaxialbmodel\      
& $  0.79 \pm 0.03  $ & $  0.06 \pm 0.07   $ & $   < 0.13    $ & $  0.79 \pm 0.04  $ & $  0.05 \pm 0.08   $ & $   < 0.14    $ & $  0.80 \pm 0.10  $ & $  0.07 \pm 0.16   $ & $   < 0.15    $ \\ [3pt]
\hline\hline
\end{tabular}
}
\tablecomments{
The best-fit parameters for the concentration to mass, minor-to-major
 axis ratio to mass, geometrical factor to mass, and triaxiality to
 mass scaling relations are listed.
Each mass scaling relation is characterized by the normalization
 $A_{\mathcal{X}}$, mass slope $B_{\mathcal{X}}$, and intrinsic
 scatter $D_{\mathcal{X}}$, where $\mathcal{X}$ runs over \Ctwooo, \qa,
 \fgeo\ and \triaxiality. 
For the concentration to mass relation, we use logarithmic observables
 (i.e., $\log\Mtwooo,\log\Ctwooo$)
 for the regression analysis.
 For
 the other scaling relations, we use linear observables without
 logarithmic transformation.
The results of \triaxialfmodel\ and \triaxialbmodel\ modeling are shown
 for each scaling relation.
Additionally, the results of \sphericalmodel\ modeling are presented for
 the concentration to mass relation. 
For ill-constrained parameters, we give $2\sigma$ upper limits.
}
\end{table*}
\begin{figure*}
\resizebox{!}{0.5\textwidth}{
\includegraphics[scale=1.0]{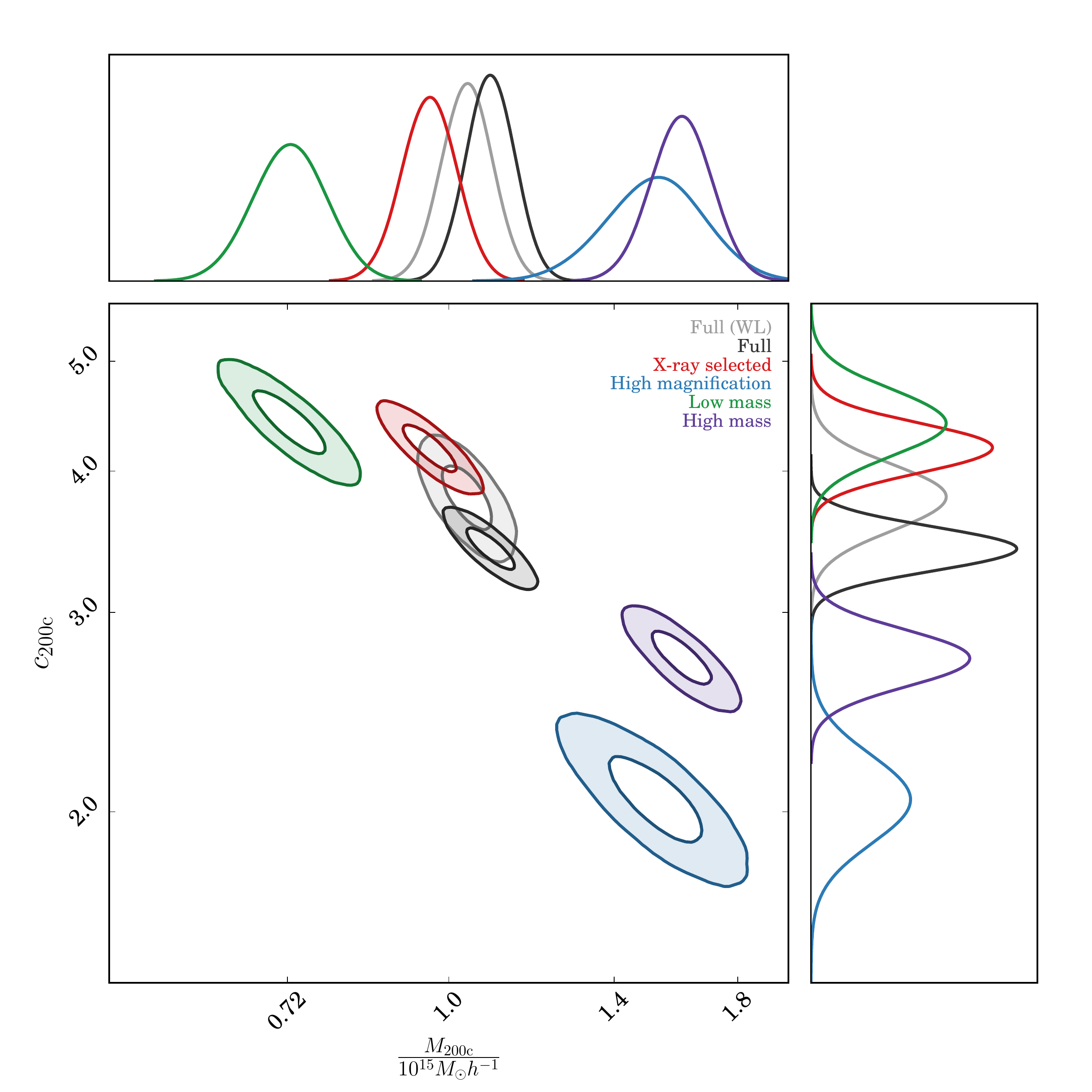}
}
\resizebox{!}{0.5\textwidth}{
\includegraphics[scale=1.0]{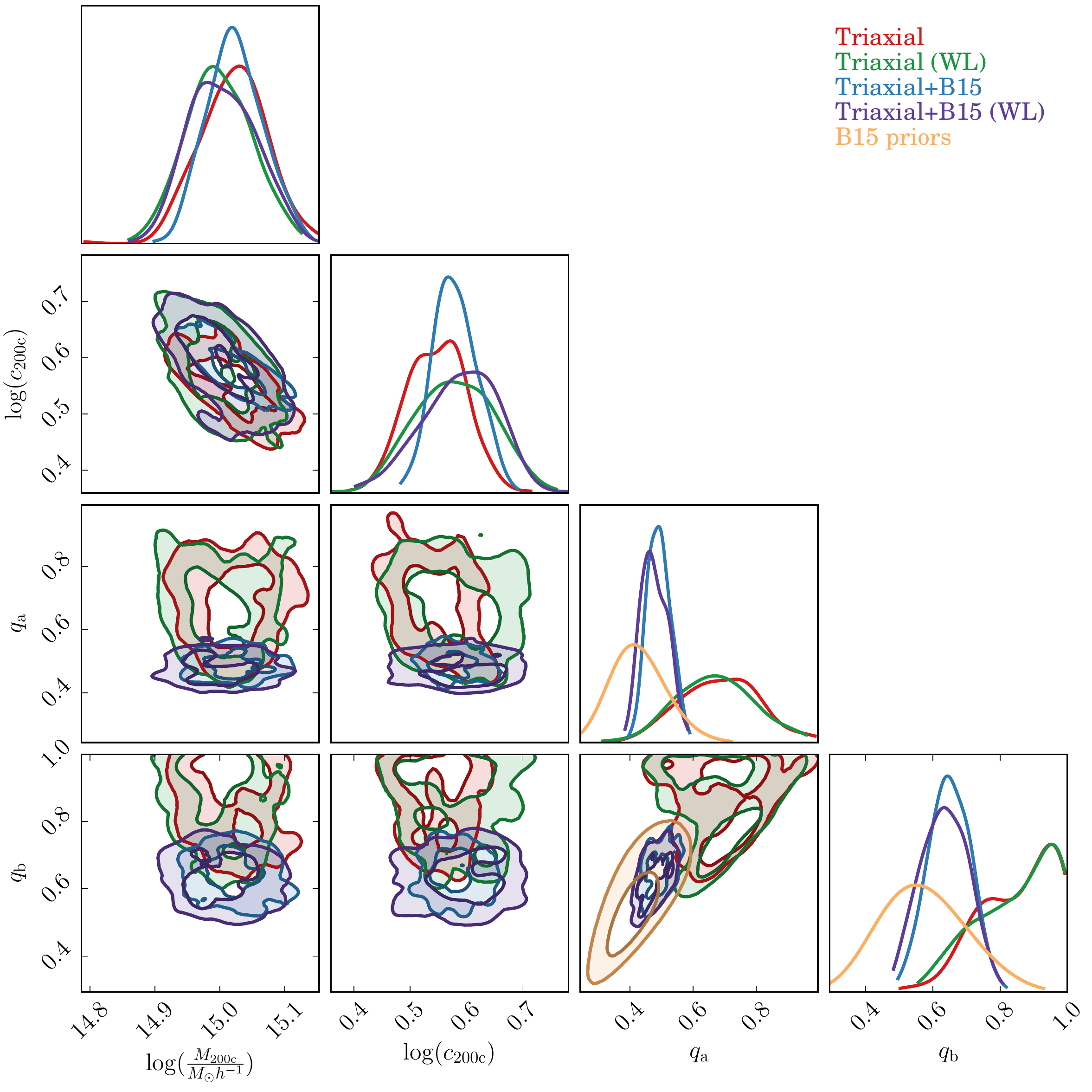}
}
\caption{ 
Joint ensemble constraints on the cluster model parameters from  
 \sphericalmodel,
 \triaxialfmodel\ and \triaxialbmodel\ modeling.
The results on the mass and concentration parameters from the
 \sphericalmodel\ modeling are presented in the left panel for
 respective subsamples. For the full sample, we show
 the combined weak+strong lensing results in black
 and the weak-lensing-only results in gray. 
 The right panel shows the joint ensemble constraints on the mass,
 concentration, and 
 two axis ratios from \triaxialfmodel\ and \triaxialbmodel\ modeling.
 The yellow contours in the right panel show the B15 prior distribution
 for clusters with $\Mtwooo=10^{15}\Msun$ at the sample median redshift,
 $\redshift = 0.377$.
For each case, we show the results from the combined weak+strong lensing
 and weak-lensing-only data. 
}
\label{fig:joint_all}
\end{figure*}

\subsection{Scaling Relation Fitting}
\label{sec:sr_fitting}

Here we describe our Bayesian regression approach to examining mass
scaling relations of various cluster observables using the results from
individual cluster modeling. Specifically, we investigate the following
four scaling relations:
(1) concentration to mass ($\Ctwooo$--$\Mtwooo$) scaling relations,
(2) minor axis ratio to mass ($\qa$--$\Mtwooo$) scaling relations,
(3) geometrical factor to mass ($\fgeo$--$\Mtwooo$) scaling relations,
(4) triaxiality to mass ($\triaxiality$--$\Mtwooo$) scaling relations.
To this end, we use the following equation:
\begin{equation}
\label{eq:sr_form}
\mathcal{X} = A_{\mathcal{X}}  \times \left( \frac{\Mtwooo}{10^{15}\Msun} \right)^{B_{\mathcal{X}}},
\end{equation}
together with the intrinsic scatter
$D_{\mathcal{X}}\equiv\sigma_{\mathcal{X}|\Mtwooo}$,
where the observable $\mathcal{X}$ runs over \qa, \fgeo
and \triaxiality, respectively.
For the concentration to mass relations, we fit
Equation~(\ref{eq:sr_form}) using logarithmic observables  (i.e.,
$\log\Ctwooo$ and $\log\Mtwooo$) 
with log-normal intrinsic scatter $D_{\Ctwooo}\equiv\sigma_{\log\Ctwooo|\Mtwooo}$.
Note that we examine these scaling relations with a pivot mass of
$\Mtwooo=10^{15}\Msun$, which is close to the median mass of our sample,
to reduce degeneracies between $A_{\mathcal{X}}$ and $B_{\mathcal{X}}$.

To derive an observable ($\mathcal{X}$) to mass relation for a given
sample of clusters, we draw 5000 random samples from MCMC posterior
distributions of the clusters. For each random realization, we fit
Equation~(\ref{eq:sr_form}) to data drawn from the posteriors and obtain
a set of the regression parameters
$\left(A_{\mathcal{X}}, B_{\mathcal{X}}, D_{\mathcal{X}}\right)$.
Finally, we derive the median values
and confidence intervals of the parameters.
In this way, we directly account for the covariance between the
observable $\mathcal{X}$ and mass \Mtwooo\ in our Bayesian regression
analysis \citep{chiu16c,gupta17,chiu17}. 
We have checked that the regression results are not sensitive to the
number of random realizations used. In our analysis, we have ignored the
intrinsic scatter between weak-lensing and true cluster mass \citep{sereno15a}.

We perform a regression analysis of observable--mass scaling relations
following this procedure for each of the \sphericalmodel,
\triaxialfmodel\ and \triaxialbmodel\ modeling approaches.

\begin{figure*}
\centering
\resizebox{!}{0.45\textwidth}{
\includegraphics[scale=1.0]{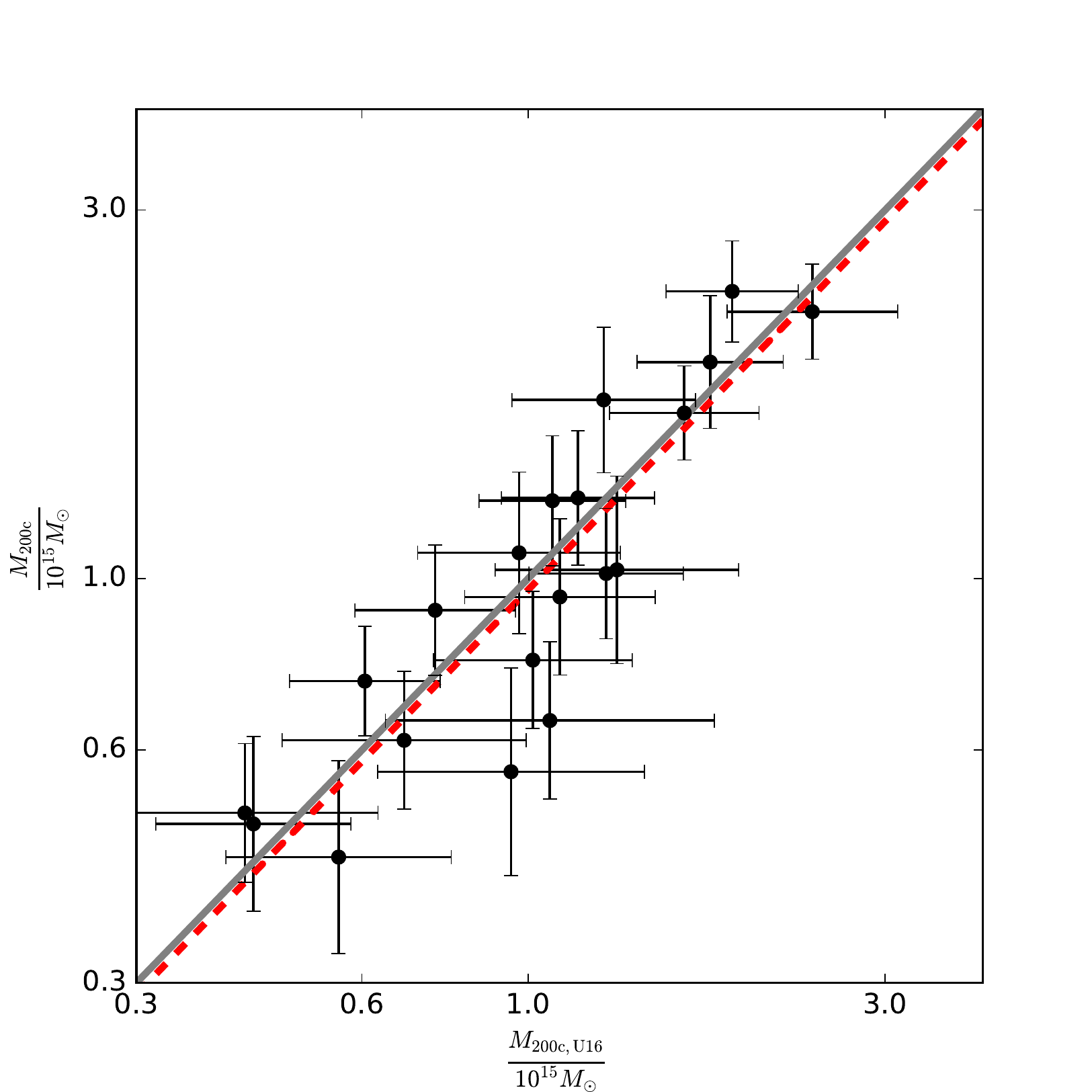}
}
\resizebox{!}{0.45\textwidth}{
\includegraphics[scale=1.0]{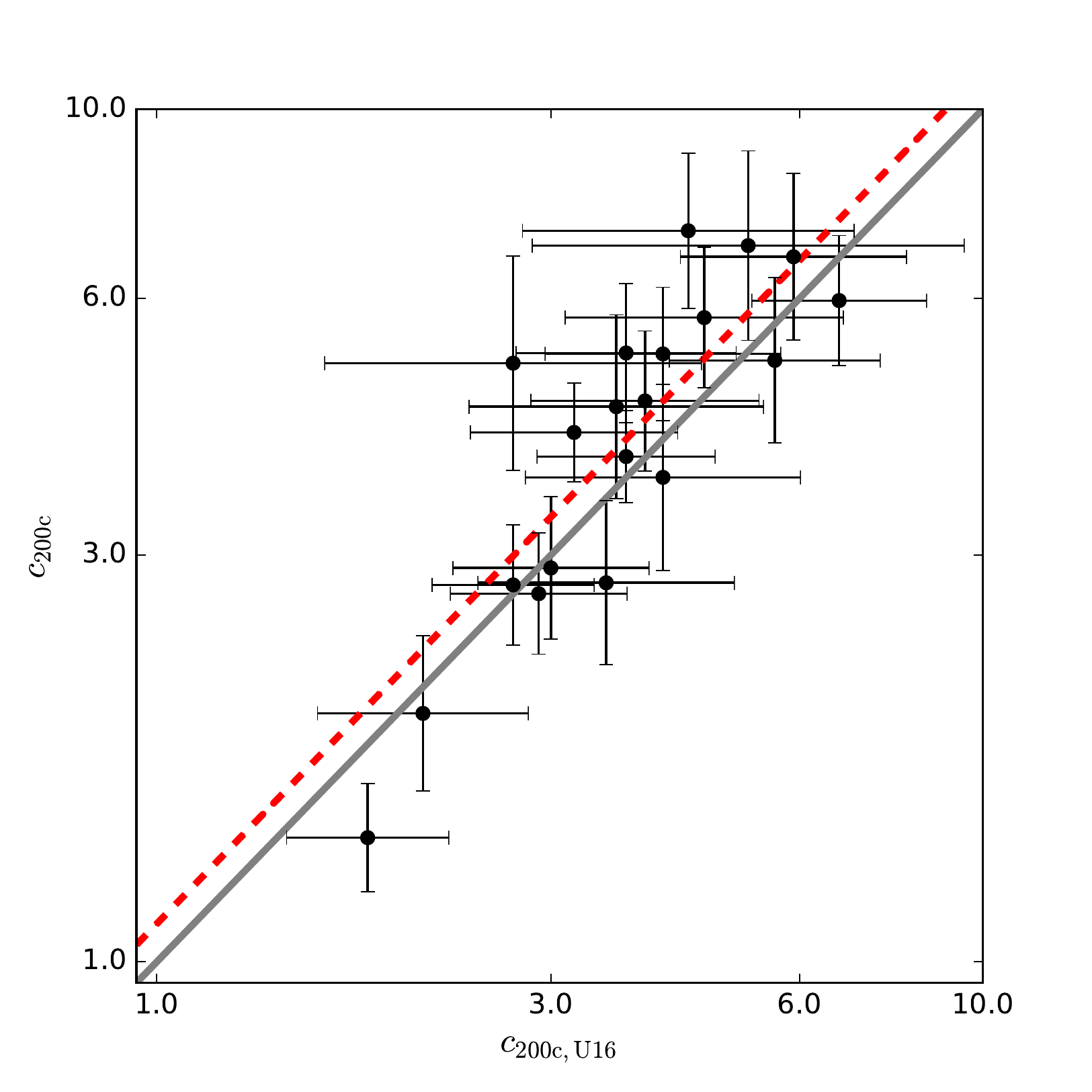}
}
\caption{
 Comparison of cluster mass (left) and concentration (right) estimates between
 U16 and
 our \sphericalmodel\ modeling. Both studies use identical sets of {\em
 HST} lensing constraintas $\Msl(<r)$ as input for their analyses. The U16
 analysis is based on azimuthally averaged weak-lensing constraints,
 while our analysis is based on 2D weak-lensing data.
For each case, the mean difference in the logarithmic observable is
 indicated by the  red dashed line.  
The gray line indicates the one-to-one relation.
}
\label{fig:du16}
\end{figure*}
\begin{figure*}
\centering
\resizebox{!}{0.45\textwidth}{
\includegraphics[scale=1.0]{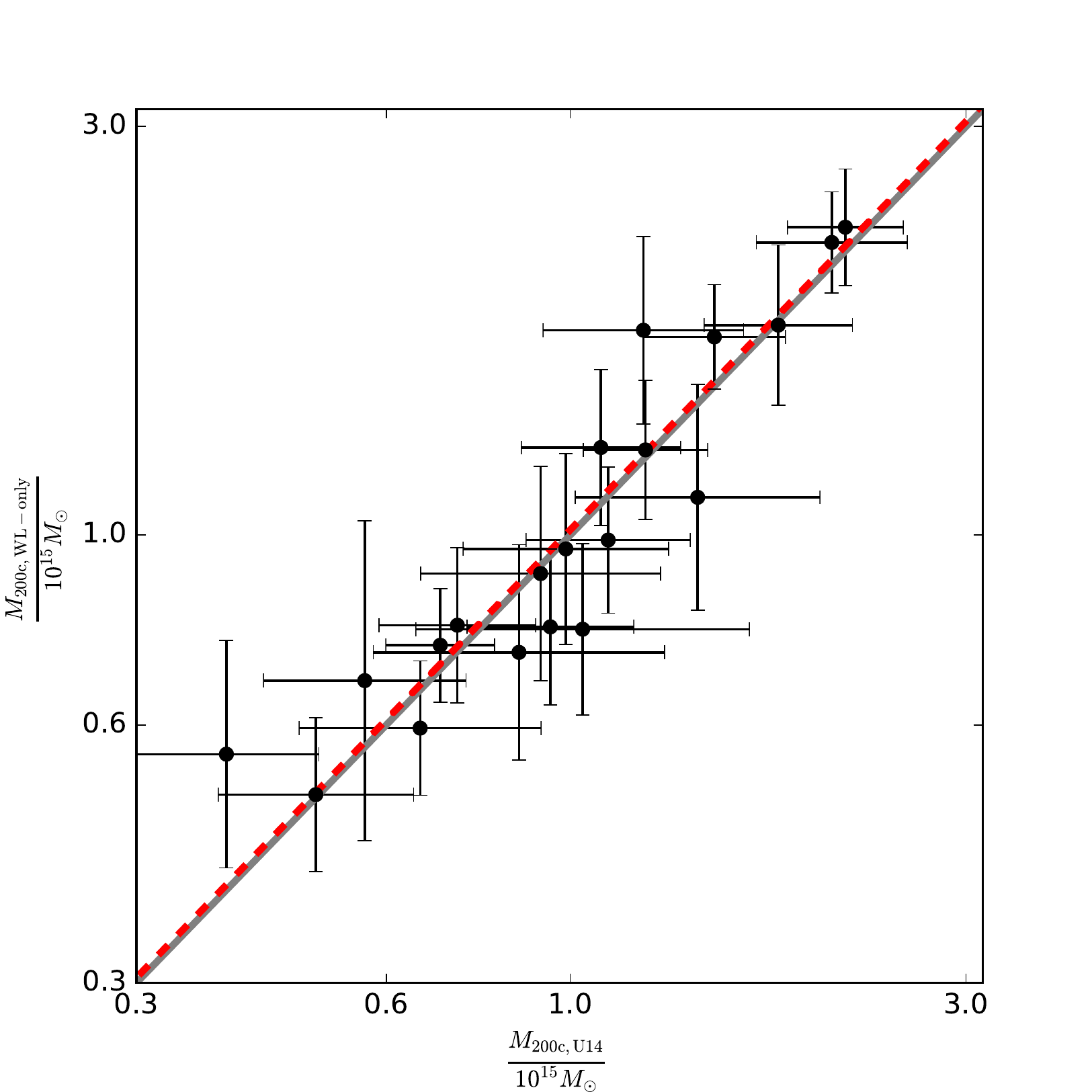}
}
\resizebox{!}{0.45\textwidth}{
\includegraphics[scale=1.0]{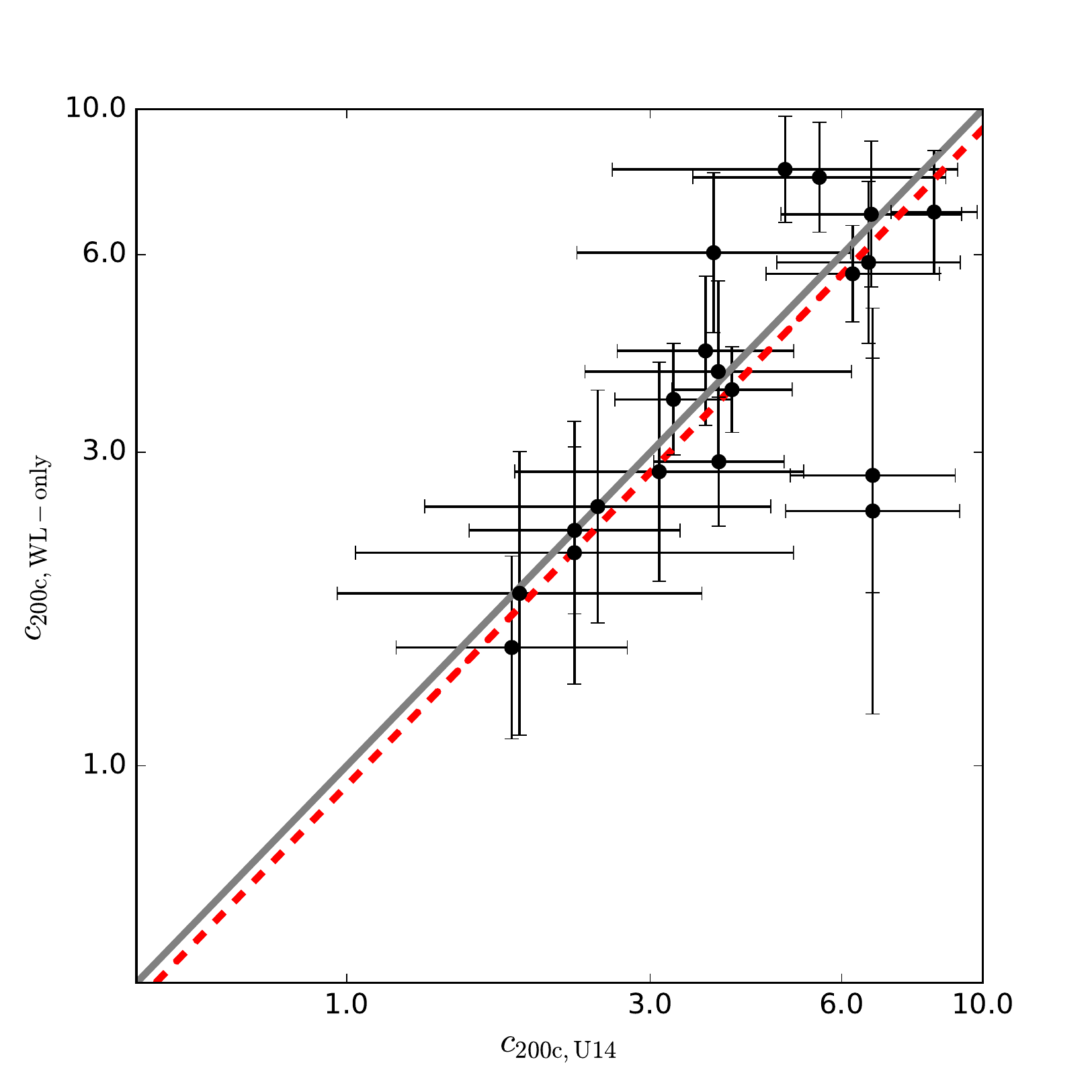}
}
\caption{
Same as Figure~\ref{fig:du16} but showing the comparison between U14 and
 our \sphericalmodel\ modeling, both using weak lensing alone.
} 
\label{fig:du14}
\end{figure*}
%

%
%

\section{Results and Discussion}
\label{sec:results_and_discussion}

In Section~\ref{sec:spherical_modeling}, we first compare our results of
\sphericalmodel\ modeling to those obtained by previous \CLASH\ work of U14
and U16.
In Sections~\ref{sec:concentration_mass_relation} to
\ref{sec:triaxiality_mass_relation}, 
we present the resulting observable--mass scaling relations based on
individual cluster modeling.
We will also discuss the results of joint ensemble modeling in these subsections.

In Table~\ref{tab:results} we summarize marginalized constraints on
the spherical and triaxial NFW model parameters derived from the
individual and ensemble modeling approaches.
Here we have employed the biweight estimators of \citet{beers90} for the central
location and scale of the marginalized posterior distributions
\citep[e.g.,][]{sereno11,umetsu14,umetsu15}\footnote{The biweight
estimator is robust against skewed distributions, because it gives a
higher weight to points that are close to the central location of a distribution.}.
The regression parameters of various mass scaling relations
are summarized in Table~\ref{tab:sr}.
In Figure~\ref{fig:joint_all}, we show the joint ensemble constraints on
cluster parameters from different modeling approaches. We also show the
results obatained with and without the {\em HST} lensing constraints
$\Msl(<r)$ to demonstrate the consistency between the weak and strong
lensing data sets. For the results from individual cluster modeling, we
show the resulting marginalized posterior distributions in
Appendix~\ref{sec:individual_constraints}.

\subsection{Consistency of Spherical Modeling}
\label{sec:spherical_modeling}

We compare our results from \sphericalmodel\ modeling of the 20 clusters
to those of U14 and U16 for a consistency check.
Both U16 and this work are based on the weak-lensing shear and
magnification data obtained by U14. U16 reconstructed azimuthally
averaged surface mass density profiles of these individual clusters by
combining the weak-lensing data of U14 with the
central {\em HST} lensing constraints $\Msl(<r)$ from \citet{zitrin15}.

In both U14 and U16, the NFW mass and concentration parameters were
derived assuming spherical symmetry, corresponding to the case of
\sphericalmodel\ modeling in this work.
Although these three studies share the same data as input to modeling,
the crucial difference of this study is that we directly fit a model
profile to the 2D surface mass density maps of \citet{umetsu18}
without azimuthal averaging. This comparison is thus useful for
validating the robustness of our reconstruction and modeling
procedures, for a given data set.

We begin with the results of individual cluster modeling. 
Comparing our \Mtwooo\ mass estimates to those
$M_{200\mathrm{c,U16}}$ from U16,
we find the mean diffrence\footnote{We use an unweighted mean here because the uncertainties 
of this work and U14/U16 are highly correlated with each other.} in
logarithmic mass of
$
\left\langle\Delta\log\Mtwooo\right\rangle=
\left\langle\log\Mtwooo - \log M_{200\mathrm{c,U16}}\right\rangle
= -0.01 \pm 0.04
$,
which meets the criterion of $<8\percent$ (or $0.035$~dex)
for consistency (see Section~\ref{sec:weak_lensing_data}).
Similarly, the mean difference in logarithmic concentration is
$
\left\langle\Delta\log\Ctwooo\right\rangle=
\left\langle\log\Ctwooo - \log c_{200\mathrm{c,U16}}\right\rangle
= 0.04 \pm 0.04.
$
Except that we observe a mild increase ($0.04$~dex or
$\approx10$\percent) in concentration with respect to U16, our results
are in satisfactory agreement with U16. 
This comparison with U16 is shown in Figure~\ref{fig:du16}.

Excluding the {\em HST} lensing constraints $\Msl(<r)$ from our
\sphericalmodel\ modeling results in mass estimates that are consistent
with those from U14 based on the one-dimensional (1D) weak-lensing
analysis, with a mean difference in logarithmic mass of
$
\left\langle\Delta\log M_{200\mathrm{c,WL}}\right\rangle=
\left\langle\log M_{200\mathrm{c,WL}} - \log M_{200\mathrm{c,U14}}\right\rangle
= 0.01 \pm 0.04.
$
This is much smaller than the systematic uncertainty in the overall mass
calibration of $8\percent$ (or $0.035$~dex).
Similarly, the mean difference in logarithmic concentration with respect to U14
is  
$
\left\langle\Delta  c_{200\mathrm{c,WL}}\right\rangle=
\left\langle\log c_{200\mathrm{c,WL}} - \log c_{200\mathrm{c,U14}}\right\rangle
= -0.03 \pm 0.05.
$
Again, no significant tension with U14 is found, as also shown in Figure~\ref{fig:du14}.

In what follows, we compare our results from joint ensemble modeling to
those from U14 and U16.
Since U14 and U16 constrained the $c$--$M$ relation only for the
X-ray-selected subsample, 
we restrict our ensemble \sphericalmodel\ modeling to the same 16 
X-ray-selected clusters for a fair comparison.

U16 constrained the NFW parameters from the stacked weak and strong
lensing profile as
$\Ctwooo=3.76^{+0.29}_{-0.27}$ 
and 
$\Mtwooo=\left(10.08\pm0.7\right)\times10^{14}\Msun$, 
respectively.
In this work, joint ensemble \sphericalmodel\ modeling with combined
weak and strong lensing yields
$\Ctwooo=4.18^{+0.20}_{-0.19}$ 
and 
$\Mtwooo=\left(9.62^{+0.49}_{-0.52}\right)\times10^{14}\Msun$, 
consistent with the 1D analysis of U16 within the quoted uncertainties.
Note that this joint ensemble constraint on \Ctwooo\ from our
\sphericalmodel\ modeling is $\approx10$\percent\ higher than that from
U16 at the $1\sigma$ level. This tendency is consistent with the case of
individual cluster modeling (see Figure~\ref{fig:du16}). 
This ensemble constraint is shown in red contours in the left panel of
Figure~\ref{fig:joint_all}.

From an NFW fit to the stacked weak-lensing profile, U14 found 
$\Ctwooo=4.01^{+0.35}_{-0.32}$ 
and 
$\Mtwooo=\left(9.4\pm0.70\right)\times10^{14}\Msun$.
Our joint ensemble \sphericalmodel\ modeling using the weak-lensing data
alone yields $\Ctwooo=4.15^{+0.29}_{-0.27}$ 
and 
$\Mtwooo=\left(9.66^{+0.53}_{-0.51}\right)\times10^{14}\Msun$, 
showing no significant discrepancy.

We further compare our spherical mass estimates to those fom \citet{sereno18}.
They obtained cluster masses $M_{200\mathrm{c,~S18}}$ from
a joint analysis of weak and strong lensing, X-ray, and the SZE data
sets, in both triaxial and spherical approaches.
For the spherical mass comparison,
we find a mean difference in logarithmic mass of 
$\left\langle M_{200\mathrm{c,~S18}}(<R) - \Mtwooo(<R)\right\rangle =\left(2\pm3\right)\percent$ and
$\left(0\pm1\right)\percent$ at
$R = 1\mathrm{Mpc}\,h^{-1}$ and $1.5\mathrm{Mpc}\,h^{-1}$, respectively.
This again demonstrates excellent agreement.

On the basis of these consistency tests, we find no significant tension
between the results using different combinations of data sets (U14, U16,
\cite{sereno18}), ensuring the robustness of our modeling procedures. 
We will discuss more results of \sphericalmodel\ modeling in
Section~\ref{sec:concentration_mass_relation}.

\subsection{Concentration to Mass Relations}
\label{sec:concentration_mass_relation}

\begin{figure}
\centering
\resizebox{!}{0.55\textwidth}{
\includegraphics[scale=1.0]{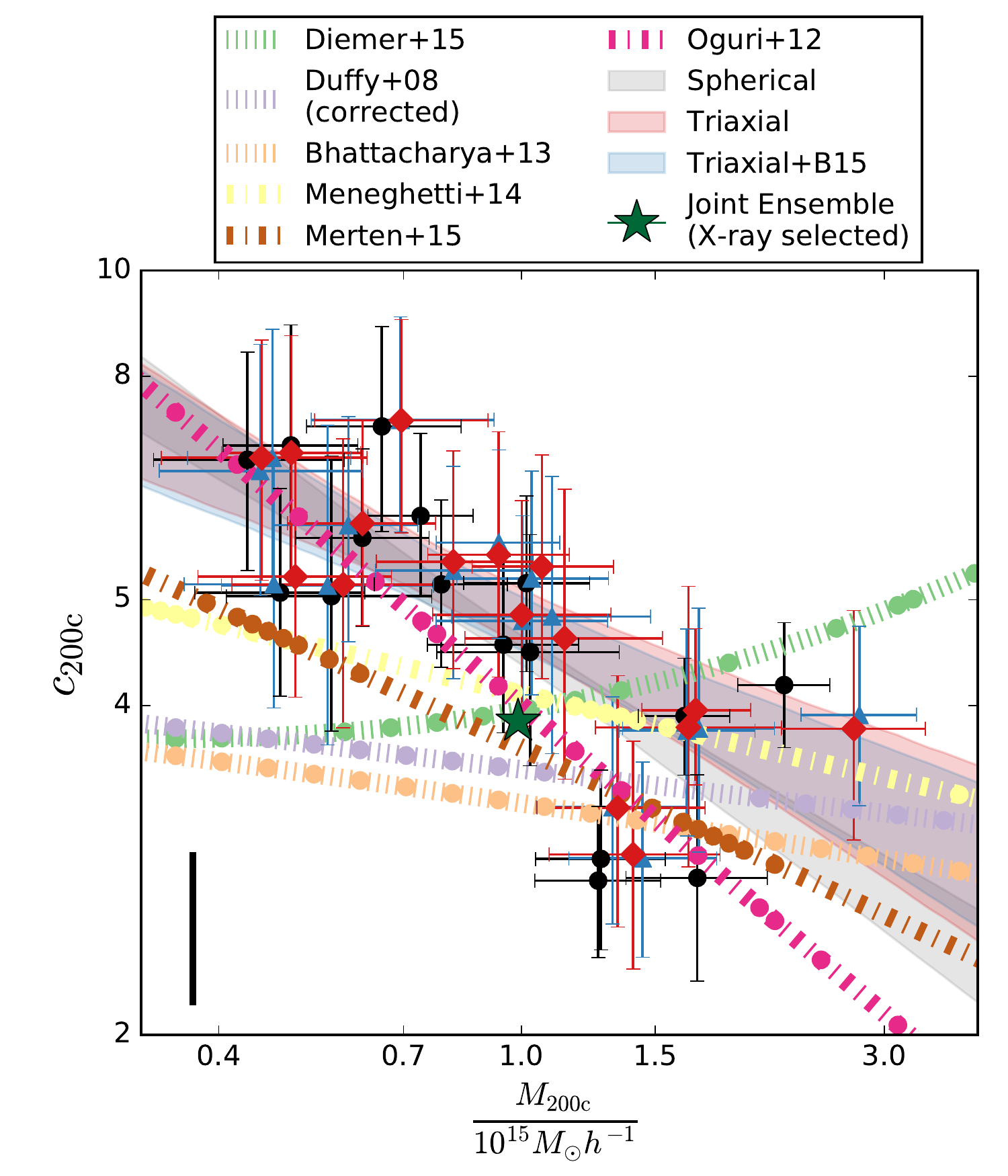}
}
\caption{
Concentration to mass scaling relation for the 16 X-ray-selected \CLASH\ clusters.
The results of \sphericalmodel, \triaxialfmodel\ and
 \triaxialbmodel\ modeling are shown by black circles, red diamonds,
 and blue triangles, respectively.
The $1\sigma$ confidence levels of the scaling relations are indicated
 by the shaded areas. 
 We also plot various results from numerical simulations
 \citep[][]{duffy08, bhattacharya13, diemer15} and 
 previous observational work 
 \citep[][]{oguri12, meneghetti14, merten15}.
 The color codes for different authors are noted in the figure.
 We multiply the normalization of the $c$-$M$ relation from 
 \cite{duffy08} by $1.2$ to account for the different cosmology
 (see the discussion in the text).
 The green star indicates our joint ensemble constraint from
 \triaxialfmodel\ modelling of the X-ray selected subsample.
 The length of the black bar at the lower-left corner indicates the
 level of log-normal intrinsic scatter 
  ($D_{\Ctwooo}\equiv\sigma_{\log\Ctwooo|\Mtwooo}$)
  of $7\percent$ at fixed mass, which is predicted by \cite{meneghetti14}
  for the X-ray selected sample in the \CLASH\ survey (see the text for more details).
}
\label{fig:cm}
\end{figure}

From Bayesian regression, we determine the $c$--$M$ relation for the 16
X-ray-selected \CLASH\ clusters as 
\begin{equation}
\label{eq:concen_sph}
\Ctwooo = \left(\numAconcenSph\right) \times \left( \frac{\Mtwooo}{10^{15}\Msun} \right)^{\numBconcenSph},
\end{equation}
\begin{equation}
\label{eq:concen_tri}
\Ctwooo = \left(\numAconcenTri\right) \times \left( \frac{\Mtwooo}{10^{15}\Msun}\right)^{\numBconcenTri},
\end{equation}
and 
\begin{equation}
\label{eq:concen_bon}
\Ctwooo = \left(\numAconcenBon\right) \times \left( \frac{\Mtwooo}{10^{15}\Msun}\right)^{\numBconcenBon},
\end{equation}
with a log-normal intrinsic scatter $D_{\Ctwooo}\equiv\sigma_{\log\Ctwooo|\Mtwooo}$ of 
\numDconcenSph, $<0.12$ ($2\sigma$ upper bound), and \numDconcenBon\
using the \sphericalmodel, \triaxialfmodel\ and \triaxialbmodel\
modeling approaches, respectively.
A redshift evolution of the NFW $c$--$M$ relation,
$c\propto (1+z)^{-0.668\pm 0.341}$,
was suggested for X-ray-selected \CLASH-like halos in $N$-body
hydrodynamical simulations \citep{meneghetti14}.
We find that including the redshift scaling in regression analysis
results in a negligible change in the inferred regression parameters
within the errors.
We thus ignore the redshift dependence of the $c$--$M$ relation in this study.
In Figure~\ref{fig:cm},
we plot the resulting scaling relations along with the individual
cluster constraints for the X-ray-selected subsample.
The scaling relations obtained for the full sample and the
high-magnification subsample are given in Table~\ref{tab:sr}.

In Figure~\ref{fig:cm}, we see a steep mass dependence 
of the $c$--$M$ relation. 
Assuming spherical symmetry, we find a mass slope of
$B_{\Ctwooo}=\numBconcenSph$ for the X-ray-selected subsample,
and an even steeper slope of $B_{\Ctwooo}=-0.65\pm0.06$ for the full
sample. Here we note that this is due in part to our fitting procedure,
in which we do not account for the underlying distribution of true
cluster masses.
That is, the steepening of the intrinsic mass function
combined with the selection function could alter the resulting
distribution of true cluster masses \citep{sereno15a}. 
Accounting for this effect, U16 found $B_{\Ctwooo}=-0.44\pm 0.19$ for
the same subsample, which is consistent with our results, but
with a much larger uncertainty.
From \triaxialfmodel\ modeling including additional shape and
orientation parameters (with uniform priors), we find a shallower mass
slope of \numBconcenTri, which is consistent with the \sphericalmodel\
modeling results within the errors.

The normalization of the $c$--$M$ relation is constrained as
$\Ctwooo=\numAconcenSph$ and $\numAconcenTri$ at the pivot mass of
$\Mtwooo=10^{15}\Msun$ for \sphericalmodel\ and \triaxialfmodel\
modeling, respectively.
We note that, by construction,
$\Ctwooo(\sphericalmodel)\le\Ctwooo(\triaxialfmodel)$ \citep[see][]{sereno18}. 
On the other hand, employing the informative shape priors from B15
in \triaxialbmodel\ modeling does not change the results in a
significant manner.
Regardless of the priors chosen, the effect of triaxiality has no
significant impact on the resulting $c$--$M$ relation, so that 
the assumption of spherical symmetry is well validated in
determining the overall density structure of the \CLASH\ clusters.

Now we discuss the results of joint ensemble modeling of the
cluster mass and concentration. 
In the left panel of Figure~\ref{fig:joint_all}, we show the
weak-lensing-only results in gray and the weak and strong lensing
results in black, both derived for the full sample of 20 clusters.
Joint ensemble constraints for the X-ray-selected, high-magnification,
low-mass, and high-mass subsamples are also shown in the same panel.
We see a clear trend of decreasing concentration with increasing mass.
In particular, the high-magnification subsample
consisting of four very massive disturbed clusters
($\Mtwooo\approx1.5\times10^{15}\Msun$) has $\Ctwooo\approx2$, 
much lower than other similar-mass clusters, indicating a selection
effect. We will further discuss this in Section~\ref{sec:comparison_cm}.

\begin{figure}
\resizebox{!}{0.5\textwidth}{
\includegraphics[scale=1.0]{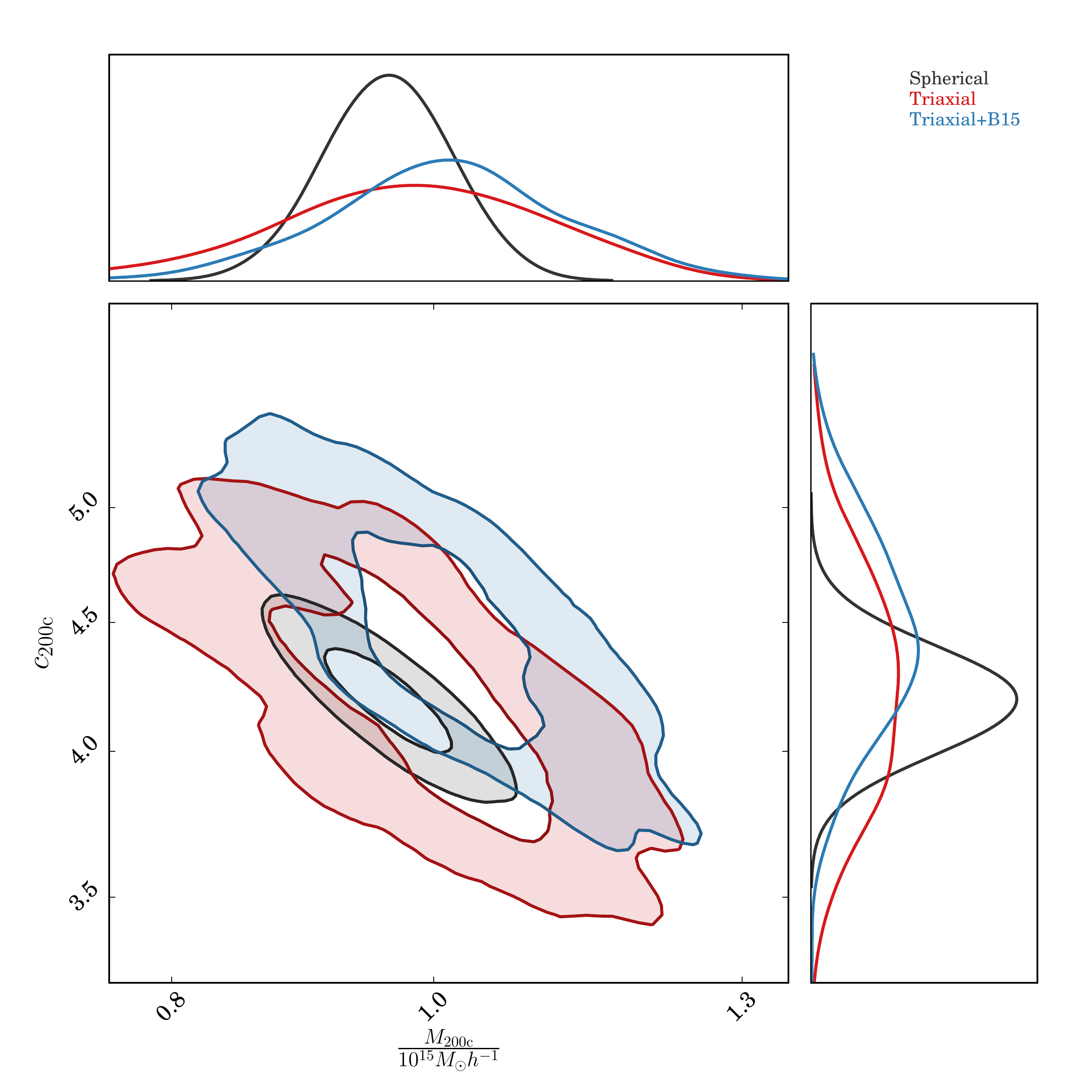}
}
\caption{
Joint ensemble constraints on the concentration and mass for the
 X-ray-selected subsample of 16 \CLASH\ clusters.
The results of \sphericalmodel, \triaxialfmodel\ and \triaxialbmodel\
 modeling are shown in black, red and blue, respectively.
}
\label{fig:joint_cm}
\end{figure}

In Figure~\ref{fig:joint_cm}, we show our joint ensemble constraints on
the mass and concentration parameters for the X-ray-selected subsample,
obtained with three different modeling approaches
(\sphericalmodel\, black;
 \triaxialfmodel\, red;
 \triaxialbmodel\, blue). In all cases, we use the weak and strong
 lensing constraints.
From \triaxialfmodel\ (\triaxialbmodel) modeling, we find
$\Ctwooo=3.87^{+0.76}_{-0.11}$ and $\Mtwooo=\left(0.99\pm0.11\right)\times10^{15}\Msun$
($\Ctwooo=4.42^{+0.41}_{-0.40}$ and $\Mtwooo=1.050^{+0.048}_{-0.144}\times10^{15}\Msun$).
Overall, triaxial modeling results in a concentration that is slightly
higher than spherical modeling at the $\approx7\percent$ level,
regardless of the chosen priors. As noted above, it is expected that
$\Ctwooo(\sphericalmodel)\le\Ctwooo(\triaxialfmodel)$ \citep{sereno18}. 
However, this level of difference is
consistent with zero within the errors.
Therefore, we conclude that the spherical symmetry is a well valid
assumption in estimating the concentration of the \CLASH\ clusters.


\subsubsection{Comparisons with Previous Work}
\label{sec:comparison_cm}

We first compare our concentration to mass scaling relations to that
obtained by U16. In U16, the $c$--$M$ relation for the X-ray-selected
subsample is constrained as
$
\Ctwooo\propto 3.98^{+0.38}_{-0.35} \times\Mtwooo^{-0.44\pm0.19}
$ 
with a log-normal intrinsic scatter of 
$
\sigma_{\log\Ctwooo|\Mtwooo}\approx0.056\pm0.026,
$ 
assuming spherical symmetry.
This is in good agreement with our \sphericalmodel\ modeling 
(see Equation~(\ref{eq:concen_sph})) at the $\lesssim1\sigma$ level, in
terms of the mass slope and intrinsic scatter.
On the other hand, we find a normalization that is $\approx13\percent$ higher than U16.
This trend is consistent with what we found in
Section~\ref{sec:spherical_modeling}.

According to \cite{meneghetti14}, it is expected that the mean
concentration of the X-ray-selected \CLASH\ subsample recovered from
projected lensing measurements is $\approx 11\percent$ higher than that for the
full population of clusters.  
Specifically, the $c$--$M$ relation predicted for this subsample
is\footnote{We use the NFW $c$--$M$--$z$ relation predicted for
CLASH-like X-ray-selected clusters taken from Table~2 of
\citet{meneghetti14}.} 
$  
\Ctwooo\propto\left(4.1\pm0.1\right)\times\Mtwooo^{-0.16\pm0.11}
$
at their median redshift of
$\left\langle\redshift\right\rangle\approx 0.35$,
with an intrinsic scatter of  $\sigma_{\log\Ctwooo|\Mtwooo}\approx0.07$.
The observed normalization ($A_{\Ctwooo}=4.51\pm 0.14$) is thus
$\left(10\pm4\right)\percent$ higher than this \CLASH\ prediction.
The derived mass slope ($B_{\Ctwooo}=-0.47\pm 0.07$)
is steeper than the \CLASH\ prediction \citep{meneghetti14} at the
$2.4\sigma$ level.
The predicted scatter is consistent with our measurements, but
considerably smaller than the typical intrinsic scatter,
$\sigma_{\log\Ctwooo|\Mtwooo}\approx0.15$ ($\approx0.11$), predicted for the
full (relaxed) population of halos in cosmological $N$-body
simulations \cite[e.g.,][]{duffy08, bhattacharya13}.
This is consistent with the expectation that the X-ray-selected
\CLASH\ sample is largely ($\approx70\percent$) composed
of regular and highly relaxed clusters \citep{meneghetti14}.
The intrinsic scatter is increased by a factor of $\approx2$ if we
include the four high-magnification \CLASH\ clusters (see
Table~\ref{tab:sr}).

Next, we compare our results to previous \CLASH\ work of
\cite{merten15}, 
who studied 19 X-ray-selected \CLASH\ clusters.
They simultaneously combined {\em HST} strong+weak lensing
constraints (specifically, {\em HST} shear catalogs plus locations and redshifts of
multiple images) with wide-field shear catalogs of \citet{umetsu14} to
reconstruct 2D mass maps of individual clusters. Weak magnification
lensing was not used in their analysis.
Cluster mass estimates were obtained from spherical NFW fits to azimuthally
averaged surface mass density profiles\footnote{
In this work, we combine the enclosed projected mass constraints from
the {\em HST} lensing analysis of \citet{zitrin15}
with the wide-field weak-lensing mass maps from \citet{umetsu18},
followed by direct model fitting without azimuthal averaging.}.
Assuming spherical symmetry, \citet{merten15} found
$
\Ctwooo\propto\left(3.66\pm0.16\right)\times\Mtwooo^{-0.32\pm0.18},
$
with a log-normal intrinsic scatter of $0.07$,
in good agreement with our results in terms of the mass slope and
intrinsic scatter. However, the normalization obtained by
\cite{merten15} is significantly lower than our results, likely arising
from the different reconstruction methods.

We then compare our results to those of \cite{sereno17b},
who carried out a 1D weak-lensing analysis to derive the $c$--$M$
relation for SZE-selected clusters from the \PLANCK\ survey.
Examining their $c$--$M$ relation with
$\Mtwooo=10^{15}\Msun$ and
the median redshift of our full sample,
$\left\langle\zd\right\rangle=0.377$,
we find 
$
\Ctwooo = 4.04^{+6.59}_{-2.50}, 
$
which is consistent with our \sphericalmodel\ modeling results
for our full sample within large errors.

In addition, we compare our results to \cite{oguri12},
who combined strong and weak lensing constraints in a 1D analysis to
derive the $c$--$M$ relation for a sample of 28 strong-lensing-selected
clusters from the Sloan Digital Sky Survey.
The best-fit $c$--$M$ relation of \cite{oguri12} is 
$
c_{\mathrm{vir}} = \left(7.7\pm0.6\right)\times\left(M_{\mathrm{vir}} /5\times10^{14}\Msun\right)^{-0.59\pm0.12},
$
with a log-normal intrinsic scatter of
$\sigma_{\log\Ctwooo|\Mtwooo}=0.12$ defined with the virial overdensity.
We convert this relation to that with $\Delta=200$
by substituting
$\Mtwooo = 0.88M_{\mathrm{vir}}$ and
$\Ctwooo = 0.83C_{\mathrm{vir}}$ at
$M_{\mathrm{vir}}=5\times10^{14}\Msun$ and
the median redshift of our full sample,
$\left\langle\zd\right\rangle=0.377$.
The resulting relation using a pivot mass of $10^{15}\Msun$ is
$
\Ctwooo = \left(4.0\pm0.3\right)\times\left(\Mtwooo/10^{15}\Msun\right)^{-0.59\pm0.12},
$
in good agreement with our full-sample results (Table~\ref{tab:sr}).

This comparison is particularly interesting because the \cite{oguri12}
sample was selected by the presence of strong-lensing features,
specifically giant arcs.
In contrast, the high-magnification \CLASH\ clusters were selected by
their high lensing magnification properties, with the goal of searching for
strongly lensed galaxies at high redshifts.
The giant-arc selection of \cite{oguri12} preferentially selects
clusters that are more centrally concentrated and/or elongated along the
line of sight, resulting in a  positive bias in the apparent degree of
concentration relative to the full population of clusters.
Importantly, this bias is predicted to be mass dependent and more
prominent for low mass clusters \citep[][e.g., estimated concentration
being biased high by $\approx80\percent$ for
$M_{\mathrm{vir}}\approx8\times10^{13}\Msun$ and $\lesssim20\percent$
for $M_{\mathrm{vir}}\gtrsim10^{15}\Msun$]{oguri12}. 
In this work, we find an opposite trend of significantly lower
concentration for high-magnification-selected clusters
(Tables~\ref{tab:results} and \ref{tab:sr}).
This is expected for typical merging, high-mass clusters, where the 
mass distribution is not as concentrated as relaxed systems.
In fact, our high-magnification clusters are found to be dynamically 
disturbed systems \citep{zitrin13,medezinski13,balestra16,jauzac17},
where complex merging events are taking place.  
Nevertheless, this comparison suggests that clusters selected by their
strong-lensing features tend to be a highly biased population in terms
of their morphology and dynamical state.

Finally, we compare our joint ensemble constraints
(Table~\ref{tab:results}) to recent simulation work of \cite{duffy08},
\cite{bhattacharya13}, and \cite{diemer15}. 
Here we compare their predictions to our results from \sphericalmodel\
modeling because they measured halo mass and concentration from 
spherically averaged density profiles of simulated halos.
Overall, our conclusions are not altered significantly if comparing to
our triaxial results, given the good agreement between the spherical and
triaxial results, regardless of the priors chosen.

\cite{duffy08} characterized the $c$--$M$ relation for both relaxed and
full populations of simulated halos at $\redshift\lesssim2$ in the \WMAP5 cosmology
 ($\OmegaM=0.258$ and $\Hnow=0.719$).
The mean concentration predicted for the full (relaxed) population of halos is 
$\Ctwooo\approx2.91$ ($\approx3.30$) at
$\Mtwooo=10^{15}\Msun$ and $\zd=0.377$,
which is lower than $\Ctwooo=3.42^{+0.14}_{-0.15}$
($4.18^{+0.20}_{-0.19}$) by $15\percent$ ($21\percent$) obtained for our
full (X-ray-selected) sample.
This is in line with the finding of \cite{dutton14} that the WMAP5
cosmology assumed in \cite{duffy08} yields a concentration that is
lower by $\approx20\percent$ relative to the \PLANCK\ cosmology.
\cite{bhattacharya13} modeled the halo concentration as a
function of the halo peak height\footnote{The average peak height of our sample
is $\approx3.8$}. Their model yields $\Ctwooo=3.59$ and $3.71$ at
$\Mtwooo=10^{15}\Msun$ for their full and relaxed samples,
respectively, with an intrinsic scatter of $\approx 0.33$.
Our result for the full sample ($\Ctwooo=3.42^{+0.14}_{-0.15}$) is in good agreement with their prediction.
Given the scatter and measurement uncertainty, these values are not in
severe tension with what we measured in this work.

We then compare our results to the model of \cite{diemer15}, who also
characterized the halo concentration as a function of the halo peak height.
Their model yields $\Ctwooo=3.73$ for the typical mass of our sample, $\Mtwooo\approx10^{15}\Msun$,
which shows no tension with our measurement (black contours
in the left panel of Figure~\ref{fig:joint_all}).

The comparisons we discussed above can be visualized in 
Figure~\ref{fig:cm}.
To sum up, our results on the $c$--$M$ relation are in satisfactory
agreement with previous lensing studies. A better agreement can be
achieved once the selection function, the cosmology adopted, and/or the
modeling systematics are taken into account.
We find that the typical values of halo concentration ($A_{\Ctwooo}$)
range from $\Ctwooo\approx3$ to $\approx4.5$ at $\Mtwooo=10^{15}\Msun$,
largely depending on the sample selection rather than the modeling
assumption on the shape of clusters.

\begin{figure}
\resizebox{!}{0.5\textwidth}{
\includegraphics[scale=1.0]{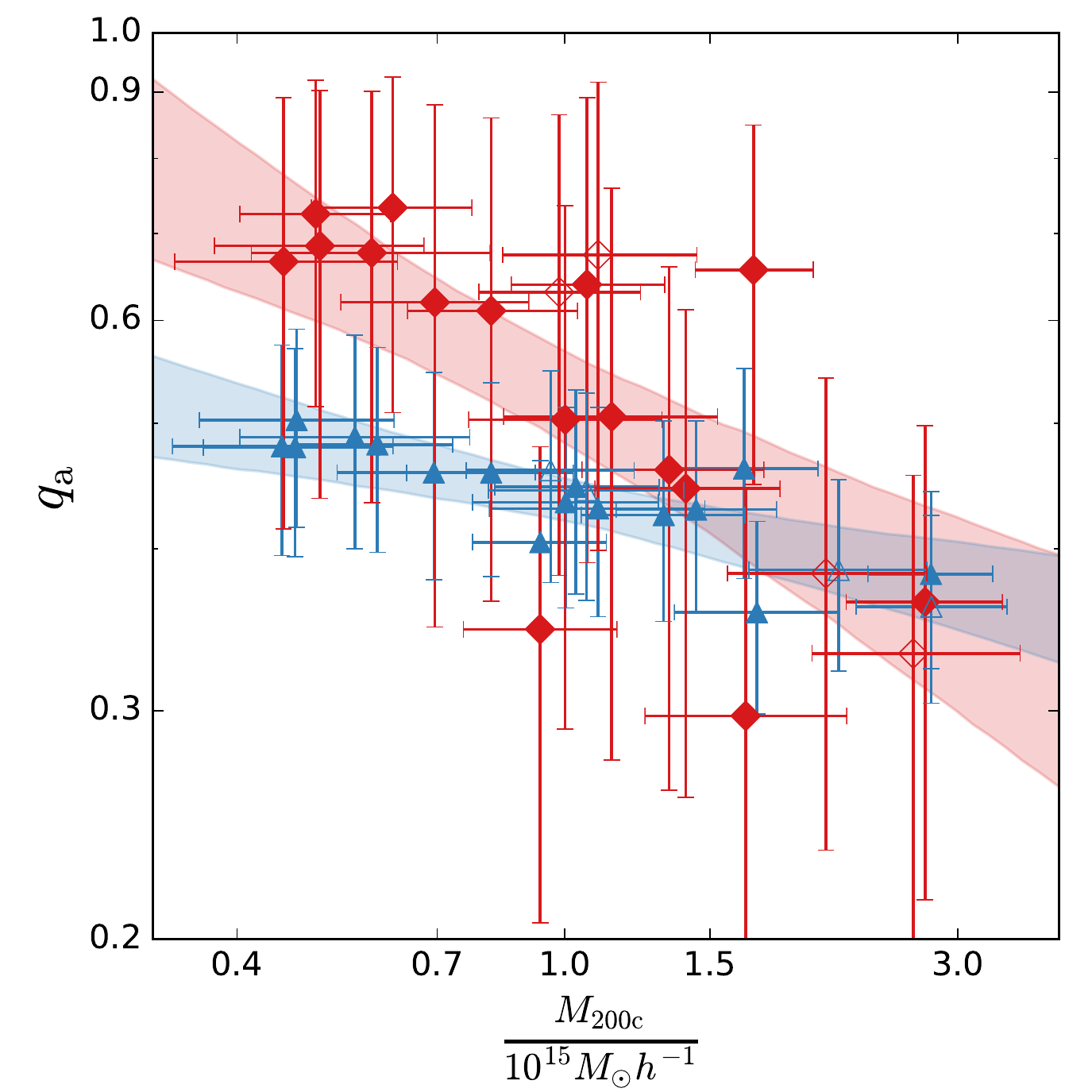}
}
\caption{
Minor-to-major axis ratio to mass scaling relation for our cluster sample.
The results of \triaxialfmodel\ and \triaxialbmodel\ modeling are shown
 by red diamonds and blue triangles, respectively.
The 16 X-ray-selected (4 high-magnification) \CLASH\ clusters are indicated
 by filled (open) markers. 
 The best-fit scaling relations and their $1\sigma$ confidence regions
 are shown by
 shaded areas.
}
\label{fig:qam}
\end{figure}
\begin{figure}
\resizebox{!}{0.5\textwidth}{
\includegraphics[scale=1.0]{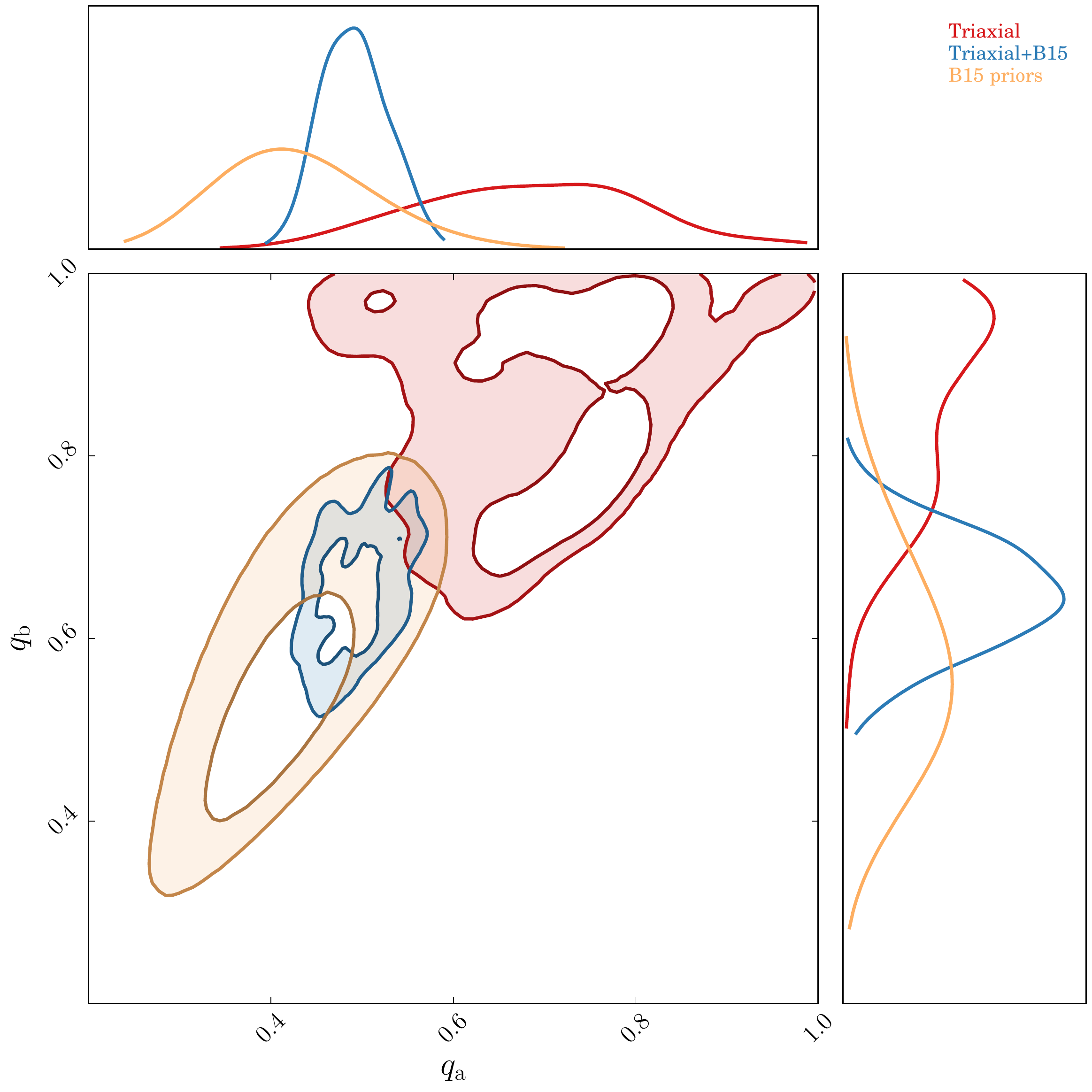}
}
\caption{
Marginalized posterior distributions of the intrinsic axis ratios \qa\
 and \qb\ derived from joint ensemble modeling of our full sample
 of 20 clusters.
The results of the \triaxialfmodel\ and \triaxialbmodel\ modeling are
 shown in red and blue, respectively.
 The B15 prior distribution evaluated at
 the typical mass $\Mtwooo=10^{15}\Msun$ and the median redshift
 $\left\langle\zd\right\rangle = 0.377$ is shown in yellow. 
}
\label{fig:joint_qab}
\end{figure}

\subsection{Axis Ratio to Mass Relations}
\label{sec:qa_mass_relation}

Here we present the minor-to-major axis ratio to mass scaling relation for
our full sample of 20 clusters derived using the \triaxialfmodel\ and
\triaxialbmodel\ modeling approaches (Equations~(\ref{eq:qa_tri}) and (\ref{eq:qa_bon})):
\begin{equation}
\label{eq:qa_tri}
\qa = \left(\numAqaTri\right) \times \left( \frac{\Mtwooo}{10^{15}\Msun} \right)^{\numBqaTri},
\end{equation}
and
\begin{equation}
\label{eq:qa_bon}
\qa = \left(\numAqaBon\right) \times \left( \frac{\Mtwooo}{10^{15}\Msun} \right)^{\numBqaBon},
\end{equation}
with an intrinsic scatter $D_{\qa|\Mtwooo}$ of \numDqaTri\ and \numDqaBon, respectively.

We plot these results in Figure~\ref{fig:qam} in a similar manner as in Figure~\ref{fig:cm}.
There is no clear difference in the resulting $\qa$--$M$ relations
between the X-ray-selected and high-magnification subsamples. We thus focus
on the results based on the full sample hereafter. 
We see from Figure~\ref{fig:qam} that in \triaxialfmodel\ modeling,
the errors of \qa\ for individual clusters are considerably large.
However, the statistical ensemble behavior  
shows that $\qa \approx\left(\numAqaTri\right)$ at
$\Mtwooo=10^{15}\Msun$ and it scales as $\qa \propto
\Mtwooo^{\numBqaTri}$ at the $2.7\sigma$ level,
indicating that
more massive clusters tend to be less spherical.
On the other hand, introducing informative shape priors in
\triaxialbmodel\ modeling yields an ensemble average of 
$\qa\approx\left(\numAqaBon\right)$ at the pivot mass
$\Mtwooo=10^{15}\Msun$ and a shallower slope of
$B_{\qa}\approx\left(\numBqaBon\right)$.
This corresponds to a marginal shfit in $A_{\qa}$ and $B_{\qa}$
at the $2\sigma$ and $1.3\sigma$ levels,
respectively, with respect to the case using uniform priors.
On the basis of the results above, we have detected a non-spherical
shape of the clusters. The average minor-to-major axis ratio \qa\ is
$\approx0.5$, depending on the priors used, and it monotonically
decreases with increasing cluster mass at the $\lesssim2.7\sigma$ level.

In Figure~\ref{fig:joint_all}, we show joint ensemble constraints on the 
concentration, mass, and axis ratios for the full sample
obtained with different modeling approaches and data sets.
It is seen in the right panel of Figure~\ref{fig:joint_all} that there
is no clear correlation between the shape parameters (i.e., axis ratios)
and the overall structural parameters (i.e., concentration and mass) regardless of the priors.
This is consistent with the implication in
Section~\ref{sec:concentration_mass_relation} that the assumption of
cluster shapes does not statistically affect the $c$--$M$ relation of
the \CLASH\ clusters.  
We see from the right panel of Figure~\ref{fig:joint_all} that
including the {\em HST} lensing data $\Msl(<r)$ results in a lower
concentration, but it does not alter the axis ratios.   
Conversely, introducing the informative shape priors from B15
has an impact on the axis ratios, but not on the mass and concentration
parameters. 

\begin{figure}
\resizebox{!}{0.5\textwidth}{
\includegraphics[scale=1.0]{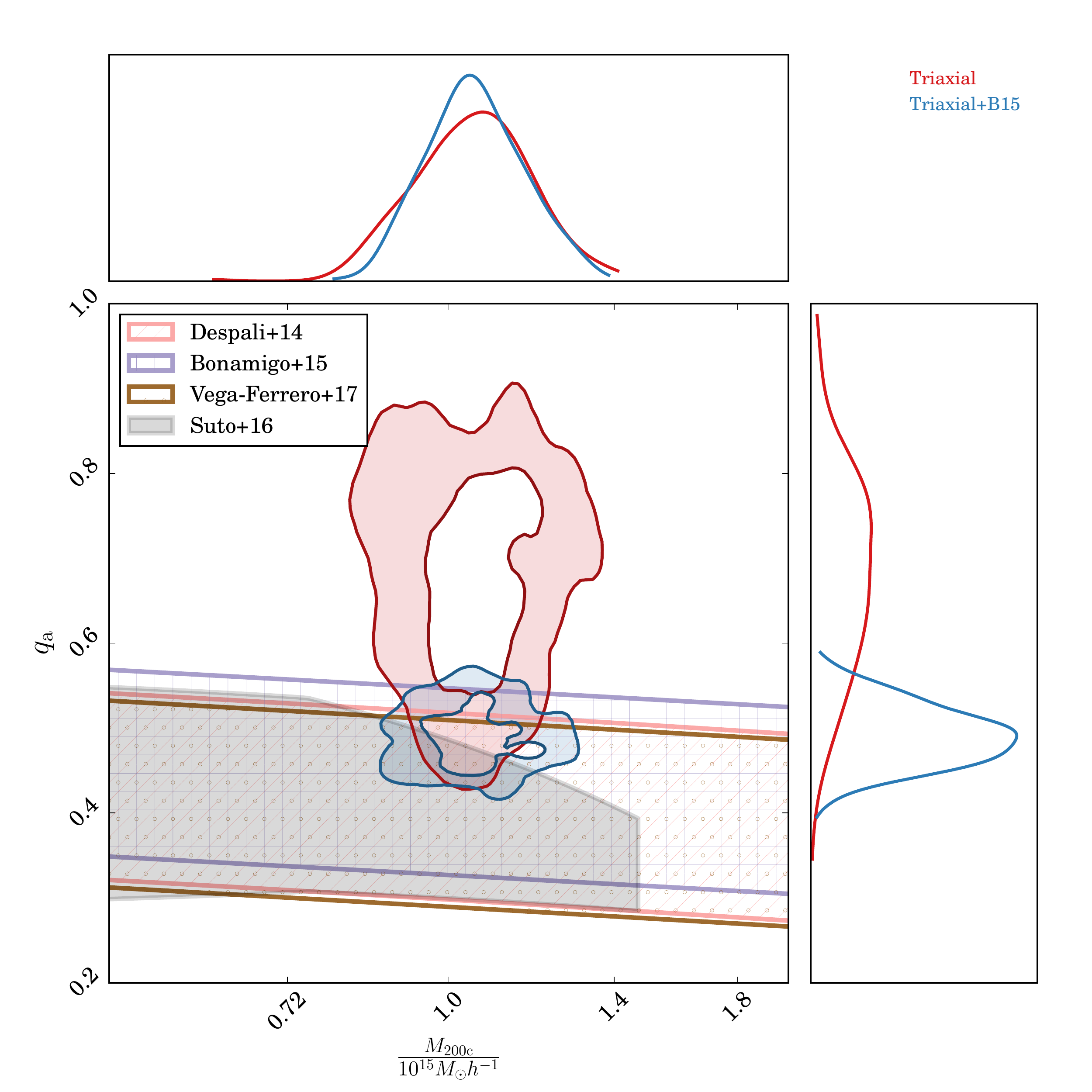}
}
\caption{
Comparison between the observational constraints on \qa\ and the theoretical
 predictions of \citet{despali14}, B15, \citet{suto16}, and
 \citet{vegaferrero17} evaluated at the median redshift of $\left\langle\zd\right\rangle=0.377$.
The shaded regions indicate the intrinsic scatter of \qa\ at fixed halo
 mass predicted by the respective theoretical models.
}
\label{fig:comp_sims_qa}
\end{figure}
\begin{figure}
\resizebox{!}{0.5\textwidth}{
\includegraphics[scale=1.0]{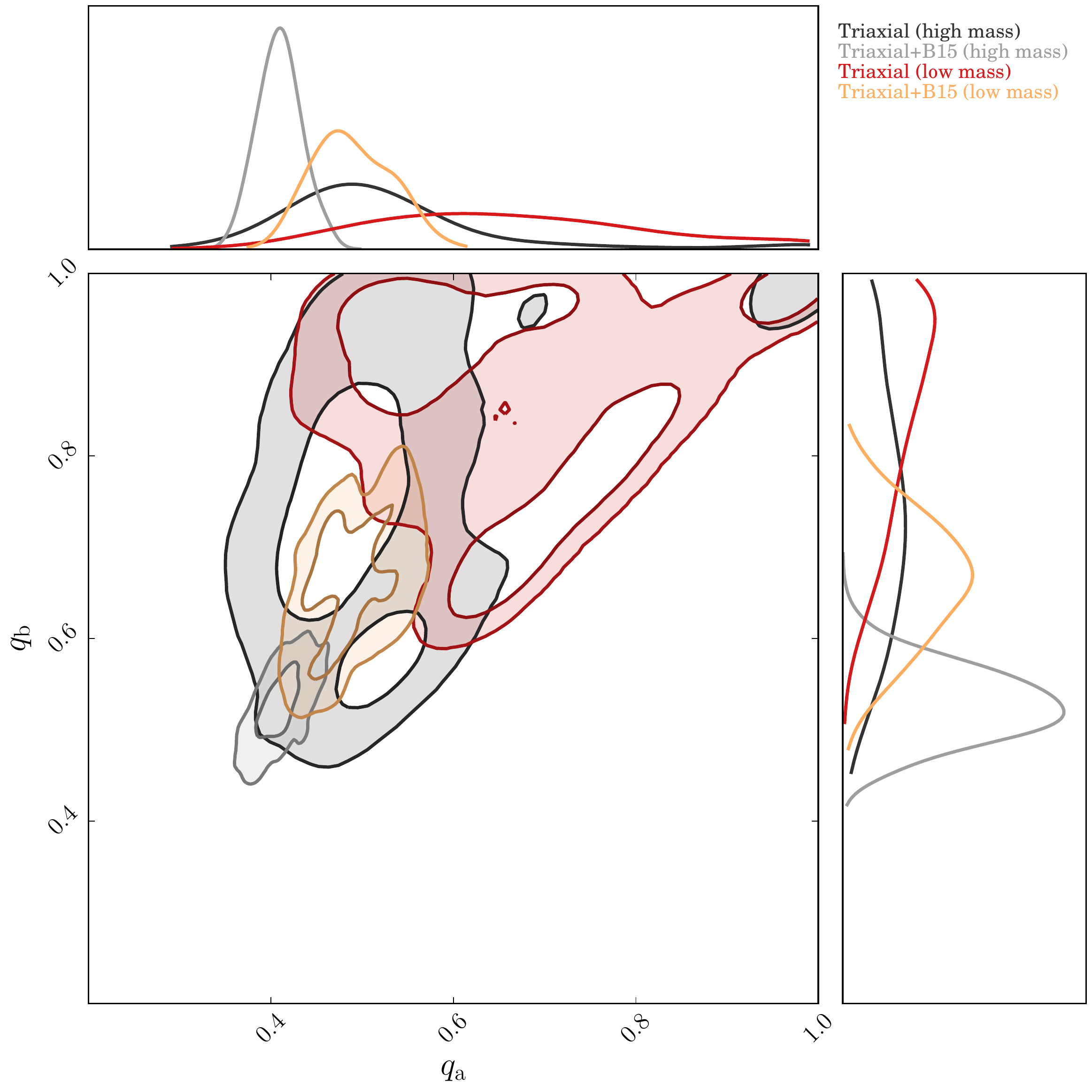}
}
\caption{
Marginalized posterior distributions of the intrinsic axis ratios \qa\
 and \qb\ derived from joint ensemble modeling of the low-mass
 and high-mass subsamples. 
The results of the \triaxialfmodel\ and \triaxialbmodel\ modeling for
 the low (high) mass subsample are shown in red and yellow (black
 and gray), respectively. 
}
\label{fig:joint_qab_subsample}
\end{figure}

We show in Figure~\ref{fig:joint_qab} our joint ensemble constraints on
the intrinsic shape parameters for our full sample of 20 clusters, along
with the B15 prior distribution. 
We constrain the minor-to-major axis ratio as
$\qa=0.652^{+0.162}_{-0.078}$ using uniform priors, which is
higher than the expectation from the B15 prior distribution,
$\qa=0.412^{+0.095}_{-0.083}$, at the $\lesssim1.5\sigma$ level. 
Employing the B15 priors in \triaxialbmodel\ modeling yields
$\qa=0.499^{+0.018}_{-0.056}$, which is consistent with the
\triaxialfmodel\ constraint at the $\lesssim1\sigma$ level,
given the long tail of the posterior distribution.

Furthermore, we compare in Figure~\ref{fig:comp_sims_qa} our joint
ensemble constraints on \qa\ for the full sample with theoretical
predictions from $N$-body numerical simulations of \citet{despali14},
B15, \citet{suto16}, and \citet{vegaferrero17}.
Note that we evaluate the theoretical predictions in Figure~\ref{fig:comp_sims_qa} at the median redshift of our cluster sample, $\left\langle\zd\right\rangle=0.377$.
We see that our constraints on \qa\ obtained using uniform
priors are in favor of the axis ratio that is higher than the
theoretical predictions. However, this trend is only at the
$\lesssim1.5\sigma$ level and not statistically significant.
It is worth mentioning that including baryonic physics in numerical
simulations results in a rounder shape of galaxy clusters
\citep[][]{kazantzidis04,bryan13,suto17}, which better agrees 
with our results based on the uniform priors than the purely $N$-body
simulations. 
With upcoming large cluster surveys to dramatically improve statistics,
this work demonstrates an opportunity to constrain the
effects of baryonic feedback on the halo shape by using gravitational
lensing.

Conversely, we do not have informative constraints on the second axis
ratio \qb\ in \triaxialfmodel\ modeling with uniform priors (see
Table~\ref{tab:results} and the right panel of
Figure~\ref{fig:joint_all}):  
We can only constrain the lower bound of \qb\ for the full sample
as $0.73$, $0.63$ and
$0.50$ at the $1\sigma$, $2\sigma$ and $3\sigma$ levels, respectively.
Introducing the B15 priors in \triaxialbmodel\ modeling gives
$\qb=0.636^{+0.078}_{-0.045}$, compared to the expectation from the B15
prior distribution, $\qb=0.55^{+0.14}_{-0.11}$. 
Taking the covariance between \qa\ and \qb\ into account, the overall
discrepancy between our lensing data and the B15 priors
is at the $\lesssim2.5\sigma$ level.
Therefore, we do not have statistically significant evidence for a
strong tension between the lensing data and the B15 simulations.

Additionally, we show in Figure~\ref{fig:joint_qab_subsample} joint
ensemble constraints on the axis ratios 
for the low-mass and high-mass subsamples.
We observe that (1) the discrepancy between the \triaxialfmodel\
modeling and \triaxialbmodel\ modeling is smaller for the high mass
samples, and (2) the constraints are significantly stronger for the high
mass sample, suggesting that the weak constraints on the shape
parameters for the full sample are likely due to the inclusion of the
low mass clusters. 

We note that we currently do not have compelling constraints on the
intrinsic shape (especially \qb) of clusters based on the lensing data
alone (using uninformative uniform priors).
Nevertheless, we observe a marginal discrepancy between the lensing data
and simulations, which can be better examined with a large statistical
sample of clusters.

\subsection{Geometrical Factor to Mass Relations}
\label{sec:fgeo_mass_relation}

We constrain the geometrical factor to mass scaling relation for our full
sample of 20 clusters from \triaxialfmodel\ and \triaxialbmodel\ modeling as
\begin{equation}
\label{eq:fgeo_tri}
\fgeo = \left(\numAfgeoTri\right) \times \left(\frac{\Mtwooo}{10^{15}\Msun}\right)^{\numBfgeoTri},
\end{equation}
and
\begin{equation}
\label{eq:fgeo_bon}
\fgeo = \left(\numAfgeoBon\right) \times \left(\frac{\Mtwooo}{10^{15}\Msun}\right)^{\numBfgeoBon},
\end{equation}
with an intrinsic scatter $D_{\fgeo|\Mtwooo}$ of \numDfgeoTri\ and \numDfgeoBon, respectively.
The geometrical factor \fgeo\ is a derived quantity from the posterior
distributions of the triaxial NFW parameters according to Equation~(\ref{eq:fgeo_def}).
Specifically, \fgeo\ is defined as the ratio of the line-of-sight half
length of a triaxial ellipsoid to the geometric-mean scale-radius
of its isodensity contour projected on the sky.
Therefore, it represents the degree of line-of-sight elongation of the mass distribution. 
A geometrical factor greater (smaller) than unity indicates a line-of-sight
excess (deficit) of mass structure. 

\begin{figure}
\resizebox{!}{0.5\textwidth}{
\includegraphics[scale=1.0]{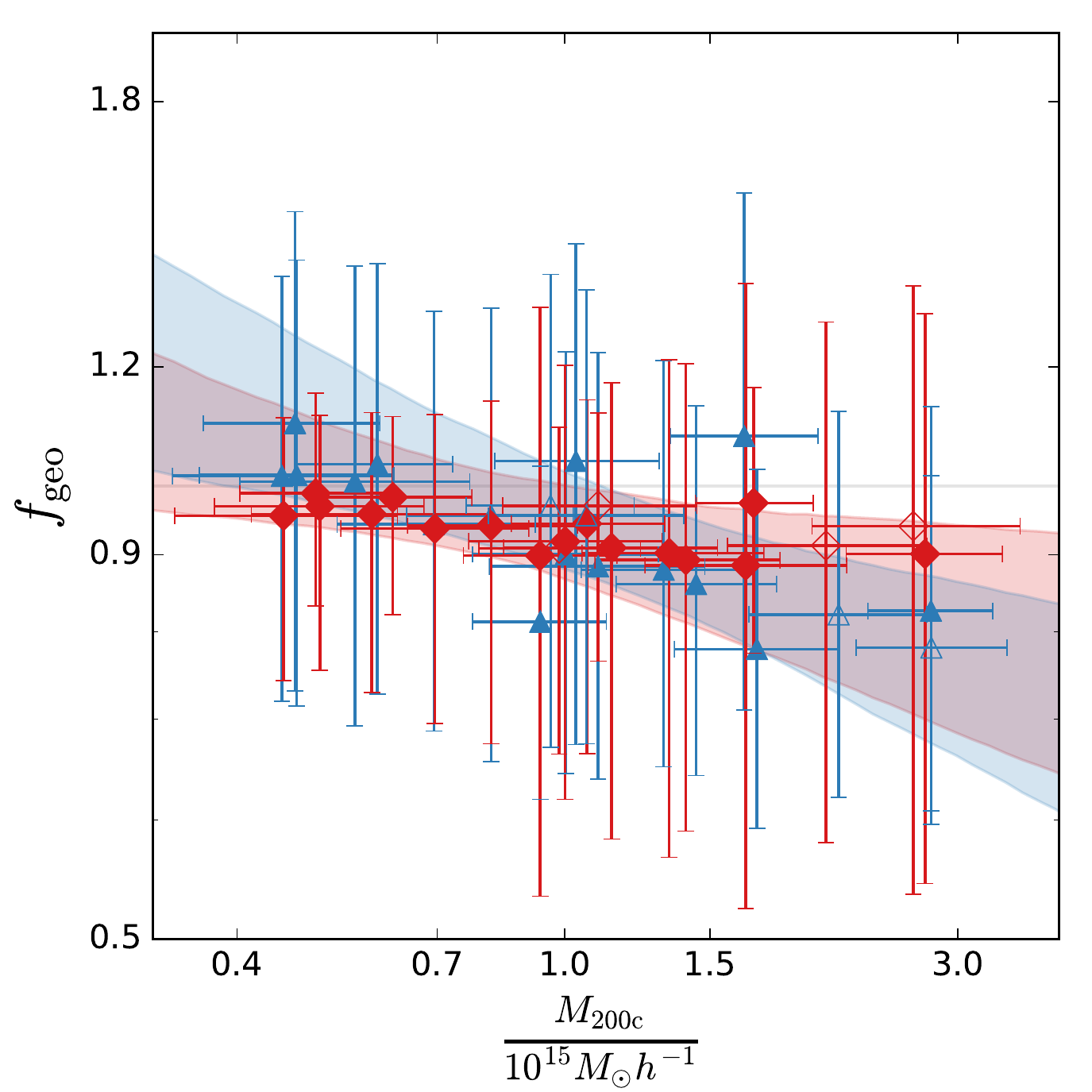}
}
\caption{
Geometrical factor ($\fgeo$) to mass scaling relation.
The results of \triaxialfmodel\ and \triaxialbmodel\ modeling are
 shown by red diamonds and blue triangles, respectively.
The 16 X-ray-selected (4 high-magnification) \CLASH\ clusters are indicated
 by filled (open) markers. 
The best-fit scaling relations and their $1\sigma$ confidence regions
 are shown by shaded areas.
The gray thin line indicates  $\fgeo=1$.
}
\label{fig:fgeom}
\end{figure}

We show the results of the geometrical factor to mass scaling relations in
Figure~\ref{fig:fgeom}.
Although the geometrical factor is a very noisy quantity for individual
clusters, the ensemble behavior from the best-fit scaling relations
suggests no significant deviation from random orientations (i.e., \fgeo\
is consistent with unity within the quoted $1\sigma$ uncertainties)
regardless of the shape priors (uniform or B15). 
We find no significant evidence of a dependence of $\fgeo$ on cluster mass.
A mild trend at the $\lesssim2\sigma$ level is found when the B15 priors
are employed.

We stress that, given the fact that lensing can only probe the
integrated mass along the line of sight, we do not have a sensitivity to 
the line-of-sight elongation of clusters using lensing data alone
\citep{dietrich14}. 
Hence, we must rely on external information (e.g., X-ray/SZE data or
simulation priors) to better constrain the orientation of clusters. 
With the B15 priors applied on the axis ratios, we extract the
biweight center location of the marginalized posterior distribution of
$\cos(\theta)$ for each cluster (Section~\ref{sec:results_and_discussion}).
The distribution of $\cos(\theta)$ spans a wide
range between $0.37$ and $0.70$, with a median value of
$\left\langle\cos(\theta)\right\rangle=0.55$ and a typical biweight
scale of $\approx0.30$.
This corresponds to a mean inclining angle of
$\cos^{-1}\left(\left\langle\cos(\theta)\right\rangle\right)\approx57\deg$,
suggesting that the orientations of our sample are nearly random
(i.e., no orientation bias). This is consistent with our results
on the $\fgeo$--$M$ relation (Table~\ref{tab:sr}) regardless of the
chosen sample, and in line with the
theoretical expectation for the X-ray-selected \CLASH\ clusters
\citep{meneghetti14}.
It is worth mentioning that a positive bias at a level of
$3\percent$--$6\percent$ 
was suggested in the mass estimates of stacked weak-lensing measurements
for optically selected clusters \citep{dietrich14}, while there is no
clear indication of orientation bias for our clusters that are largely
selected by their X-ray properties.

\subsection{Triaxiality to Mass Relations}
\label{sec:triaxiality_mass_relation}

The degree of triaxiality \triaxiality\ is a quantity derived from the
posterior distributions of the intrinsic axis ratios 
(Equation~(\ref{eq:triaxiality_def})). 
A prolate mass distribution (i.e., $\qa=\qb$) has $\triaxiality=1$,
while a oblate shape (i.e., $\qa<\qb=1$) has $\triaxiality=0$.

\begin{figure}
\resizebox{!}{0.5\textwidth}{
\includegraphics[scale=1.0]{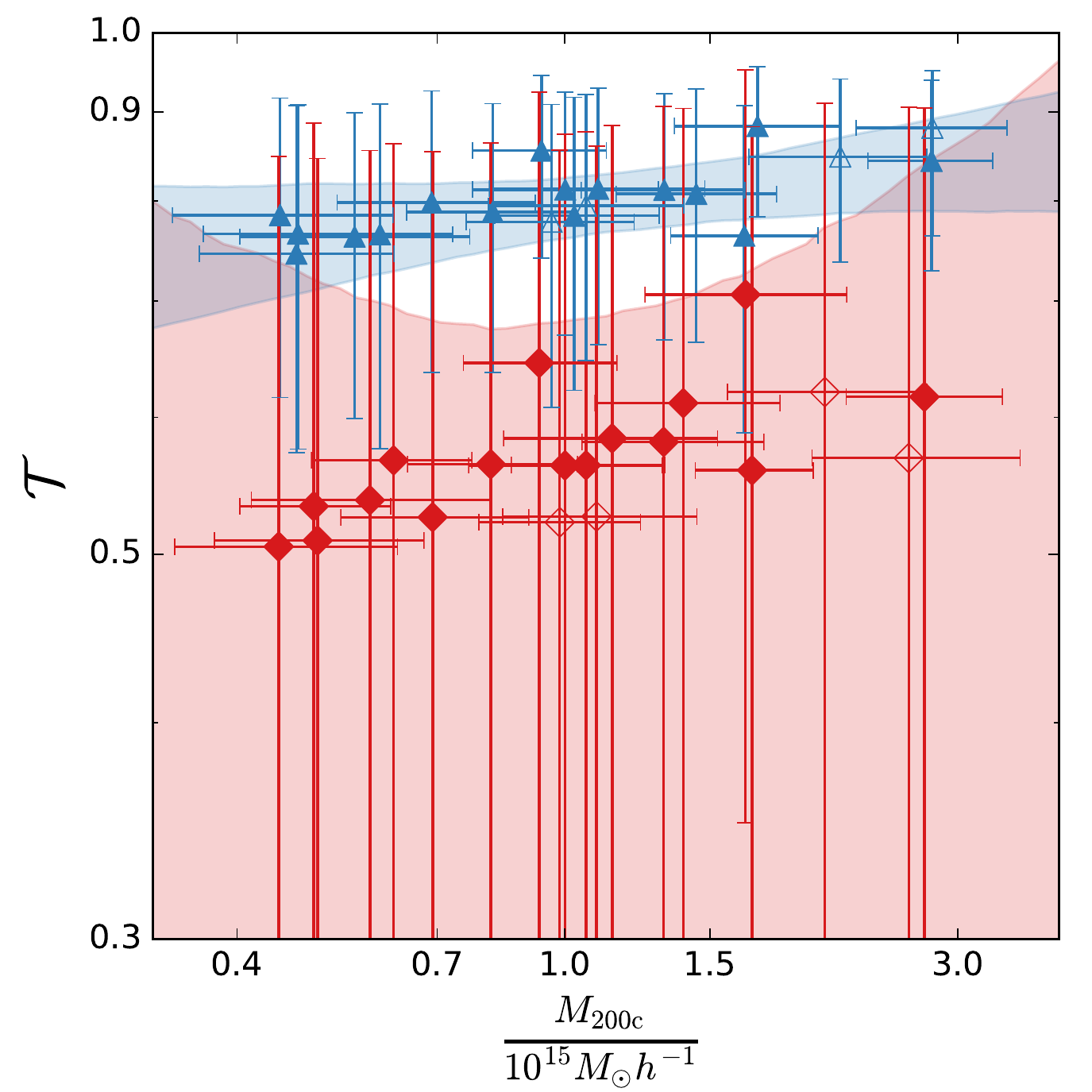}
}
\caption{
Triaxiality to mass scaling relation.
The results of \triaxialfmodel\ and \triaxialbmodel\ modeling are
 shown by red diamonds and blue triangles, respectively.
The 16 X-ray-selected (4 high-magnification) \CLASH\ clusters are indicated
 by filled (open) markers. 
The $1\sigma$ confidence region of the \triaxialbmodel\ modeling is
 shown by the blue shaded area.
The $2\sigma$ upper bound of the scaling relation from the
 \triaxialfmodel\ modeling is indicated by the red shaded area.
}
\label{fig:triaxialitym}
\end{figure}

We stress again that we can only constrain the lower bound of the second
axis ratio \qb\ and thus the upper bound on the degree of triaxiality
\triaxiality\ from \triaxialfmodel\ modeling when using uniform priors. 
Accordingly, we only present the $2\sigma$ upper bound for the results
from \triaxialfmodel\ modeling.
The $2\sigma$ upper bound on the $\triaxiality$--$M$ relation from
\triaxialfmodel\ modeling is   
\begin{equation}
\label{eq:triaxiality_tri}
\triaxiality < 0.69 \times \left(\frac{\Mtwooo}{10^{15}\Msun}\right)^{\numBtriaxialityTri},
\end{equation}
while the best-fit $\triaxiality$--$M$ relation from \triaxialbmodel\ modeling is
\begin{equation}
\label{eq:triaxiality_bon}
\triaxiality = \left(\numAtriaxialityBon\right) \times \left(\frac{\Mtwooo}{10^{15}\Msun}\right)^{\numBtriaxialityBon},
\end{equation}
with an intrinsic scatter of $\numDtriaxialityBon$. 
We show the results of the $\triaxiality$--$M$ relation as well as the
individual cluster constraints in Figure~\ref{fig:triaxialitym}.

\begin{figure}
\resizebox{!}{0.5\textwidth}{
\includegraphics[scale=1.0]{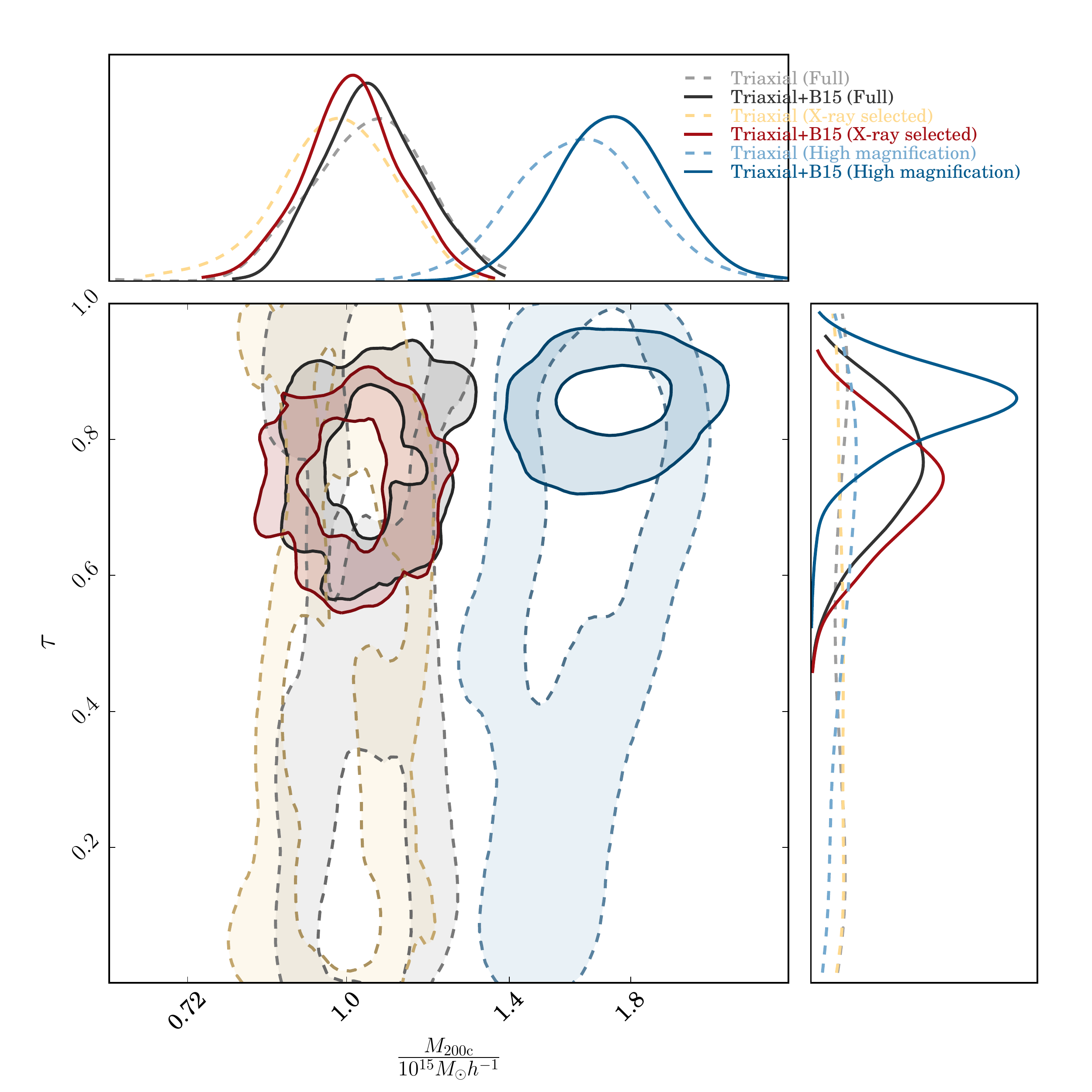}
}
\caption{
Constraints on the degree of triaxiality \triaxiality\ and cluster mass
 \Mtwooo\ from joint ensemble modeling of different cluster subsamples.
 Using uninformative uniform priors, the constraints on \triaxiality\ from
 \triaxialfmodel\ modeling are shown in gray (yellow, light blue) for the
 full (X-ray selected, high-magnification) sample. 
Using the B15 priors, the constraints on \triaxiality\ are shown in
 black (red, dark blue) for the full (X-ray selected, high-magnification)
 sample.
}
\label{fig:joint_triaxiality}
\end{figure}

In Figure~\ref{fig:triaxialitym}, a clear offset in the normalization
$A_{\triaxiality}$ is seen between the \triaxialfmodel\ and
\triaxialbmodel\ modeling results:  
The degree of triaxiality is constrained as
$\triaxiality \approx\numAtriaxialityBon$
at the pivot mass $\Mtwooo=10^{15}\Msun$ using the B15 priors,
and the offset in the normalization relative to the \triaxialfmodel\ results
is at the $\approx3\sigma$ level. 
This discrepancy is strongly driven by the fact that we can only
constrain the lower bound of the intermediate-to-major axis ratio \qb\
from \triaxialfmodel\ modeling, resulting in a nearly flat distribution of
\triaxiality.
This can be further seen in Figure~\ref{fig:joint_triaxiality}, where we
plot \triaxiality\ against cluster mass using the posteriors 
joint ensemble modeling (full, X-ray-selected, and high-magnification samples).
Without the B15 priors, \triaxiality\ is essentially unconstrained by
the lensing data alone. On the other hand, employing the B15 priors gives
$\triaxiality\approx0.8$, implying a prolate
configuration of the CLASH clusters.
We find that the posterior constraints on the mass slope are not
sensitive to the chosen prior ($B_{\triaxiality}\approx0.07$),
although the uncertainties are too large to claim a significant mass
dependence.
Even though the \CLASH\ clusters exhibit non-spherical shapes,
we echo that spherical symmetry is a well-validated assumption in
estimating the cluster mass and concentration (see
Section~\ref{sec:concentration_mass_relation}). 

%

%
%

\section{Conclusions}
\label{sec:conclusions}

In this paper, we have combined the wide-field weak-lensing mass maps
obtained by \cite{umetsu18} with the central  CLASH-{\em HST} 
lensing constraints \citep{zitrin15,umetsu16} to perform
three-dimensional modeling of the intrinsic mass 
distribution for a sample of 16 X-ray-selected and 4
high-magnification clusters targeted by the \CLASH\ survey.
These clusters span a mass range of
$4\times10^{14}\Msun\lesssim\Mtwooo<20\times10^{14}\Msun$ 
and a redshift range of
$0.18<\redshift<0.7$, with a median redshift of $\approx0.377$.
Specifically, we have forward-modeled these lensing data sets assuming a
triaxial NFW halo in a Bayesian
framework, and constrained the mass \Mtwooo, concentration \Ctwooo,
intrinsic axis ratios (minor-to-major ratio \qa\ and intermediate-to-major
ratio \qb), and orientation angles.
For the case of triaxial modeling, we considered either uniform priors on all
parameters or, alternatively, informative shape priors on \qa\ and \qb\ taken
from cosmological $N$-bod simulations  \citep{bonamigo15}.
We have also performed spherical NFW modeling with the \Mtwooo\
and \Ctwooo\ parameters,
while fixing the other parameters to the spherical configuration.
We performed Bayesian modeling of both
individual and ensemble clusters using the combined weak and strong
lensing data sets. 
With the observed constraints on each individual cluster, we have
investigated mass-scaling relations of the halo concentration \Ctwooo,
the minor-to-major axis ratio \qa, the geometrical factor \fgeo, and the
degree of triaxiality \triaxiality.

Our results show that the halo concentration decreases with increasing
mass, as found by previous work assuming spherical symmetry. The results
are insensitive to both the assumed cluster geometry (spherical or
triaxial) and the chosen shape prior.
However, we find that the selection of clusters plays an important role.
The four high-magnification \CLASH\ clusters 
($\Mtwooo\approx1.5\times10^{15}\Msun$) have  
a significantly low concentration,
compared to the X-ray-selected \CLASH\ subsample.
For the 16 X-ray-selected \CLASH\ clusters, we find a mean concentration
of 
$\Ctwooo=\numAconcenTri$ at the pivot mass $\Mtwooo=10^{15}\Msun$, and it scales as
$\Mtwooo^{\numBconcenTri}$ according to triaxial modeling with uniform
priors. 
On the other hand,
jointly modeling this subsample assuming a triaxial NFW halo,
we obtain joint ensemble constraints of $\Ctwooo=3.87^{+0.76}_{-0.11}$ and
$\Mtwooo=\left(0.99\pm0.11\right)\times10^{15}\Msun$.
Our results are consistent with previous work from observations and
simulations. A better agreement can be achieved if accounting for the
sample selection, geometry of clusters, the background cosmology adopted, and the choice of the priors.
The results from triaxial modeling are in good agreement with those from
spherical modeling within the errors, suggesting that the
assumption of spherical symmetry is well validated in estimating the
overall mass profile of the \CLASH\ clusters, even though we do observe
evidence of aspherical shapes of clusters.

When using uninformative uniform priors, we obtain joint ensemble constraints on the
minor-to-major axis ratio of  $\qa=0.652^{+0.162}_{-0.078}$ at the
typical mass $\Mtwooo=10^{15}\Msun$ for our full sample of 20 \CLASH\ clusters.
Conversely, only a lower bound on the intermediate-to-major axis ratio
\qb\ is obtained as $\qb>0.632$ at the $2\sigma$ level. 
Using the B15 priors gives improved joint ensemble constraints of
$\qa=0.499^{+0.018}_{-0.056}$ and $\qb=0.636^{+0.078}_{-0.045}$, respectively.
The resulting $\qa$--$M$ relation suggests that \qa\ decreases with
increasing halo mass as $\Mtwooo^{\numBqaTri}$ and
$\Mtwooo^{\numBqaBon}$ based on the results with the uniform and B15 priors,
respectively. 
Overall, no significant tension is seen between the lensing data and the
numerical predictions from B15 in terms of the intrinsic cluster axis
ratios.
Our results suggest that we currently do not have strong constraints
($\lesssim3\sigma$) on the intrinsic shape of clusters based on
gravitational lensing alone, unless informative shape priors are
employed.

We have also studied the geometrical factor \fgeo, an indicator of the
line-of-sight elongation of cluster mass distributions. 
We find that our sample shows no significant deviation from isotropic,
random orientations: $\fgeo=\numAfgeoTri$ and
$\fgeo=\numAfgeoBon$ based on the uniform and B15 priors, respectively.
The results are in agreement with the theoretical expectation for the
CLASH clusters dominated by relaxed systems \citep{meneghetti14}. 
No significant mass dependence of $\fgeo$ is seen
regardless of the chosen prior.
The average inclination angle $\theta$ between the cluster major axis and the
line of sight is
$\left\langle\theta\right\rangle\approx57\deg$, suggesting again that
there is no evidence of orientation bias for the \CLASH\ clusters.

Finally, the degree of triaxiality for our sample is
constrained as $\triaxiality = \numAtriaxialityBon$ at the pivot mass
$\Mtwooo=10^{15}\Msun$ using the B15 priors, suggesting that the
geometry of our sample is close 
to the prolate configuration ($\triaxiality=1$) rather than the oblate
one ($\triaxiality=0$). 
However, we stress that this result strongly depends on the choice of the
shape priors. With the uniform priors, we can only constrain the upper
bound of \triaxiality\ as $\triaxiality < 0.69$ at the $2\sigma$ level. 
No significant mass trend of triaxiality is observed in our sample
regardless of the priors. 

We have presented a statistical three-dimensional analysis of a sizable
sample of high-mass galaxy clusters using high-quality weak and strong
lensing data sets.  
We observed clear evidence of a departure from spherical symmetry in our
sample of 20 clusters.
On the other hand, we find that the assumption of spherical symmetry is
still well validated in terms of determining the overall mass profile
(such as concentration and mass) if the sample is free from orientation
bias. 
We find increasingly promising constraints on the intrinsic shape
parameters with increasing halo mass or with increasing size of cluster   
sample. Therefore, it will be very desirable to extend this type of
analysis to large, well-controlled samples of clusters defined from
ongoing large-sky surveys, such as the Subaru Hyper Suprime-Cam survey
\citep{miyazaki15} and the Dark Energy Survey \citep{flaugher05}.
 
%
%

\section*{Acknowledgments}

We thank Tetsu Kitayama and Daichi Suto for providing us with simulated
data points that are presented in Figure~\ref{fig:comp_sims_qa}.
We thank the anonymous referee for providing constructive suggestions that lead to the improvement of this paper.
KU acknowledges support from the Ministry of Science and Technology of
Taiwan (grants MoST 103-2112-M-001-030-MY3 and MoST 106-2628-M-001-003-MY3)
and from the Academia Sinica Investigator Award.
MS acknowledges financial support from the contracts ASI-INAF
I/009/10/0, NARO15 ASI-INAF I/037/12/0, ASI 2015-046-R.0 and ASI-INAF
n.2017-14-H.0. 
JS was supported by NSF/AST-1617022.
MM, MS, SE, JS acknowledge support from the Italian Ministry of Foreign Affairs and International Cooperation, Directorate General for Country Promotion (Project "Crack the lens").
This work was made possible by the availability of high-quality lensing
data produced by the \CLASH\ survey.

This paper made use of the code \texttt{colossus} \citep{diemer17} and the packages from \cite{bocquet16b} and \cite{hinton16} for plotting.
This work made use of the IPython package \citep{PER-GRA:2007}, SciPy \citep{jones_scipy_2001}, TOPCAT, an interactive graphical viewer and editor for tabular data \citep{2005ASPC..347...29T}, matplotlib, a Python library for publication quality graphics \citep{Hunter:2007}, Astropy, a community-developed core Python package for Astronomy \citep{2013A&A...558A..33A}, NumPy \citep{van2011numpy}. 

%
%

\bibliography{literature}

\begin{thebibliography}{125}
\expandafter\ifx\csname natexlab\endcsname\relax\def\natexlab#1{#1}\fi

\bibitem[{{Allgood} {et~al.}(2006){Allgood}, {Flores}, {Primack}, {Kravtsov},
  {Wechsler}, {Faltenbacher}, \& {Bullock}}]{allgood06}
{Allgood}, B., {Flores}, R.~A., {Primack}, J.~R., {et~al.} 2006,
  \href{http://dx.doi.org/10.1111/j.1365-2966.2006.10094.x}{\mnras, 367, 1781}

\bibitem[{{Annunziatella} {et~al.}(2014){Annunziatella}, {Biviano}, {Mercurio},
  {Nonino}, {Rosati}, {Balestra}, {Presotto}, {Girardi}, {Gobat}, {Grillo},
  {Kelson}, {Medezinski}, {Postman}, {Scodeggio}, {Brescia}, {Demarco},
  {Fritz}, {Koekemoer}, {Lemze}, {Lombardi}, {Sartoris}, {Umetsu}, {Vanzella},
  {Bradley}, {Coe}, {Donahue}, {Infante}, {Kuchner}, {Maier}, {Reg{\H o}s},
  {Verdugo}, \& {Ziegler}}]{annunziatella14}
{Annunziatella}, M., {Biviano}, A., {Mercurio}, A., {et~al.} 2014,
  \href{http://dx.doi.org/10.1051/0004-6361/201424102}{\aap, 571, A80}

\bibitem[{{Astropy Collaboration} {et~al.}(2013){Astropy Collaboration},
  {Robitaille}, {Tollerud}, {Greenfield}, {Droettboom}, {Bray}, {Aldcroft},
  {Davis}, {Ginsburg}, {Price-Whelan}, {Kerzendorf}, {Conley}, {Crighton},
  {Barbary}, {Muna}, {Ferguson}, {Grollier}, {Parikh}, {Nair}, {Unther},
  {Deil}, {Woillez}, {Conseil}, {Kramer}, {Turner}, {Singer}, {Fox}, {Weaver},
  {Zabalza}, {Edwards}, {Azalee Bostroem}, {Burke}, {Casey}, {Crawford},
  {Dencheva}, {Ely}, {Jenness}, {Labrie}, {Lim}, {Pierfederici}, {Pontzen},
  {Ptak}, {Refsdal}, {Servillat}, \& {Streicher}}]{2013A&A...558A..33A}
{Astropy Collaboration}, {Robitaille}, T.~P., {Tollerud}, E.~J., {et~al.} 2013,
  \href{http://dx.doi.org/10.1051/0004-6361/201322068}{\aap, 558, A33}

\bibitem[{{Bailin} \& {Steinmetz}(2005)}]{bailin05}
{Bailin}, J., \& {Steinmetz}, M. 2005,
  \href{http://dx.doi.org/10.1086/430397}{\apj, 627, 647}

\bibitem[{{Balestra} {et~al.}(2013){Balestra}, {Vanzella}, {Rosati}, {Monna},
  {Grillo}, {Nonino}, {Mercurio}, {Biviano}, {Bradley}, {Coe}, {Fritz},
  {Postman}, {Seitz}, {Scodeggio}, {Tozzi}, {Zheng}, {Ziegler}, {Zitrin},
  {Annunziatella}, {Bartelmann}, {Benitez}, {Broadhurst}, {Bouwens}, {Czoske},
  {Donahue}, {Ford}, {Girardi}, {Infante}, {Jouvel}, {Kelson}, {Koekemoer},
  {Kuchner}, {Lemze}, {Lombardi}, {Maier}, {Medezinski}, {Melchior},
  {Meneghetti}, {Merten}, {Molino}, {Moustakas}, {Presotto}, {Smit}, \&
  {Umetsu}}]{balestra13}
{Balestra}, I., {Vanzella}, E., {Rosati}, P., {et~al.} 2013,
  \href{http://dx.doi.org/10.1051/0004-6361/201322620}{\aap, 559, L9}

\bibitem[{{Balestra} {et~al.}(2016){Balestra}, {Mercurio}, {Sartoris},
  {Girardi}, {Grillo}, {Nonino}, {Rosati}, {Biviano}, {Ettori}, {Forman},
  {Jones}, {Koekemoer}, {Medezinski}, {Merten}, {Ogrean}, {Tozzi}, {Umetsu},
  {Vanzella}, {van Weeren}, {Zitrin}, {Annunziatella}, {Caminha}, {Broadhurst},
  {Coe}, {Donahue}, {Fritz}, {Frye}, {Kelson}, {Lombardi}, {Maier},
  {Meneghetti}, {Monna}, {Postman}, {Scodeggio}, {Seitz}, \&
  {Ziegler}}]{balestra16}
{Balestra}, I., {Mercurio}, A., {Sartoris}, B., {et~al.} 2016,
  \href{http://dx.doi.org/10.3847/0067-0049/224/2/33}{\apjs, 224, 33}

\bibitem[{Bartelmann \& Schneider(2001)}]{bartelmann01}
Bartelmann, M., \& Schneider, P. 2001,
  \href{http://dx.doi.org/10.1016/S0370-1573(00)00082-X}{\physrep, 340, 291}

\bibitem[{Battaglia {et~al.}(2011)Battaglia, Bond, Pfrommer, \&
  Sievers}]{battaglia11b}
Battaglia, N., Bond, J., Pfrommer, C., \& Sievers, J. 2011, ArXiv:1109.3709,
  \href{http://arxiv.org/abs/1109.3709}{{\sffamily arXiv:1109.3709
  [astro-ph.CO]}}

\bibitem[{Beers {et~al.}(1990)Beers, Flynn, \& Gebhardt}]{beers90}
Beers, T., Flynn, K., \& Gebhardt, K. 1990, \aj, 100, 32

\bibitem[{{Bett} {et~al.}(2007){Bett}, {Eke}, {Frenk}, {Jenkins}, {Helly}, \&
  {Navarro}}]{bett07}
{Bett}, P., {Eke}, V., {Frenk}, C.~S., {et~al.} 2007,
  \href{http://dx.doi.org/10.1111/j.1365-2966.2007.11432.x}{\mnras, 376, 215}

\bibitem[{{Bhattacharya} {et~al.}(2013){Bhattacharya}, {Habib}, {Heitmann}, \&
  {Vikhlinin}}]{bhattacharya13}
{Bhattacharya}, S., {Habib}, S., {Heitmann}, K., \& {Vikhlinin}, A. 2013,
  \href{http://dx.doi.org/10.1088/0004-637X/766/1/32}{\apj, 766, 32}

\bibitem[{Bocquet \& Carter(2016)}]{bocquet16b}
Bocquet, S., \& Carter, F.~W. 2016,
  \href{http://dx.doi.org/10.21105/joss.00046}{The Journal of Open Source
  Software, 1}

\bibitem[{Bocquet {et~al.}(2015)Bocquet, Saro, Mohr, Aird, Ashby, Bautz,
  Bayliss, Bazin, Benson, Bleem, Brodwin, Carlstrom, Chang, Chiu, Cho,
  Clocchiatti, Crawford, Crites, Desai, de~Haan, Dietrich, Dobbs, Foley,
  Forman, Gangkofner, George, Gladders, Gonzalez, Halverson, Hennig,
  Hlavacek-Larrondo, Holder, Holzapfel, Hrubes, Jones, Keisler, Knox, Lee,
  Leitch, Liu, Lueker, Luong-Van, Marrone, McDonald, McMahon, Meyer, Mocanu,
  Murray, Padin, Pryke, Reichardt, Rest, Ruel, Ruhl, Saliwanchik, Sayre,
  Schaffer, Shirokoff, Spieler, Stalder, Stanford, Staniszewski, Stark, Story,
  Stubbs, Vanderlinde, Vieira, Vikhlinin, Williamson, Zahn, \&
  Zenteno}]{bocquet15}
Bocquet, S., Saro, A., Mohr, J., {et~al.} 2015,
  \href{http://dx.doi.org/10.1088/0004-637X/799/2/214}{\apj, 799, 214}

\bibitem[{{Bonamigo} {et~al.}(2015){Bonamigo}, {Despali}, {Limousin}, {Angulo},
  {Giocoli}, \& {Soucail}}]{bonamigo15}
{Bonamigo}, M., {Despali}, G., {Limousin}, M., {et~al.} 2015,
  \href{http://dx.doi.org/10.1093/mnras/stv417}{\mnras, 449, 3171}

\bibitem[{Brada{\v{c}} {et~al.}(2006)Brada{\v{c}}, Clowe, Gonzalez, Marshall,
  Forman, Jones, Markevitch, Randall, Schrabback, \& Zaritsky}]{bradac06}
Brada{\v{c}}, M., Clowe, D., Gonzalez, A., {et~al.} 2006,
  \href{http://dx.doi.org/10.1086/508601}{\apj, 652, 937}

\bibitem[{{Bridle} {et~al.}(2010){Bridle}, {Balan}, {Bethge}, {Gentile},
  {Harmeling}, {Heymans}, {Hirsch}, {Hosseini}, {Jarvis}, {Kirk}, {Kitching},
  {Kuijken}, {Lewis}, {Paulin-Henriksson}, {Sch{\"o}lkopf}, {Velander},
  {Voigt}, {Witherick}, {Amara}, {Bernstein}, {Courbin}, {Gill}, {Heavens},
  {Mandelbaum}, {Massey}, {Moghaddam}, {Rassat}, {R{\'e}fr{\'e}gier}, {Rhodes},
  {Schrabback}, {Shawe-Taylor}, {Shmakova}, {van Waerbeke}, \&
  {Wittman}}]{bridle10}
{Bridle}, S., {Balan}, S.~T., {Bethge}, M., {et~al.} 2010,
  \href{http://dx.doi.org/10.1111/j.1365-2966.2010.16598.x}{\mnras, 405, 2044}

\bibitem[{{Broadhurst} {et~al.}(2005{\natexlab{a}}){Broadhurst}, {Takada},
  {Umetsu}, {Kong}, {Arimoto}, {Chiba}, \& {Futamase}}]{broadhurst05b}
{Broadhurst}, T., {Takada}, M., {Umetsu}, K., {et~al.} 2005{\natexlab{a}},
  \href{http://dx.doi.org/10.1086/428122}{\apjl, 619, L143}

\bibitem[{Broadhurst {et~al.}(1995)Broadhurst, Taylor, \&
  Peacock}]{broadhurst95}
Broadhurst, T., Taylor, A., \& Peacock, J. 1995,
  \href{http://dx.doi.org/10.1086/175053}{\apj, 438, 49}

\bibitem[{{Broadhurst} {et~al.}(2005{\natexlab{b}}){Broadhurst},
  {Ben{\'{\i}}tez}, {Coe}, {Sharon}, {Zekser}, {White}, {Ford}, {Bouwens},
  {Blakeslee}, {Clampin}, {Cross}, {Franx}, {Frye}, {Hartig}, {Illingworth},
  {Infante}, {Menanteau}, {Meurer}, {Postman}, {Ardila}, {Bartko}, {Brown},
  {Burrows}, {Cheng}, {Feldman}, {Golimowski}, {Goto}, {Gronwall}, {Herranz},
  {Holden}, {Homeier}, {Krist}, {Lesser}, {Martel}, {Miley}, {Rosati},
  {Sirianni}, {Sparks}, {Steindling}, {Tran}, {Tsvetanov}, \&
  {Zheng}}]{broadhurst05}
{Broadhurst}, T., {Ben{\'{\i}}tez}, N., {Coe}, D., {et~al.} 2005{\natexlab{b}},
  \href{http://dx.doi.org/10.1086/426494}{\apj, 621, 53}

\bibitem[{{Bryan} {et~al.}(2013){Bryan}, {Kay}, {Duffy}, {Schaye}, {Dalla
  Vecchia}, \& {Booth}}]{bryan13}
{Bryan}, S.~E., {Kay}, S.~T., {Duffy}, A.~R., {et~al.} 2013,
  \href{http://dx.doi.org/10.1093/mnras/sts587}{\mnras, 429, 3316}

\bibitem[{Chiu {et~al.}(2016{\natexlab{a}})Chiu, Dietrich, Mohr, Applegate,
  Benson, Bleem, Bayliss, Bocquet, Carlstrom, Capasso, Desai, Gangkofner,
  Gonzalez, Gupta, Hennig, Hoekstra, von~der Linden, Liu, McDonald, Reichardt,
  Saro, Schrabback, Strazzullo, Stubbs, \& Zenteno}]{chiu16b}
Chiu, I., Dietrich, J., Mohr, J., {et~al.} 2016{\natexlab{a}},
  \href{http://dx.doi.org/10.1093/mnras/stw190}{\mnras, 457, 3050}

\bibitem[{Chiu {et~al.}(2016{\natexlab{b}})Chiu, Saro, Mohr, Desai, Bocquet,
  Capasso, Gangkofner, Gupta, \& Liu}]{chiu16c}
Chiu, I., Saro, A., Mohr, J., {et~al.} 2016{\natexlab{b}},
  \href{http://dx.doi.org/10.1093/mnras/stw292}{\mnras, 458, 379}

\bibitem[{{Chiu} {et~al.}(2017){Chiu}, {Mohr}, {McDonald}, {Bocquet}, {Desai},
  {Klein}, {Israel}, {Ashby}, {Stanford}, {Benson}, {Brodwin}, {Abbott},
  {Abdalla}, {Allam}, {Annis}, {Bayliss}, {Benoit-L{\'e}vy}, {Bertin}, {Bleem},
  {Brooks}, {Buckley-Geer}, {Bulbul}, {Capasso}, {Carlstrom}, {Carnero Rosell},
  {Carretero}, {Castander}, {Cunha}, {D'Andrea}, {da Costa}, {Davis}, {Diehl},
  {Dietrich}, {Doel}, {Drlica-Wagner}, {Eifler}, {Evrard}, {Flaugher},
  {Garc{\'{\i}}a-Bellido}, {Garmire}, {Gaztanaga}, {Gerdes}, {Gonzalez},
  {Gruen}, {Gruendl}, {Gschwend}, {Gupta}, {Gutierrez}, {Hlavacek-L.},
  {Honscheid}, {James}, {Jeltema}, {Kraft}, {Krause}, {Kuehn}, {Kuhlmann},
  {Kuropatkin}, {Lahav}, {Lima}, {Maia}, {Marshall}, {Melchior}, {Menanteau},
  {Miquel}, {Murray}, {Nord}, {Ogando}, {Plazas}, {Rapetti}, {Reichardt},
  {Romer}, {Roodman}, {Sanchez}, {Saro}, {Scarpine}, {Schindler}, {Schubnell},
  {Sharon}, {Smith}, {Smith}, {Soares-Santos}, {Sobreira}, {Stalder}, {Stern},
  {Strazzullo}, {Suchyta}, {Swanson}, {Tarle}, {Vikram}, {Walker}, {Weller}, \&
  {Zhang}}]{chiu17}
{Chiu}, I., {Mohr}, J.~J., {McDonald}, M., {et~al.} 2017, ArXiv e-prints,
  \href{http://arxiv.org/abs/1711.00917}{{\sffamily arXiv:1711.00917}}

\bibitem[{Chiu \& Molnar(2012)}]{chiu12}
Chiu, I.-N., \& Molnar, S. 2012,
  \href{http://dx.doi.org/10.1088/0004-637X/756/1/1}{\apj, 756, 1}

\bibitem[{{Cialone} {et~al.}(2017){Cialone}, {De Petris}, {Sembolini}, {Yepes},
  {Baldi}, \& {Rasia}}]{cialone17}
{Cialone}, G., {De Petris}, M., {Sembolini}, F., {et~al.} 2017, ArXiv e-prints,
  \href{http://arxiv.org/abs/1708.03325}{{\sffamily arXiv:1708.03325}}

\bibitem[{{Coe} {et~al.}(2013){Coe}, {Zitrin}, {Carrasco}, {Shu}, {Zheng},
  {Postman}, {Bradley}, {Koekemoer}, {Bouwens}, {Broadhurst}, {Monna}, {Host},
  {Moustakas}, {Ford}, {Moustakas}, {van der Wel}, {Donahue}, {Rodney},
  {Ben{\'{\i}}tez}, {Jouvel}, {Seitz}, {Kelson}, \& {Rosati}}]{coe13}
{Coe}, D., {Zitrin}, A., {Carrasco}, M., {et~al.} 2013,
  \href{http://dx.doi.org/10.1088/0004-637X/762/1/32}{\apj, 762, 32}

\bibitem[{Corless {et~al.}(2008)Corless, King, \& Clowe}]{corless08b}
Corless, V., King, L., \& Clowe, D. 2008, \mnras, 393, 1235

\bibitem[{{Corless} {et~al.}(2009){Corless}, {King}, \& {Clowe}}]{corless09b}
{Corless}, V.~L., {King}, L.~J., \& {Clowe}, D. 2009,
  \href{http://dx.doi.org/10.1111/j.1365-2966.2008.14294.x}{\mnras, 393, 1235}

\bibitem[{{Czakon} {et~al.}(2015){Czakon}, {Sayers}, {Mantz}, {Golwala},
  {Downes}, {Koch}, {Lin}, {Molnar}, {Moustakas}, {Mroczkowski}, {Pierpaoli},
  {Shitanishi}, {Siegel}, \& {Umetsu}}]{czakon15}
{Czakon}, N.~G., {Sayers}, J., {Mantz}, A., {et~al.} 2015,
  \href{http://dx.doi.org/10.1088/0004-637X/806/1/18}{\apj, 806, 18}

\bibitem[{{De Filippis} {et~al.}(2005){De Filippis}, {Sereno}, {Bautz}, \&
  {Longo}}]{defilippis05}
{De Filippis}, E., {Sereno}, M., {Bautz}, M.~W., \& {Longo}, G. 2005,
  \href{http://dx.doi.org/10.1086/429401}{\apj, 625, 108}

\bibitem[{de~Haan {et~al.}(2016)de~Haan, Benson, Bleem, Allen, Applegate,
  Ashby, Bautz, Bayliss, Bocquet, Brodwin, Carlstrom, Chang, Chiu, Cho,
  Clocchiatti, Crawford, Crites, Desai, Dietrich, Dobbs, Doucouliagos, Foley,
  Forman, Garmire, George, Gladders, Gonzalez, Gupta, Halverson,
  Hlavacek-Larrondo, Hoekstra, Holder, Holzapfel, Hou, Hrubes, Huang, Jones,
  Keisler, Knox, Lee, Leitch, von~der Linden, Luong-Van, Mantz, Marrone,
  McDonald, McMahon, Meyer, Mocanu, Mohr, Murray, Padin, Pryke, Rapetti,
  Reichardt, Rest, Ruel, Ruhl, Saliwanchik, Saro, Sayre, Schaffer, Schrabback,
  Shirokoff, Song, Spieler, Stalder, Stanford, Staniszewski, Stark, Story,
  Stubbs, Vanderlinde, Vieira, Vikhlinin, Williamson, \& Zenteno}]{dehaan16}
de~Haan, T., Benson, B., Bleem, L., {et~al.} 2016,
  \href{http://dx.doi.org/10.3847/0004-637X/832/1/95}{\apj, 832, 95}

\bibitem[{{DeMaio} {et~al.}(2015){DeMaio}, {Gonzalez}, {Zabludoff}, {Zaritsky},
  \& {Brada{\v c}}}]{demaio15}
{DeMaio}, T., {Gonzalez}, A.~H., {Zabludoff}, A., {Zaritsky}, D., \& {Brada{\v
  c}}, M. 2015, \href{http://dx.doi.org/10.1093/mnras/stv033}{\mnras, 448,
  1162}

\bibitem[{{Despali} {et~al.}(2014){Despali}, {Giocoli}, \&
  {Tormen}}]{despali14}
{Despali}, G., {Giocoli}, C., \& {Tormen}, G. 2014,
  \href{http://dx.doi.org/10.1093/mnras/stu1393}{\mnras, 443, 3208}

\bibitem[{{Diemer}(2017)}]{diemer17}
{Diemer}, B. 2017, ArXiv e-prints,
  \href{http://arxiv.org/abs/1712.04512}{{\sffamily arXiv:1712.04512}}

\bibitem[{{Diemer} \& {Kravtsov}(2015)}]{diemer15}
{Diemer}, B., \& {Kravtsov}, A.~V. 2015,
  \href{http://dx.doi.org/10.1088/0004-637X/799/1/108}{\apj, 799, 108}

\bibitem[{{Dietrich} {et~al.}(2014){Dietrich}, {Zhang}, {Song}, {Davis},
  {McKay}, {Baruah}, {Becker}, {Benoist}, {Busha}, {da Costa}, {Hao}, {Maia},
  {Miller}, {Ogando}, {Romer}, {Rozo}, {Rykoff}, \& {Wechsler}}]{dietrich14}
{Dietrich}, J.~P., {Zhang}, Y., {Song}, J., {et~al.} 2014,
  \href{http://dx.doi.org/10.1093/mnras/stu1282}{\mnras, 443, 1713}

\bibitem[{Donahue {et~al.}(2014)Donahue, Voit, Mahdavi, Umetsu, Ettori, Merten,
  Postman, Hoffer, Baldi, Coe, Czakon, Bartelmann, Benitez, Bouwens, Bradley,
  Broadhurst, Ford, Gastaldello, Grillo, Infante, Jouvel, Koekemoer, Kelson,
  Lahav, Lemze, Medezinski, Melchior, Meneghetti, Molino, Moustakas, Moustakas,
  Nonino, Rosati, Sayers, Seitz, {Van der Wel}, Zheng, \& Zitrin}]{donahue14}
Donahue, M., Voit, G., Mahdavi, A., {et~al.} 2014,
  \href{http://dx.doi.org/10.1088/0004-637X/794/2/136}{\apj, 794, 136}

\bibitem[{Dressler(1980)}]{dressler80}
Dressler, A. 1980, \href{http://dx.doi.org/10.1086/157753}{\apj, 236, 351}

\bibitem[{{Dubinski} \& {Carlberg}(1991)}]{dubinski1991}
{Dubinski}, J., \& {Carlberg}, R.~G. 1991,
  \href{http://dx.doi.org/10.1086/170451}{\apj, 378, 496}

\bibitem[{Duffy {et~al.}(2008)Duffy, Schaye, Kay, \& {Dalla Vecchia}}]{duffy08}
Duffy, A., Schaye, J., Kay, S., \& {Dalla Vecchia}, C. 2008,
  \href{http://dx.doi.org/10.1111/j.1745-3933.2008.00537.x}{\mnras, 390, L64}

\bibitem[{{Dutton} \& {Macci{\`o}}(2014)}]{dutton14}
{Dutton}, A.~A., \& {Macci{\`o}}, A.~V. 2014,
  \href{http://dx.doi.org/10.1093/mnras/stu742}{\mnras, 441, 3359}

\bibitem[{Flaugher(2005)}]{flaugher05}
Flaugher, B. 2005,
  \href{http://dx.doi.org/10.1142/S0217751X05025917}{International Journal of
  Modern Physics A, 20, 3121}

\bibitem[{{Flores} {et~al.}(2007){Flores}, {Allgood}, {Kravtsov}, {Primack},
  {Buote}, \& {Bullock}}]{flores05}
{Flores}, R.~A., {Allgood}, B., {Kravtsov}, A.~V., {et~al.} 2007,
  \href{http://dx.doi.org/10.1111/j.1365-2966.2007.11658.x}{\mnras, 377, 883}

\bibitem[{Ford {et~al.}(2012)Ford, Hildebrandt, {Van Waerbeke}, Leauthaud,
  Capak, Finoguenov, Tanaka, George, \& Rhodes}]{ford12}
Ford, J., Hildebrandt, H., {Van Waerbeke}, L., {et~al.} 2012,
  \href{http://dx.doi.org/10.1088/0004-637X/754/2/143}{\apj, 754, 143}

\bibitem[{Foreman-Mackey {et~al.}(2013)Foreman-Mackey, Hogg, Lang, \&
  Goodman}]{foreman13}
Foreman-Mackey, D., Hogg, D., Lang, D., \& Goodman, J. 2013,
  \href{http://dx.doi.org/10.1086/670067}{\pasp, 125, 306}

\bibitem[{{Frenk} {et~al.}(1988){Frenk}, {White}, {Davis}, \&
  {Efstathiou}}]{frenk1988}
{Frenk}, C.~S., {White}, S.~D.~M., {Davis}, M., \& {Efstathiou}, G. 1988,
  \href{http://dx.doi.org/10.1086/166213}{\apj, 327, 507}

\bibitem[{{Grillo} {et~al.}(2015){Grillo}, {Suyu}, {Rosati}, {Mercurio},
  {Balestra}, {Munari}, {Nonino}, {Caminha}, {Lombardi}, {De Lucia}, {Borgani},
  {Gobat}, {Biviano}, {Girardi}, {Umetsu}, {Coe}, {Koekemoer}, {Postman},
  {Zitrin}, {Halkola}, {Broadhurst}, {Sartoris}, {Presotto}, {Annunziatella},
  {Maier}, {Fritz}, {Vanzella}, \& {Frye}}]{grillo15}
{Grillo}, C., {Suyu}, S.~H., {Rosati}, P., {et~al.} 2015,
  \href{http://dx.doi.org/10.1088/0004-637X/800/1/38}{\apj, 800, 38}

\bibitem[{{Gruen} {et~al.}(2013){Gruen}, {Brimioulle}, {Seitz}, {Lee}, {Young},
  {Koppenhoefer}, {Eichner}, {Riffeser}, {Vikram}, {Weidinger}, \&
  {Zenteno}}]{gruen13}
{Gruen}, D., {Brimioulle}, F., {Seitz}, S., {et~al.} 2013,
  \href{http://dx.doi.org/10.1093/mnras/stt566}{\mnras, 432, 1455}

\bibitem[{{Gupta} {et~al.}(2016){Gupta}, {Yuan}, {Tran}, {Martizzi}, {Taylor},
  \& {Kewley}}]{gupta16}
{Gupta}, A., {Yuan}, T., {Tran}, K.-V.~H., {et~al.} 2016,
  \href{http://dx.doi.org/10.3847/0004-637X/831/1/104}{\apj, 831, 104}

\bibitem[{{Gupta} {et~al.}(2017){Gupta}, {Saro}, {Mohr}, {Benson}, {Bocquet},
  {Capasso}, {Carlstrom}, {Chiu}, {Crawford}, {de Haan}, {Dietrich},
  {Gangkofner}, {Holzapfel}, {McDonald}, {Rapetti}, \& {Reichardt}}]{gupta17}
{Gupta}, N., {Saro}, A., {Mohr}, J.~J., {et~al.} 2017,
  \href{http://dx.doi.org/10.1093/mnras/stx095}{\mnras, 467, 3737}

\bibitem[{Hildebrandt {et~al.}(2009)Hildebrandt, Pielorz, Erben, van Waerbeke,
  Simon, \& Capak}]{hildebrandt09}
Hildebrandt, H., Pielorz, J., Erben, T., {et~al.} 2009,
  \href{http://dx.doi.org/10.1051/0004-6361/200811042}{\aap, 498, 725}

\bibitem[{Hinton(2016)}]{hinton16}
Hinton, S. 2016, \href{http://dx.doi.org/10.21105/joss.00045}{{JOSS}, 1}

\bibitem[{Hoekstra {et~al.}(2013)Hoekstra, Bartelmann, Dahle, Israel, Limousin,
  \& Meneghetti}]{hoekstra13}
Hoekstra, H., Bartelmann, M., Dahle, H., {et~al.} 2013,
  \href{http://dx.doi.org/10.1007/s11214-013-9978-5}{\ssr, 177, 75}

\bibitem[{Hoekstra {et~al.}(2015)Hoekstra, Herbonnet, Muzzin, Babul, Mahdavi,
  Viola, \& Cacciato}]{hoekstra15}
Hoekstra, H., Herbonnet, R., Muzzin, A., {et~al.} 2015,
  \href{http://dx.doi.org/10.1093/mnras/stv275}{\mnras, 449, 685}

\bibitem[{{Hopkins} {et~al.}(2005){Hopkins}, {Bahcall}, \& {Bode}}]{hopkins05}
{Hopkins}, P.~F., {Bahcall}, N.~A., \& {Bode}, P. 2005,
  \href{http://dx.doi.org/10.1086/425993}{\apj, 618, 1}

\bibitem[{Hunter(2007)}]{Hunter:2007}
Hunter, J.~D. 2007, Computing In Science \& Engineering, 9, 90

\bibitem[{{Jauzac} {et~al.}(2017){Jauzac}, {Eckert}, {Schaller}, {Schwinn},
  {Massey}, {Bah{\'e}}, {Baugh}, {Barnes}, {Dalla Vecchia}, {Ebeling},
  {Harvey}, {Jullo}, {Kay}, {Kneib}, {Limousin}, {Medezinski}, {Natarajan},
  {Nonino}, {Robertson}, {Tam}, \& {Umetsu}}]{jauzac17}
{Jauzac}, M., {Eckert}, D., {Schaller}, M., {et~al.} 2017, ArXiv e-prints,
  \href{http://arxiv.org/abs/1711.01324}{{\sffamily arXiv:1711.01324}}

\bibitem[{{Jing} \& {Suto}(2002)}]{jing02}
{Jing}, Y.~P., \& {Suto}, Y. 2002,
  \href{http://dx.doi.org/10.1086/341065}{\apj, 574, 538}

\bibitem[{Johnston {et~al.}(2007)Johnston, Sheldon, Wechsler, Rozo, Koester,
  Frieman, McKay, Evrard, Becker, \& Annis}]{johnston07}
Johnston, D., Sheldon, E., Wechsler, R., {et~al.} 2007, ArXiv e-prints,
  \href{http://arxiv.org/abs/0709.1159}{{\sffamily arXiv:0709.1159}}

\bibitem[{Jones {et~al.}(2001)Jones, Oliphant, Peterson,
  {et~al.}}]{jones_scipy_2001}
Jones, E., Oliphant, T., Peterson, P., {et~al.} 2001, {SciPy}: Open source
  scientific tools for Python

\bibitem[{{Kasun} \& {Evrard}(2005)}]{kasun05}
{Kasun}, S.~F., \& {Evrard}, A.~E. 2005,
  \href{http://dx.doi.org/10.1086/430811}{\apj, 629, 781}

\bibitem[{{Kazantzidis} {et~al.}(2004){Kazantzidis}, {Kravtsov}, {Zentner},
  {Allgood}, {Nagai}, \& {Moore}}]{kazantzidis04}
{Kazantzidis}, S., {Kravtsov}, A.~V., {Zentner}, A.~R., {et~al.} 2004,
  \href{http://dx.doi.org/10.1086/423992}{\apjl, 611, L73}

\bibitem[{{Kitching} {et~al.}(2012){Kitching}, {Balan}, {Bridle}, {Cantale},
  {Courbin}, {Eifler}, {Gentile}, {Gill}, {Harmeling}, {Heymans}, {Hirsch},
  {Honscheid}, {Kacprzak}, {Kirkby}, {Margala}, {Massey}, {Melchior},
  {Nurbaeva}, {Patton}, {Rhodes}, {Rowe}, {Taylor}, {Tewes}, {Viola},
  {Witherick}, {Voigt}, {Young}, \& {Zuntz}}]{kitching12}
{Kitching}, T.~D., {Balan}, S.~T., {Bridle}, S., {et~al.} 2012,
  \href{http://dx.doi.org/10.1111/j.1365-2966.2012.21095.x}{\mnras, 423, 3163}

\bibitem[{Lau {et~al.}(2009)Lau, Kravtsov, \& Nagai}]{lau09}
Lau, E., Kravtsov, A., \& Nagai, D. 2009,
  \href{http://dx.doi.org/10.1088/0004-637X/705/2/1129}{\apj, 705, 1129}

\bibitem[{{Limousin} {et~al.}(2013){Limousin}, {Morandi}, {Sereno},
  {Meneghetti}, {Ettori}, {Bartelmann}, \& {Verdugo}}]{limousin13}
{Limousin}, M., {Morandi}, A., {Sereno}, M., {et~al.} 2013,
  \href{http://dx.doi.org/10.1007/s11214-013-9980-y}{\ssr, 177, 155}

\bibitem[{{Mandelbaum} {et~al.}(2015){Mandelbaum}, {Rowe}, {Armstrong}, {Bard},
  {Bertin}, {Bosch}, {Boutigny}, {Courbin}, {Dawson}, {Donnarumma}, {Fenech
  Conti}, {Gavazzi}, {Gentile}, {Gill}, {Hogg}, {Huff}, {Jee}, {Kacprzak},
  {Kilbinger}, {Kuntzer}, {Lang}, {Luo}, {March}, {Marshall}, {Meyers},
  {Miller}, {Miyatake}, {Nakajima}, {Ngol{\'e} Mboula}, {Nurbaeva}, {Okura},
  {Paulin-Henriksson}, {Rhodes}, {Schneider}, {Shan}, {Sheldon}, {Simet},
  {Starck}, {Sureau}, {Tewes}, {Zarb Adami}, {Zhang}, \&
  {Zuntz}}]{mandelbaum15}
{Mandelbaum}, R., {Rowe}, B., {Armstrong}, R., {et~al.} 2015,
  \href{http://dx.doi.org/10.1093/mnras/stv781}{\mnras, 450, 2963}

\bibitem[{{Mantz} {et~al.}(2015){Mantz}, {von der Linden}, {Allen},
  {Applegate}, {Kelly}, {Morris}, {Rapetti}, {Schmidt}, {Adhikari}, {Allen},
  {Burchat}, {Burke}, {Cataneo}, {Donovan}, {Ebeling}, {Shandera}, \&
  {Wright}}]{mantz15}
{Mantz}, A.~B., {von der Linden}, A., {Allen}, S.~W., {et~al.} 2015,
  \href{http://dx.doi.org/10.1093/mnras/stu2096}{\mnras, 446, 2205}

\bibitem[{{McLeod} {et~al.}(2016){McLeod}, {McLure}, \& {Dunlop}}]{mcleod16}
{McLeod}, D.~J., {McLure}, R.~J., \& {Dunlop}, J.~S. 2016,
  \href{http://dx.doi.org/10.1093/mnras/stw904}{\mnras, 459, 3812}

\bibitem[{{Medezinski} {et~al.}(2013){Medezinski}, {Umetsu}, {Nonino},
  {Merten}, {Zitrin}, {Broadhurst}, {Donahue}, {Sayers}, {Waizmann},
  {Koekemoer}, {Coe}, {Molino}, {Melchior}, {Mroczkowski}, {Czakon}, {Postman},
  {Meneghetti}, {Lemze}, {Ford}, {Grillo}, {Kelson}, {Bradley}, {Moustakas},
  {Bartelmann}, {Ben{\'{\i}}tez}, {Biviano}, {Bouwens}, {Golwala}, {Graves},
  {Infante}, {Jim{\'e}nez-Teja}, {Jouvel}, {Lahav}, {Moustakas}, {Ogaz},
  {Rosati}, {Seitz}, \& {Zheng}}]{medezinski13}
{Medezinski}, E., {Umetsu}, K., {Nonino}, M., {et~al.} 2013,
  \href{http://dx.doi.org/10.1088/0004-637X/777/1/43}{\apj, 777, 43}

\bibitem[{{Medezinski} {et~al.}(2017){Medezinski}, {Battaglia}, {Umetsu},
  {Oguri}, {Miyatake}, {Nishizawa}, {Sif{\'o}n}, {Spergel}, {Chiu}, {Lin},
  {Bahcall}, \& {Komiyama}}]{medezinski17}
{Medezinski}, E., {Battaglia}, N., {Umetsu}, K., {et~al.} 2017,
  \href{http://dx.doi.org/10.1093/pasj/psx128}{\pasj},
  \href{http://arxiv.org/abs/1706.00434}{{\sffamily arXiv:1706.00434}}

\bibitem[{{Melchior} {et~al.}(2017){Melchior}, {Gruen}, {McClintock}, {Varga},
  {Sheldon}, {Rozo}, {Amara}, {Becker}, {Benson}, {Bermeo}, {Bridle},
  {Clampitt}, {Dietrich}, {Hartley}, {Hollowood}, {Jain}, {Jarvis}, {Jeltema},
  {Kacprzak}, {MacCrann}, {Rykoff}, {Saro}, {Suchyta}, {Troxel}, {Zuntz},
  {Bonnett}, {Plazas}, {Abbott}, {Abdalla}, {Annis}, {Benoit-L{\'e}vy},
  {Bernstein}, {Bertin}, {Brooks}, {Buckley-Geer}, {Carnero Rosell}, {Carrasco
  Kind}, {Carretero}, {Cunha}, {D'Andrea}, {da Costa}, {Desai}, {Eifler},
  {Flaugher}, {Fosalba}, {Garc{\'{\i}}a-Bellido}, {Gaztanaga}, {Gerdes},
  {Gruendl}, {Gschwend}, {Gutierrez}, {Honscheid}, {James}, {Kirk}, {Krause},
  {Kuehn}, {Kuropatkin}, {Lahav}, {Lima}, {Maia}, {March}, {Martini},
  {Menanteau}, {Miller}, {Miquel}, {Mohr}, {Nichol}, {Ogando}, {Romer},
  {Sanchez}, {Scarpine}, {Sevilla-Noarbe}, {Smith}, {Soares-Santos},
  {Sobreira}, {Swanson}, {Tarle}, {Thomas}, {Walker}, {Weller}, \&
  {Zhang}}]{melchior17}
{Melchior}, P., {Gruen}, D., {McClintock}, T., {et~al.} 2017,
  \href{http://dx.doi.org/10.1093/mnras/stx1053}{\mnras, 469, 4899}

\bibitem[{{Meneghetti} {et~al.}(2014){Meneghetti}, {Rasia}, {Vega}, {Merten},
  {Postman}, {Yepes}, {Sembolini}, {Donahue}, {Ettori}, {Umetsu}, {Balestra},
  {Bartelmann}, {Ben{\'{\i}}tez}, {Biviano}, {Bouwens}, {Bradley},
  {Broadhurst}, {Coe}, {Czakon}, {De Petris}, {Ford}, {Giocoli},
  {Gottl{\"o}ber}, {Grillo}, {Infante}, {Jouvel}, {Kelson}, {Koekemoer},
  {Lahav}, {Lemze}, {Medezinski}, {Melchior}, {Mercurio}, {Molino},
  {Moscardini}, {Monna}, {Moustakas}, {Moustakas}, {Nonino}, {Rhodes},
  {Rosati}, {Sayers}, {Seitz}, {Zheng}, \& {Zitrin}}]{meneghetti14}
{Meneghetti}, M., {Rasia}, E., {Vega}, J., {et~al.} 2014,
  \href{http://dx.doi.org/10.1088/0004-637X/797/1/34}{\apj, 797, 34}

\bibitem[{{Merten} {et~al.}(2015){Merten}, {Meneghetti}, {Postman}, {Umetsu},
  {Zitrin}, {Medezinski}, {Nonino}, {Koekemoer}, {Melchior}, {Gruen},
  {Moustakas}, {Bartelmann}, {Host}, {Donahue}, {Coe}, {Molino}, {Jouvel},
  {Monna}, {Seitz}, {Czakon}, {Lemze}, {Sayers}, {Balestra}, {Rosati},
  {Ben{\'{\i}}tez}, {Biviano}, {Bouwens}, {Bradley}, {Broadhurst}, {Carrasco},
  {Ford}, {Grillo}, {Infante}, {Kelson}, {Lahav}, {Massey}, {Moustakas},
  {Rasia}, {Rhodes}, {Vega}, \& {Zheng}}]{merten15}
{Merten}, J., {Meneghetti}, M., {Postman}, M., {et~al.} 2015,
  \href{http://dx.doi.org/10.1088/0004-637X/806/1/4}{\apj, 806, 4}

\bibitem[{{Miyazaki}(2015)}]{miyazaki15}
{Miyazaki}, S. 2015, IAU General Assembly, 22, 2255916

\bibitem[{Miyazaki {et~al.}(2012)Miyazaki, Komiyama, Nakaya, Kamata, Doi,
  Hamana, Karoji, Furusawa, Kawanomoto, Morokuma, Ishizuka, Nariai, Tanaka,
  Uraguchi, Utsumi, Obuchi, Okura, Oguri, Takata, Tomono, Kurakami, Namikawa,
  Usuda, Yamanoi, Terai, Uekiyo, Yamada, Koike, Aihara, Fujimori, Mineo,
  Miyatake, Yasuda, Nishizawa, Saito, Tanaka, Uchida, Katayama, Wang, Chen,
  Lupton, Loomis, Bickerton, Price, Gunn, Suzuki, Miyazaki, Muramatsu,
  Yamamoto, Endo, Ezaki, Itoh, Miwa, Yokota, Matsuda, Ebinuma, \&
  Takeshi}]{miyazaki12}
Miyazaki, S., Komiyama, Y., Nakaya, H., {et~al.} 2012,
  \href{http://dx.doi.org/10.1117/12.926844}{in Society of Photo-Optical
  Instrumentation Engineers (SPIE) Conference Series, Vol. 8446, Society of
  Photo-Optical Instrumentation Engineers (SPIE) Conference Series}

\bibitem[{Molnar {et~al.}(2010)Molnar, Chiu, Umetsu, Chen, Hearn, Broadhurst,
  Bryan, \& Shang}]{molnar10}
Molnar, S., Chiu, I.-N., Umetsu, K., {et~al.} 2010,
  \href{http://dx.doi.org/10.1088/2041-8205/724/1/L1}{\apjl, 724, L1}

\bibitem[{{Monna} {et~al.}(2014){Monna}, {Seitz}, {Greisel}, {Eichner},
  {Drory}, {Postman}, {Zitrin}, {Coe}, {Halkola}, {Suyu}, {Grillo}, {Rosati},
  {Lemze}, {Balestra}, {Snigula}, {Bradley}, {Umetsu}, {Koekemoer}, {Kuchner},
  {Moustakas}, {Bartelmann}, {Ben{\'{\i}}tez}, {Bouwens}, {Broadhurst},
  {Donahue}, {Ford}, {Host}, {Infante}, {Jimenez-Teja}, {Jouvel}, {Kelson},
  {Lahav}, {Medezinski}, {Melchior}, {Meneghetti}, {Merten}, {Molino},
  {Moustakas}, {Nonino}, \& {Zheng}}]{monna14}
{Monna}, A., {Seitz}, S., {Greisel}, N., {et~al.} 2014,
  \href{http://dx.doi.org/10.1093/mnras/stt2284}{\mnras, 438, 1417}

\bibitem[{{Morandi} \& {Limousin}(2012)}]{morandi12}
{Morandi}, A., \& {Limousin}, M. 2012,
  \href{http://dx.doi.org/10.1111/j.1365-2966.2012.20537.x}{\mnras, 421, 3147}

\bibitem[{Navarro {et~al.}(1996)Navarro, Frenk, \& White}]{navarro96}
Navarro, J., Frenk, C., \& White, S. 1996,
  \href{http://dx.doi.org/10.1086/177173}{\apj, 462, 563}

\bibitem[{Navarro {et~al.}(1997)Navarro, Frenk, \& White}]{navarro1997}
---. 1997, \apj, 490, 493

\bibitem[{{Newman} {et~al.}(2013){Newman}, {Treu}, {Ellis}, {Sand}, {Nipoti},
  {Richard}, \& {Jullo}}]{newman13}
{Newman}, A.~B., {Treu}, T., {Ellis}, R.~S., {et~al.} 2013,
  \href{http://dx.doi.org/10.1088/0004-637X/765/1/24}{\apj, 765, 24}

\bibitem[{Oguri {et~al.}(2012)Oguri, Bayliss, Dahle, Sharon, Gladders,
  Natarajan, Hennawi, \& Koester}]{oguri12}
Oguri, M., Bayliss, M., Dahle, H., {et~al.} 2012,
  \href{http://dx.doi.org/10.1111/j.1365-2966.2011.20248.x}{\mnras, 420, 3213}

\bibitem[{{Oguri} {et~al.}(2010){Oguri}, {Takada}, {Okabe}, \&
  {Smith}}]{oguri10b}
{Oguri}, M., {Takada}, M., {Okabe}, N., \& {Smith}, G.~P. 2010,
  \href{http://dx.doi.org/10.1111/j.1365-2966.2010.16622.x}{\mnras, 405, 2215}

\bibitem[{{Oguri} {et~al.}(2005){Oguri}, {Takada}, {Umetsu}, \&
  {Broadhurst}}]{oguri05}
{Oguri}, M., {Takada}, M., {Umetsu}, K., \& {Broadhurst}, T. 2005,
  \href{http://dx.doi.org/10.1086/452629}{\apj, 632, 841}

\bibitem[{{Okabe} \& {Smith}(2016)}]{okabe16}
{Okabe}, N., \& {Smith}, G.~P. 2016,
  \href{http://dx.doi.org/10.1093/mnras/stw1539}{\mnras, 461, 3794}

\bibitem[{Okabe {et~al.}(2010)Okabe, Zhang, Finoguenov, Takada, Smith, Umetsu,
  \& Futamase}]{okabe10}
Okabe, N., Zhang, Y.-Y., Finoguenov, A., {et~al.} 2010,
  \href{http://dx.doi.org/10.1088/0004-637X/721/1/875}{\apj, 721, 875}

\bibitem[{P\'erez \& Granger(2007)}]{PER-GRA:2007}
P\'erez, F., \& Granger, B.~E. 2007,
  \href{http://dx.doi.org/10.1109/MCSE.2007.53}{Computing in Science and
  Engineering, 9, 21}

\bibitem[{{Planck Collaboration} {et~al.}(2015){Planck Collaboration}, Ade,
  Aghanim, Arnaud, Ashdown, Aumont, Baccigalupi, Banday, Barreiro, Bartlett, \&
  al.}]{planck2015clstrcnts}
{Planck Collaboration}, Ade, P., Aghanim, N., {et~al.} 2015, ArXiv e-prints,
  \href{http://arxiv.org/abs/1502.01597}{{\sffamily arXiv:1502.01597}}

\bibitem[{{Postman} {et~al.}(2012){Postman}, {Coe}, {Ben{\'{\i}}tez},
  {Bradley}, {Broadhurst}, {Donahue}, {Ford}, {Graur}, {Graves}, {Jouvel},
  {Koekemoer}, {Lemze}, {Medezinski}, {Molino}, {Moustakas}, {Ogaz}, {Riess},
  {Rodney}, {Rosati}, {Umetsu}, {Zheng}, {Zitrin}, {Bartelmann}, {Bouwens},
  {Czakon}, {Golwala}, {Host}, {Infante}, {Jha}, {Jimenez-Teja}, {Kelson},
  {Lahav}, {Lazkoz}, {Maoz}, {McCully}, {Melchior}, {Meneghetti}, {Merten},
  {Moustakas}, {Nonino}, {Patel}, {Reg{\"o}s}, {Sayers}, {Seitz}, \& {Van der
  Wel}}]{postman12}
{Postman}, M., {Coe}, D., {Ben{\'{\i}}tez}, N., {et~al.} 2012,
  \href{http://dx.doi.org/10.1088/0067-0049/199/2/25}{\apjs, 199, 25}

\bibitem[{{Richard} {et~al.}(2010){Richard}, {Smith}, {Kneib}, {Ellis},
  {Sanderson}, {Pei}, {Targett}, {Sand}, {Swinbank}, {Dannerbauer}, {Mazzotta},
  {Limousin}, {Egami}, {Jullo}, {Hamilton-Morris}, \& {Moran}}]{richard10}
{Richard}, J., {Smith}, G.~P., {Kneib}, J.-P., {et~al.} 2010,
  \href{http://dx.doi.org/10.1111/j.1365-2966.2009.16274.x}{\mnras, 404, 325}

\bibitem[{{Rosati} {et~al.}(2014){Rosati}, {Balestra}, {Grillo}, {Mercurio},
  {Nonino}, {Biviano}, {Girardi}, {Vanzella}, \& {Clash-VLT Team}}]{rosati14}
{Rosati}, P., {Balestra}, I., {Grillo}, C., {et~al.} 2014, The Messenger, 158,
  48

\bibitem[{{Schrabback} {et~al.}(2018){Schrabback}, {Applegate}, {Dietrich},
  {Hoekstra}, {Bocquet}, {Gonzalez}, {von der Linden}, {McDonald}, {Morrison},
  {Raihan}, {Allen}, {Bayliss}, {Benson}, {Bleem}, {Chiu}, {Desai}, {Foley},
  {de Haan}, {High}, {Hilbert}, {Mantz}, {Massey}, {Mohr}, {Reichardt}, {Saro},
  {Simon}, {Stern}, {Stubbs}, \& {Zenteno}}]{schrabback18}
{Schrabback}, T., {Applegate}, D., {Dietrich}, J.~P., {et~al.} 2018,
  \href{http://dx.doi.org/10.1093/mnras/stx2666}{\mnras, 474, 2635}

\bibitem[{{Sereno} {et~al.}(2017{\natexlab{a}}){Sereno}, {Covone}, {Izzo},
  {Ettori}, {Coupon}, \& {Lieu}}]{sereno17b}
{Sereno}, M., {Covone}, G., {Izzo}, L., {et~al.} 2017{\natexlab{a}},
  \href{http://dx.doi.org/10.1093/mnras/stx2085}{\mnras, 472, 1946}

\bibitem[{{Sereno} {et~al.}(2006){Sereno}, {De Filippis}, {Longo}, \&
  {Bautz}}]{sereno06}
{Sereno}, M., {De Filippis}, E., {Longo}, G., \& {Bautz}, M.~W. 2006,
  \href{http://dx.doi.org/10.1086/503198}{\apj, 645, 170}

\bibitem[{{Sereno} \& {Ettori}(2015)}]{sereno15a}
{Sereno}, M., \& {Ettori}, S. 2015,
  \href{http://dx.doi.org/10.1093/mnras/stv810}{\mnras, 450, 3633}

\bibitem[{{Sereno} {et~al.}(2017{\natexlab{b}}){Sereno}, {Ettori},
  {Meneghetti}, {Sayers}, {Umetsu}, {Merten}, {Chiu}, \& {Zitrin}}]{sereno17}
{Sereno}, M., {Ettori}, S., {Meneghetti}, M., {et~al.} 2017{\natexlab{b}},
  \href{http://dx.doi.org/10.1093/mnras/stx326}{\mnras, 467, 3801}

\bibitem[{{Sereno} {et~al.}(2010){Sereno}, {Lubini}, \& {Jetzer}}]{sereno10}
{Sereno}, M., {Lubini}, M., \& {Jetzer}, P. 2010,
  \href{http://dx.doi.org/10.1051/0004-6361/200913843}{\aap, 518, A55}

\bibitem[{{Sereno} \& {Umetsu}(2011)}]{sereno11}
{Sereno}, M., \& {Umetsu}, K. 2011,
  \href{http://dx.doi.org/10.1111/j.1365-2966.2011.19274.x}{\mnras, 416, 3187}

\bibitem[{{Sereno} {et~al.}(2018){Sereno}, {Umetsu}, {Ettori}, {Sayers},
  {Chiu}, {Meneghetti}, {Vega-Ferrero}, \& {Zitrin}}]{sereno18}
{Sereno}, M., {Umetsu}, K., {Ettori}, S., {et~al.} 2018, ArXiv e-prints,
  \href{http://arxiv.org/abs/1804.00667}{{\sffamily arXiv:1804.00667}}

\bibitem[{{Stark}(1977)}]{stark77}
{Stark}, A.~A. 1977, \href{http://dx.doi.org/10.1086/155164}{\apj, 213, 368}

\bibitem[{Sunyaev \& Zel'dovich(1970)}]{sunyaev70}
Sunyaev, R., \& Zel'dovich, Y. 1970, Comments on Astrophysics and Space
  Physics, 2, 66

\bibitem[{Sunyaev \& Zel'dovich(1972)}]{sunyaev72}
---. 1972, Comments on Astrophysics and Space Physics, 4, 173

\bibitem[{{Suto} {et~al.}(2016){Suto}, {Kitayama}, {Nishimichi}, {Sasaki}, \&
  {Suto}}]{suto16}
{Suto}, D., {Kitayama}, T., {Nishimichi}, T., {Sasaki}, S., \& {Suto}, Y. 2016,
  \href{http://dx.doi.org/10.1093/pasj/psw088}{\pasj, 68, 97}

\bibitem[{{Suto} {et~al.}(2017){Suto}, {Peirani}, {Dubois}, {Kitayama},
  {Nishimichi}, {Sasaki}, \& {Suto}}]{suto17}
{Suto}, D., {Peirani}, S., {Dubois}, Y., {et~al.} 2017,
  \href{http://dx.doi.org/10.1093/pasj/psw118}{\pasj, 69, 14}

\bibitem[{Taylor {et~al.}(1998)Taylor, Dye, Broadhurst, Benitez, \& van
  Kampen}]{taylor98}
Taylor, A., Dye, S., Broadhurst, T., Benitez, N., \& van Kampen, E. 1998,
  \href{http://dx.doi.org/10.1086/305827}{\apj, 501, 539}

\bibitem[{{Taylor}(2005)}]{2005ASPC..347...29T}
{Taylor}, M.~B. 2005, in Astronomical Society of the Pacific Conference Series,
  Vol. 347, Astronomical Data Analysis Software and Systems XIV, ed.
  P.~{Shopbell}, M.~{Britton}, \& R.~{Ebert}, 29

\bibitem[{{Tudorica} {et~al.}(2017){Tudorica}, {Hildebrandt}, {Tewes},
  {Hoekstra}, {Morrison}, {Muzzin}, {Wilson}, {Yee}, {Lidman}, {Hicks},
  {Nantais}, {Erben}, {van der Burg}, \& {Demarco}}]{tudorica17}
{Tudorica}, A., {Hildebrandt}, H., {Tewes}, M., {et~al.} 2017,
  \href{http://dx.doi.org/10.1051/0004-6361/201731267}{\aap, 608, A141}

\bibitem[{{Umetsu}(2010)}]{umetsu10}
{Umetsu}, K. 2010, ArXiv e-prints,
  \href{http://arxiv.org/abs/1002.3952}{{\sffamily arXiv:1002.3952
  [astro-ph.CO]}}

\bibitem[{Umetsu(2013)}]{umetsu13}
Umetsu, K. 2013, \href{http://dx.doi.org/10.1088/0004-637X/769/1/13}{\apj, 769,
  13}

\bibitem[{Umetsu \& Broadhurst(2008)}]{umetsu08}
Umetsu, K., \& Broadhurst, T. 2008,
  \href{http://dx.doi.org/10.1086/589683}{\apj, 684, 177}

\bibitem[{Umetsu {et~al.}(2011{\natexlab{a}})Umetsu, Broadhurst, Zitrin,
  Medezinski, Coe, \& Postman}]{umetsu11b}
Umetsu, K., Broadhurst, T., Zitrin, A., {et~al.} 2011{\natexlab{a}},
  \href{http://dx.doi.org/10.1088/0004-637X/738/1/41}{\apj, 738, 41}

\bibitem[{Umetsu {et~al.}(2011{\natexlab{b}})Umetsu, Broadhurst, Zitrin,
  Medezinski, \& Hsu}]{umetsu11}
Umetsu, K., Broadhurst, T., Zitrin, A., Medezinski, E., \& Hsu, L.-Y.
  2011{\natexlab{b}},
  \href{http://dx.doi.org/10.1088/0004-637X/729/2/127}{\apj, 729, 127}

\bibitem[{{Umetsu} \& {Diemer}(2017)}]{umetsu17}
{Umetsu}, K., \& {Diemer}, B. 2017,
  \href{http://dx.doi.org/10.3847/1538-4357/aa5c90}{\apj, 836, 231}

\bibitem[{{Umetsu} {et~al.}(2016){Umetsu}, {Zitrin}, {Gruen}, {Merten},
  {Donahue}, \& {Postman}}]{umetsu16}
{Umetsu}, K., {Zitrin}, A., {Gruen}, D., {et~al.} 2016,
  \href{http://dx.doi.org/10.3847/0004-637X/821/2/116}{\apj, 821, 116}

\bibitem[{Umetsu {et~al.}(2012)Umetsu, Medezinski, Nonino, Merten, Zitrin,
  Molino, Grillo, Carrasco, Donahue, Mahdavi, Coe, Postman, Koekemoer, Czakon,
  Sayers, Mroczkowski, Golwala, Koch, Lin, Molnar, Rosati, Balestra, Mercurio,
  Scodeggio, Biviano, Anguita, Infante, Seidel, Sendra, Jouvel, Host, Lemze,
  Broadhurst, Meneghetti, Moustakas, Bartelmann, Benitez, Bouwens, Bradley,
  Ford, Jim{\'{e}}nez-Teja, Kelson, Lahav, Melchior, Moustakas, Ogaz, Seitz, \&
  Zheng}]{umetsu12}
Umetsu, K., Medezinski, E., Nonino, M., {et~al.} 2012,
  \href{http://dx.doi.org/10.1088/0004-637X/755/1/56}{\apj, 755, 56}

\bibitem[{Umetsu {et~al.}(2014)Umetsu, Medezinski, Nonino, Merten, Postman,
  Meneghetti, Donahue, Czakon, Molino, Seitz, Gruen, Lemze, Balestra, Benitez,
  Biviano, Broadhurst, Ford, Grillo, Koekemoer, Melchior, Mercurio, Moustakas,
  Rosati, \& Zitrin}]{umetsu14}
---. 2014, \href{http://dx.doi.org/10.1088/0004-637X/795/2/163}{\apj, 795, 163}

\bibitem[{{Umetsu} {et~al.}(2015){Umetsu}, {Sereno}, {Medezinski}, {Nonino},
  {Mroczkowski}, {Diego}, {Ettori}, {Okabe}, {Broadhurst}, \&
  {Lemze}}]{umetsu15}
{Umetsu}, K., {Sereno}, M., {Medezinski}, E., {et~al.} 2015,
  \href{http://dx.doi.org/10.1088/0004-637X/806/2/207}{\apj, 806, 207}

\bibitem[{{Umetsu} {et~al.}(2018){Umetsu}, {Sereno}, {Tam}, {Chiu}, {Fan},
  {Ettori}, {Gruen}, {Okumura}, {Medezinski}, {Donahue}, {Meneghetti}, {Frye},
  {Koekemoer}, {Broadhurst}, {Zitrin}, {Balestra}, {Benitez}, {Higuchi},
  {Melchior}, {Mercurio}, {Merten}, {Molino}, {Nonino}, {Postman}, {Rosati},
  {Sayers}, \& {Seitz}}]{umetsu18}
{Umetsu}, K., {Sereno}, M., {Tam}, S.-I., {et~al.} 2018, ArXiv e-prints,
  \href{http://arxiv.org/abs/1804.00664}{{\sffamily arXiv:1804.00664}}

\bibitem[{Van Der~Walt {et~al.}(2011)Van Der~Walt, Colbert, \&
  Varoquaux}]{van2011numpy}
Van Der~Walt, S., Colbert, S.~C., \& Varoquaux, G. 2011, Computing in Science
  \& Engineering, 13, 22

\bibitem[{{Vega-Ferrero} {et~al.}(2017){Vega-Ferrero}, {Yepes}, \&
  {Gottl{\"o}ber}}]{vegaferrero17}
{Vega-Ferrero}, J., {Yepes}, G., \& {Gottl{\"o}ber}, S. 2017,
  \href{http://dx.doi.org/10.1093/mnras/stx282}{\mnras, 467, 3226}

\bibitem[{von~der Linden {et~al.}(2014)von~der Linden, Allen, Applegate, Kelly,
  Allen, Ebeling, Burchat, Burke, Donovan, Morris, Blandford, Erben, \&
  Mantz}]{vonderlinden14a}
von~der Linden, A., Allen, M., Applegate, D., {et~al.} 2014,
  \href{http://dx.doi.org/10.1093/mnras/stt1945}{\mnras, 439, 2}

\bibitem[{{Warren} {et~al.}(1992){Warren}, {Quinn}, {Salmon}, \&
  {Zurek}}]{warren92}
{Warren}, M.~S., {Quinn}, P.~J., {Salmon}, J.~K., \& {Zurek}, W.~H. 1992,
  \href{http://dx.doi.org/10.1086/171937}{\apj, 399, 405}

\bibitem[{{Zheng} {et~al.}(2012){Zheng}, {Postman}, {Zitrin}, {Moustakas},
  {Shu}, {Jouvel}, {H{\o}st}, {Molino}, {Bradley}, {Coe}, {Moustakas},
  {Carrasco}, {Ford}, {Ben{\'{\i}}tez}, {Lauer}, {Seitz}, {Bouwens},
  {Koekemoer}, {Medezinski}, {Bartelmann}, {Broadhurst}, {Donahue}, {Grillo},
  {Infante}, {Jha}, {Kelson}, {Lahav}, {Lemze}, {Melchior}, {Meneghetti},
  {Merten}, {Nonino}, {Ogaz}, {Rosati}, {Umetsu}, \& {van der Wel}}]{zheng12}
{Zheng}, W., {Postman}, M., {Zitrin}, A., {et~al.} 2012,
  \href{http://dx.doi.org/10.1038/nature11446}{\nat, 489, 406}

\bibitem[{{Zitrin} {et~al.}(2013){Zitrin}, {Meneghetti}, {Umetsu},
  {Broadhurst}, {Bartelmann}, {Bouwens}, {Bradley}, {Carrasco}, {Coe}, {Ford},
  {Kelson}, {Koekemoer}, {Medezinski}, {Moustakas}, {Moustakas}, {Nonino},
  {Postman}, {Rosati}, {Seidel}, {Seitz}, {Sendra}, {Shu}, {Vega}, \&
  {Zheng}}]{zitrin13}
{Zitrin}, A., {Meneghetti}, M., {Umetsu}, K., {et~al.} 2013,
  \href{http://dx.doi.org/10.1088/2041-8205/762/2/L30}{\apjl, 762, L30}

\bibitem[{{Zitrin} {et~al.}(2015){Zitrin}, {Fabris}, {Merten}, {Melchior},
  {Meneghetti}, {Koekemoer}, {Coe}, {Maturi}, {Bartelmann}, {Postman},
  {Umetsu}, {Seidel}, {Sendra}, {Broadhurst}, {Balestra}, {Biviano}, {Grillo},
  {Mercurio}, {Nonino}, {Rosati}, {Bradley}, {Carrasco}, {Donahue}, {Ford},
  {Frye}, \& {Moustakas}}]{zitrin15}
{Zitrin}, A., {Fabris}, A., {Merten}, J., {et~al.} 2015,
  \href{http://dx.doi.org/10.1088/0004-637X/801/1/44}{\apj, 801, 44}

\end{thebibliography}

%
%

\appendix

\section{Individual Posterior Constraints}
\label{sec:individual_constraints}

For each cluster, we show in Figure~\ref{fig:indi} the marginalized
posterior distributions of model parameters from \sphericalmodel,
\triaxialfmodel\ and \triaxialbmodel\ modeling approaches (indicated by red, blue,
and green areas, respectively).

\begin{figure}
\centering
\resizebox{!}{0.485\textwidth}{\includegraphics[scale=1.0]{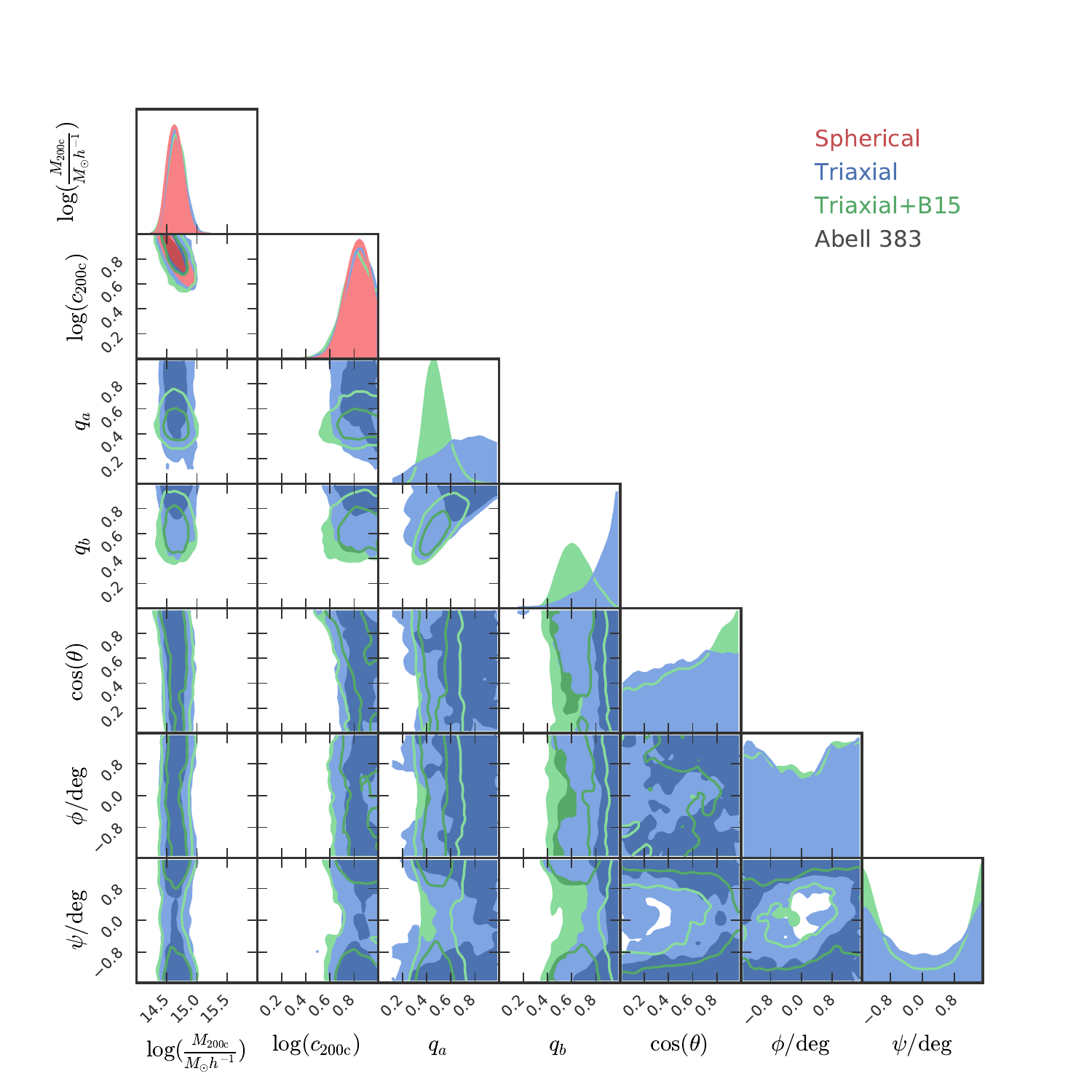}}
\resizebox{!}{0.485\textwidth}{\includegraphics[scale=1.0]{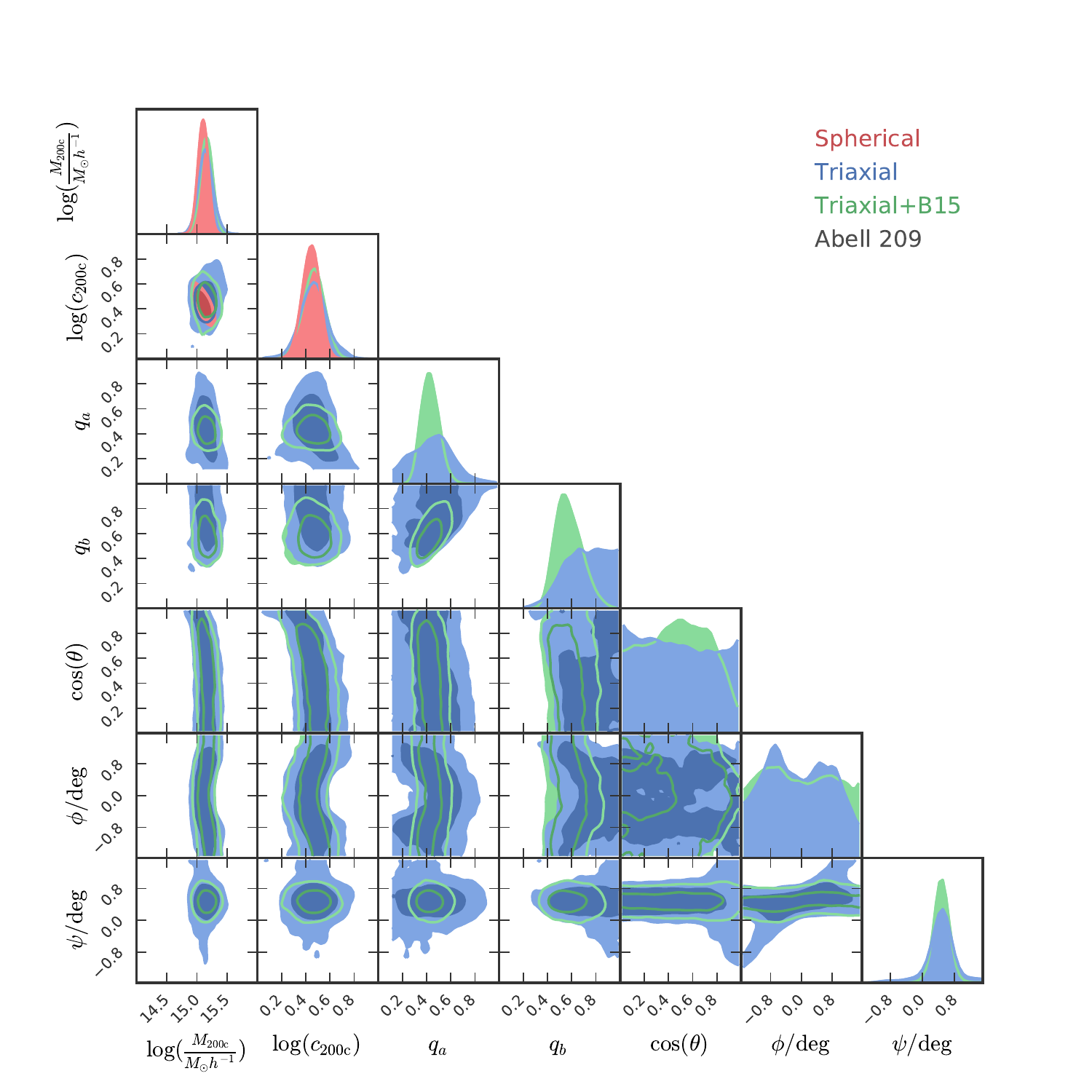}}
\resizebox{!}{0.485\textwidth}{\includegraphics[scale=1.0]{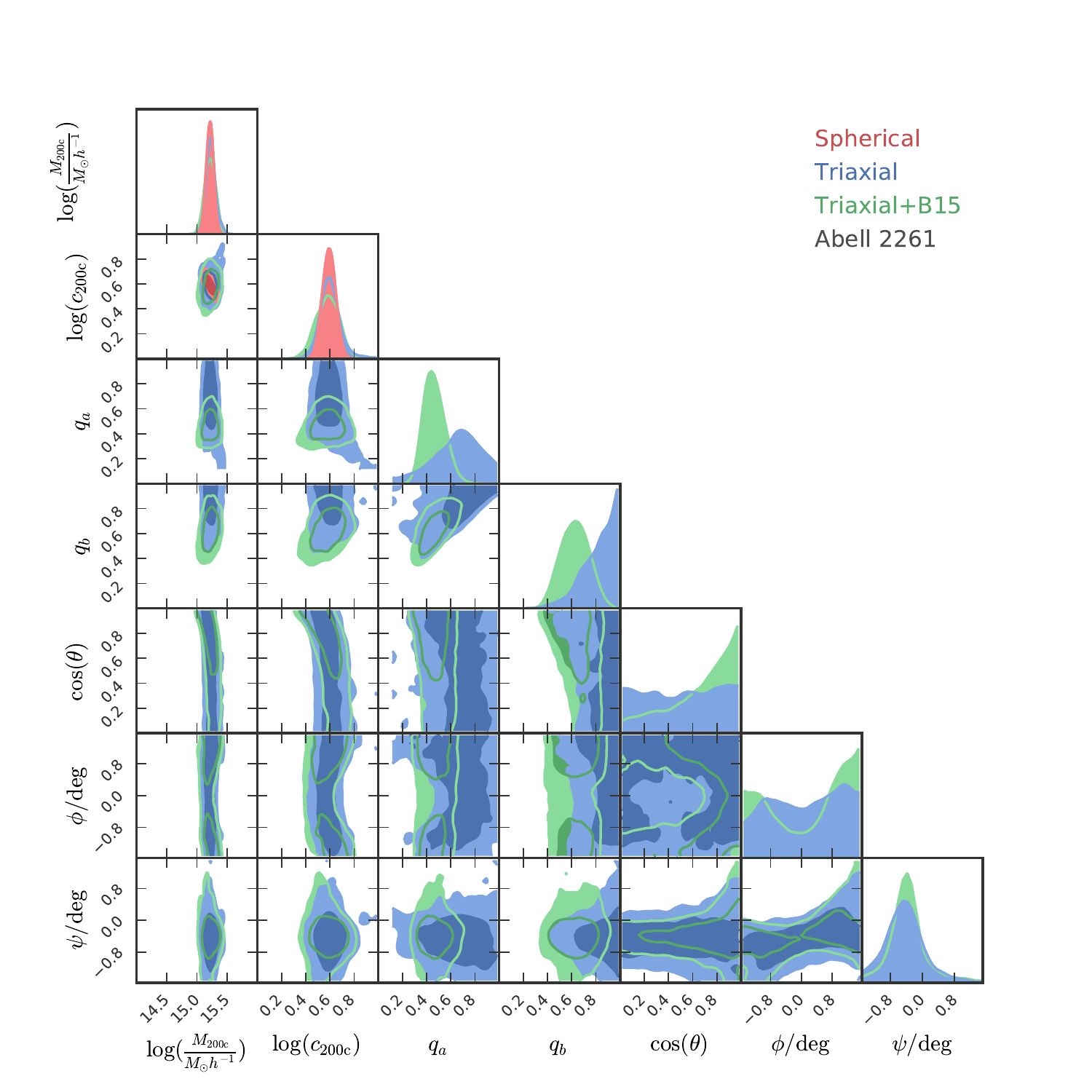}}
\resizebox{!}{0.485\textwidth}{\includegraphics[scale=1.0]{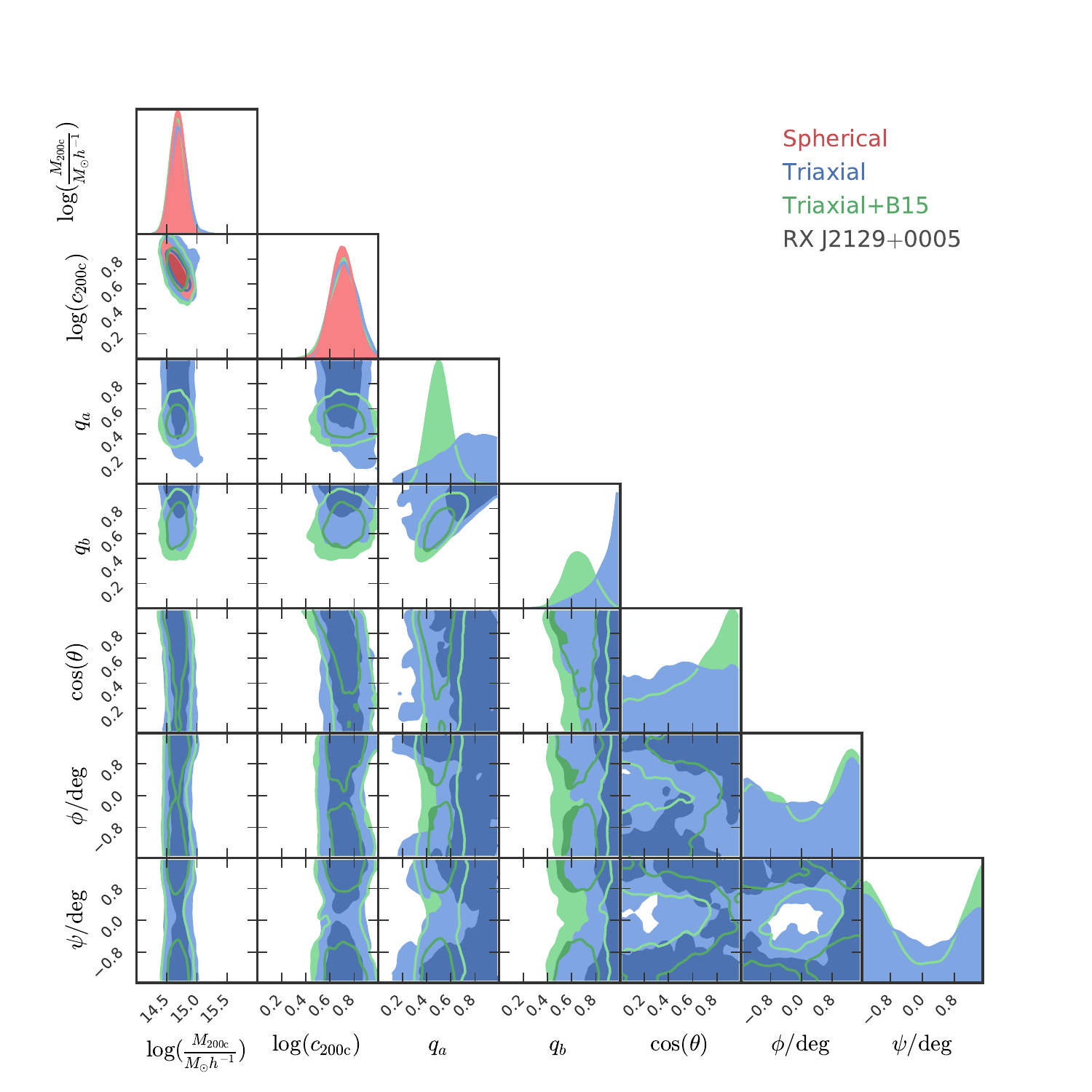}}
 \caption{
Constraints on the cluster model parameters derived for
 each individual cluster with the \sphericalmodel, \triaxialfmodel\, and
 \triaxialbmodel modeling approaches,
 showing marginalized 1D (histograms) and 2D ($68\percent$ and $95\percent$ confidence
 level contour plots) posterior distributions.  
 Seven parameters $(\Mtwooo, \Ctwooo, \qa, \qb,
 \cos\theta, \phi, \psi)$ are shown for the triaxial cases, while only the mass and
 concentration parameters (\Mtwooo\ and \Ctwooo) are presented for
 the \sphericalmodel\ modeling.  
}
\label{fig:indi}
\end{figure}
\begin{figure}
\addtocounter{figure}{-1}
\centering
\resizebox{!}{0.485\textwidth}{\includegraphics[scale=1.0]{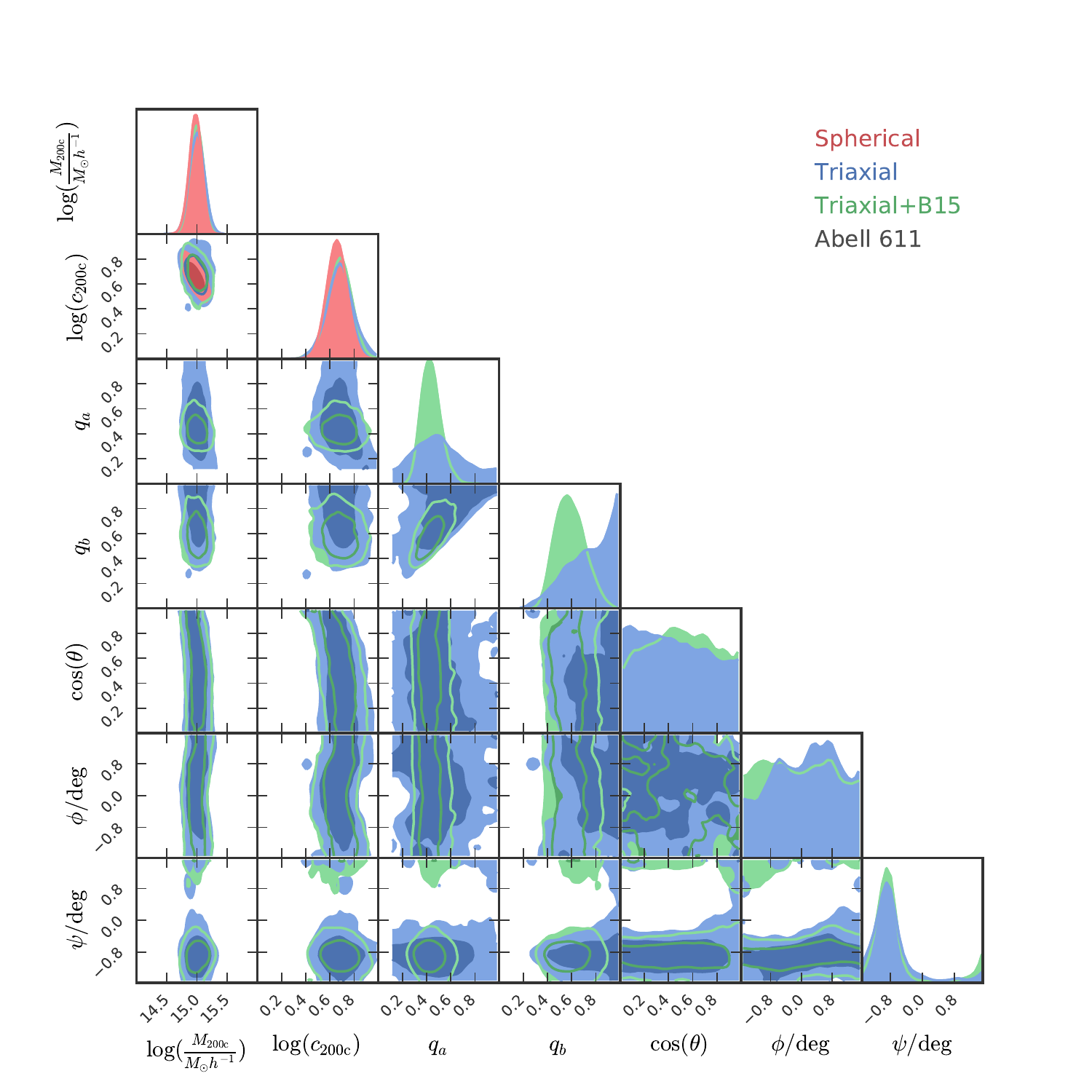}}
\resizebox{!}{0.485\textwidth}{\includegraphics[scale=1.0]{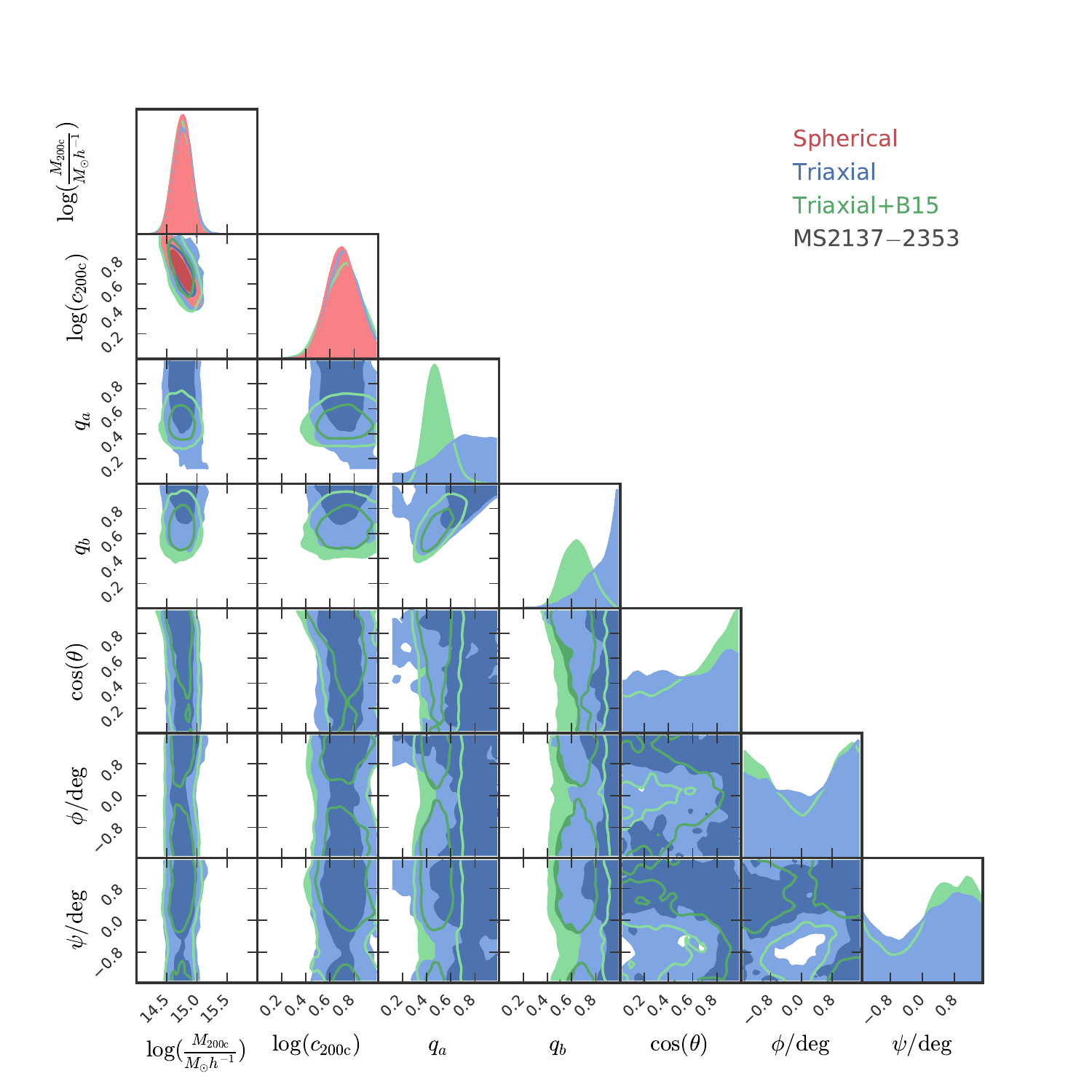}}
\resizebox{!}{0.485\textwidth}{\includegraphics[scale=1.0]{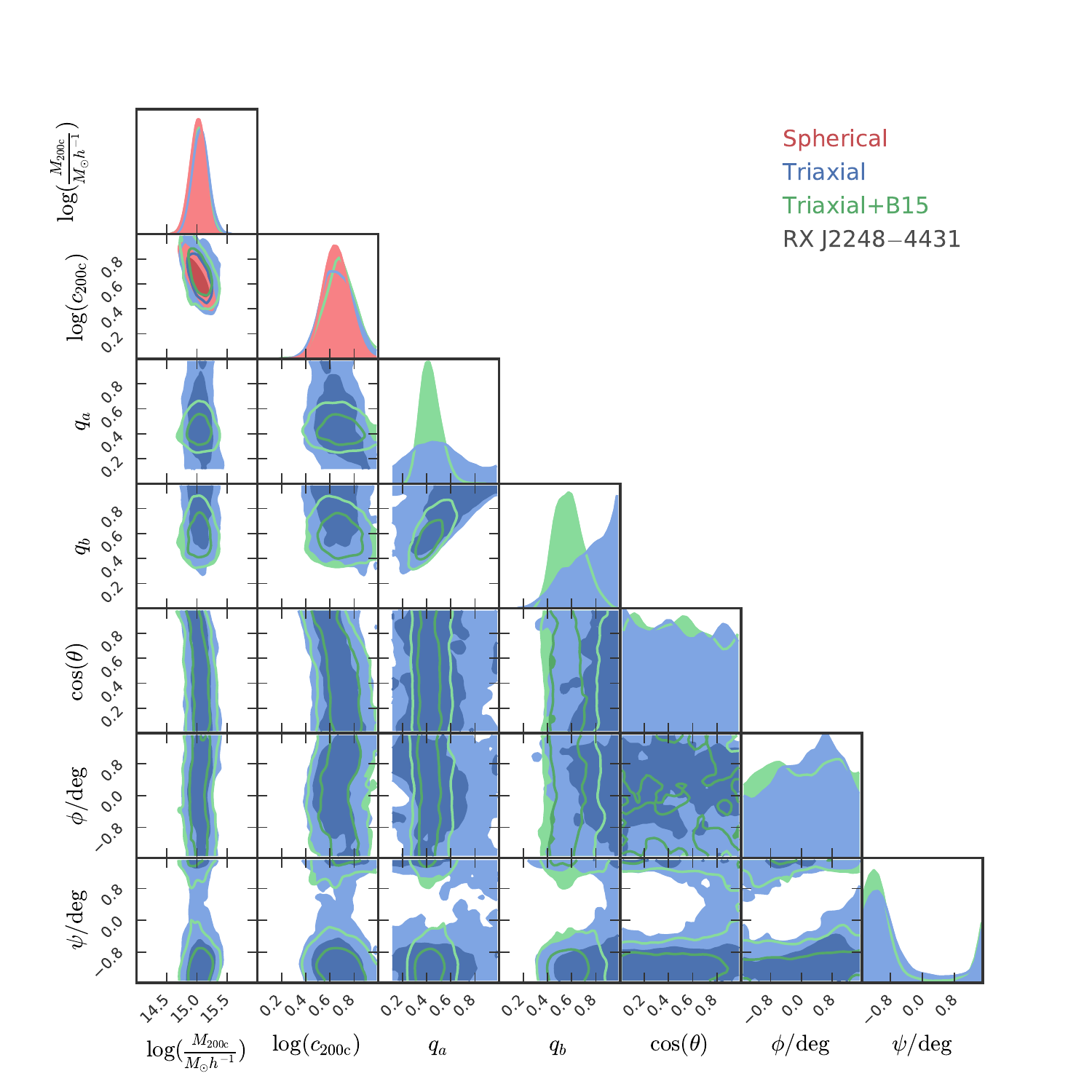}}
\resizebox{!}{0.485\textwidth}{\includegraphics[scale=1.0]{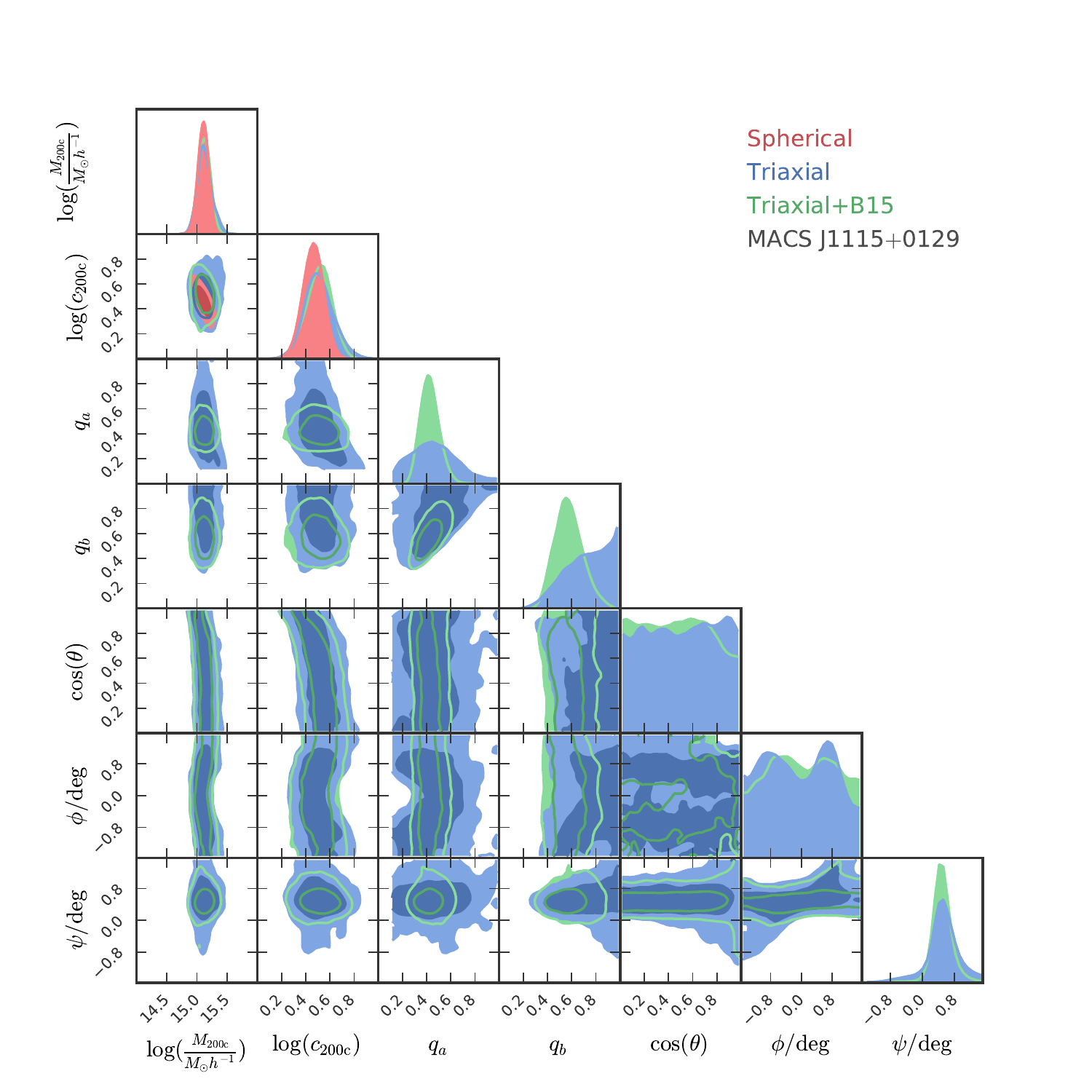}}
\caption{
Continued. 
}
\end{figure}
\begin{figure}
\addtocounter{figure}{-1}
\centering
\resizebox{!}{0.485\textwidth}{\includegraphics[scale=1.0]{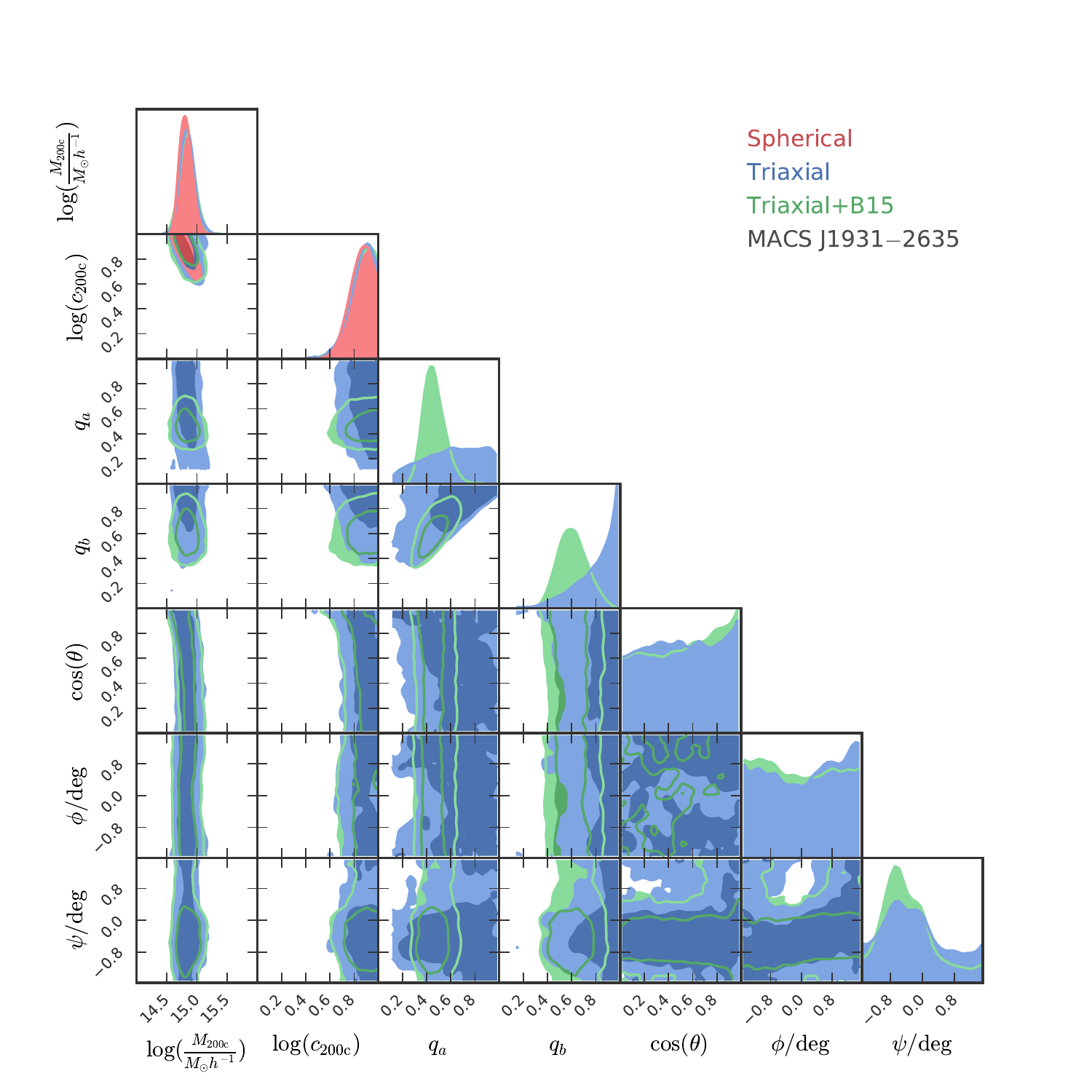}}
\resizebox{!}{0.485\textwidth}{\includegraphics[scale=1.0]{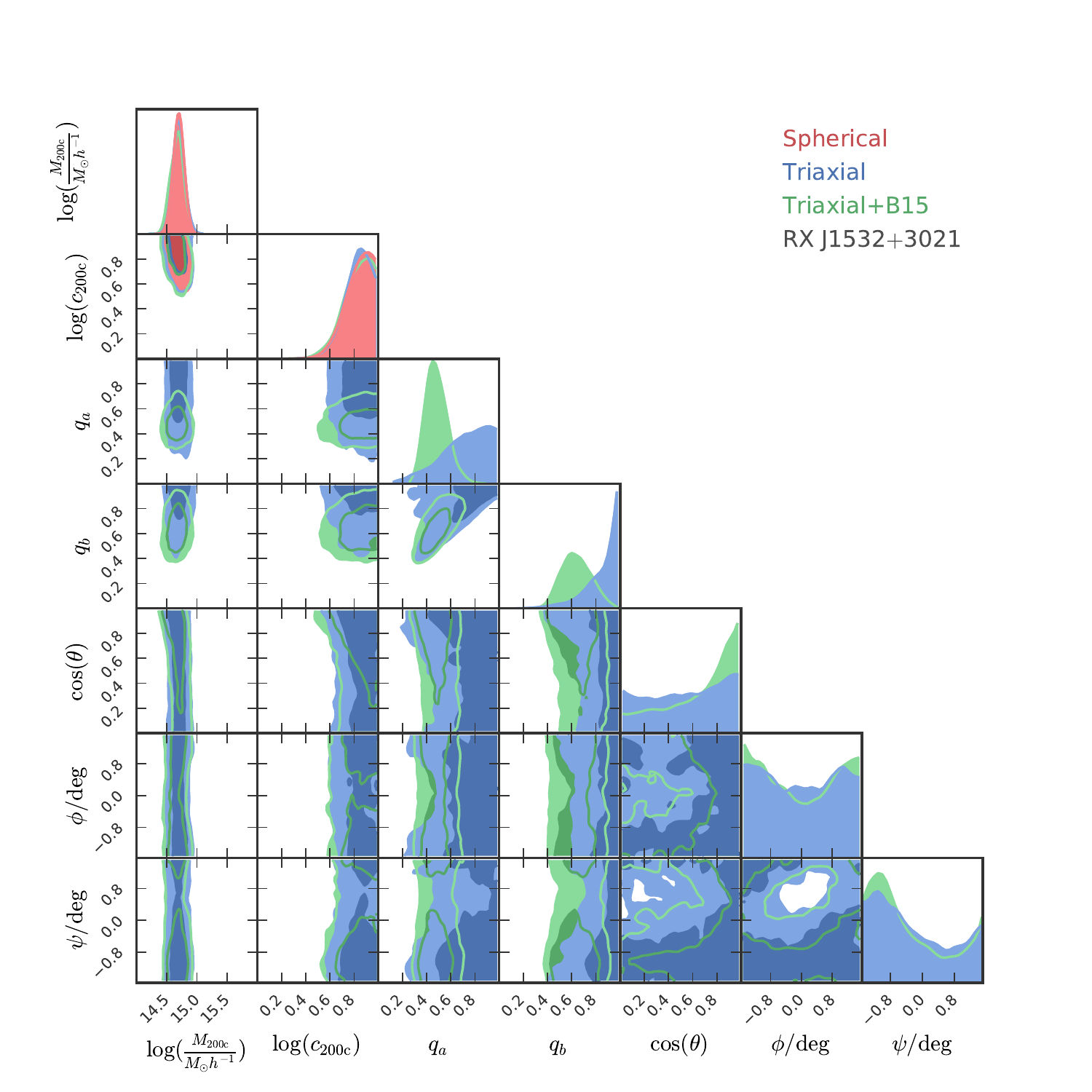}}
\resizebox{!}{0.485\textwidth}{\includegraphics[scale=1.0]{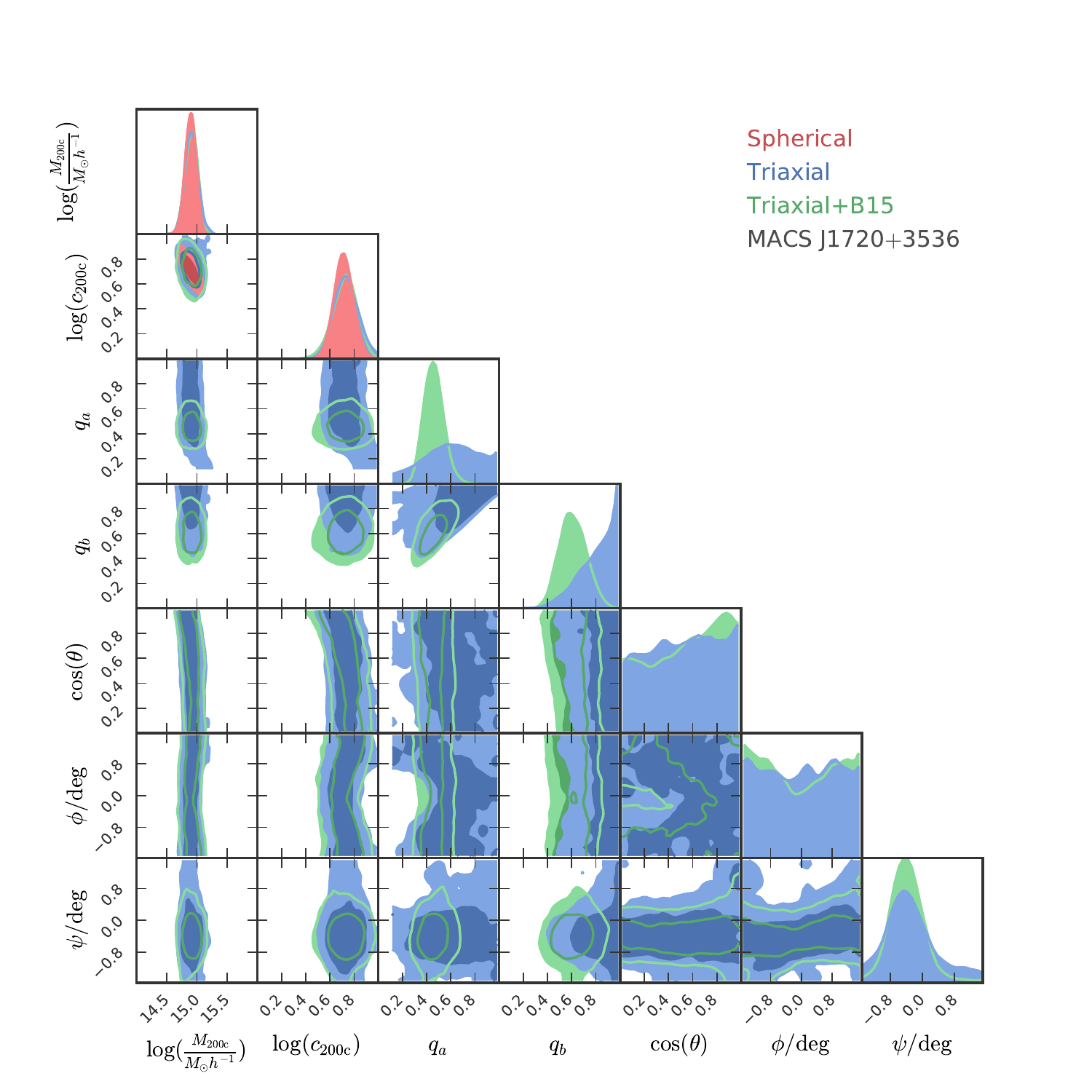}}
\resizebox{!}{0.485\textwidth}{\includegraphics[scale=1.0]{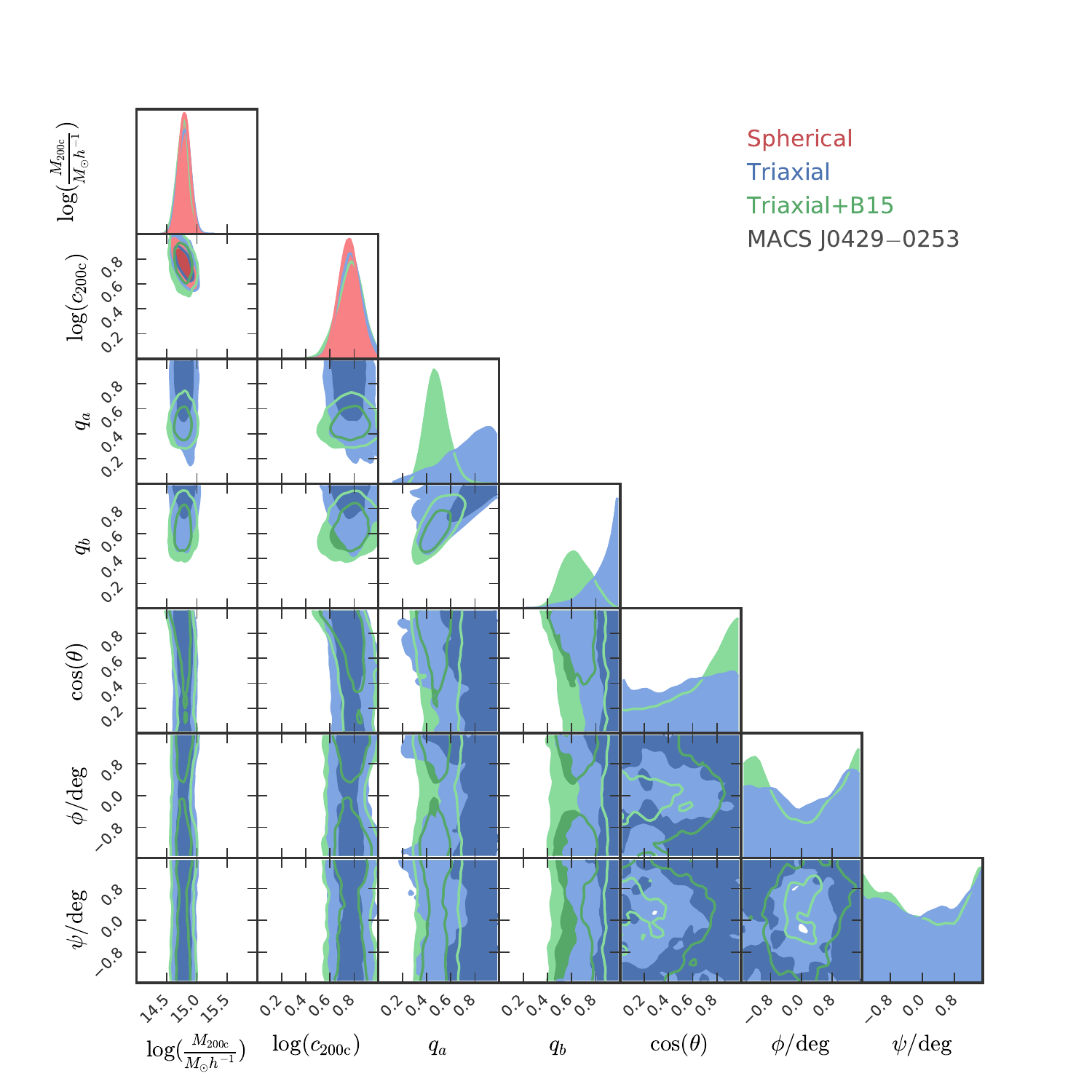}}
\caption{
Continued.
}
\end{figure}
\begin{figure}
\addtocounter{figure}{-1}
\centering
\resizebox{!}{0.485\textwidth}{\includegraphics[scale=1.0]{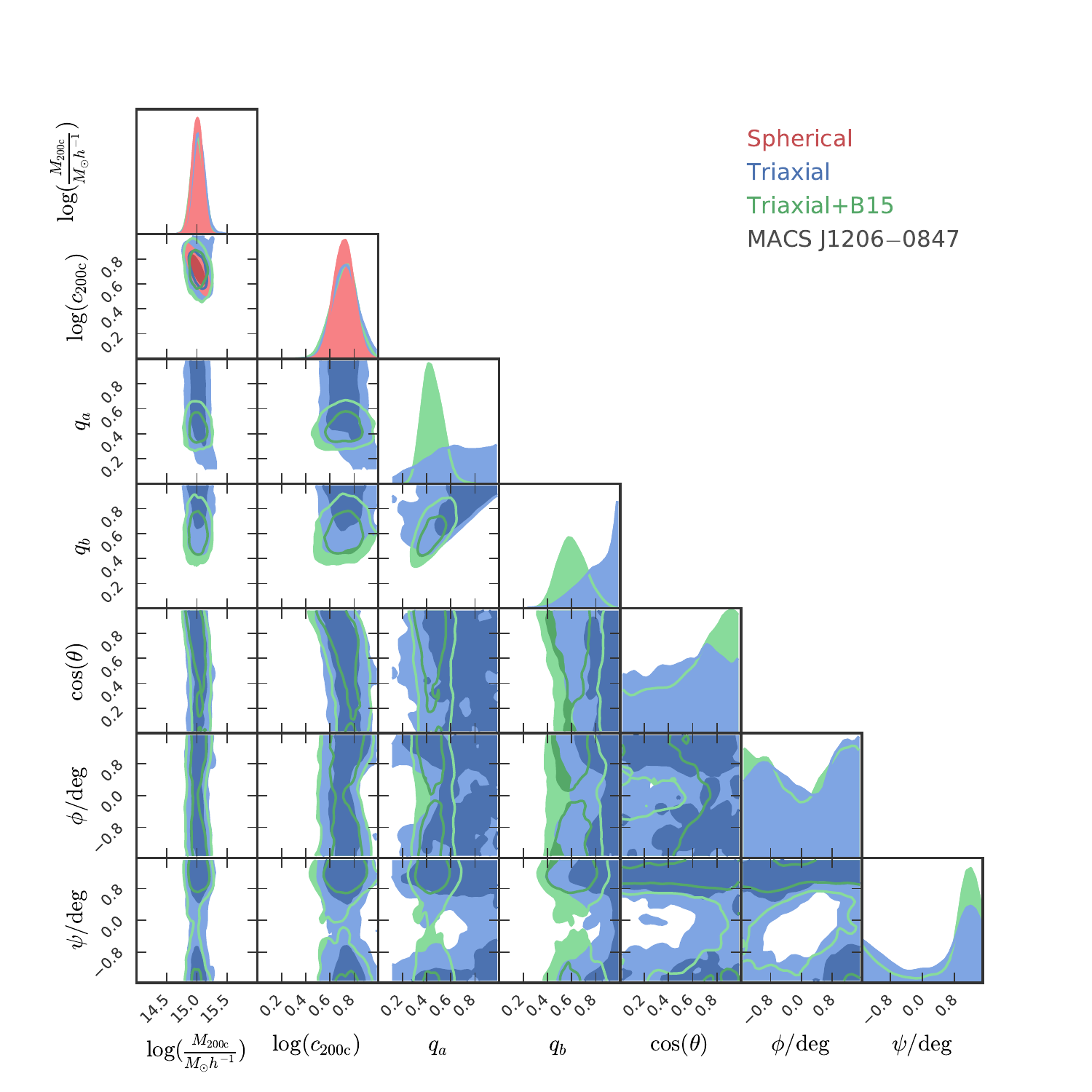}}
\resizebox{!}{0.485\textwidth}{\includegraphics[scale=1.0]{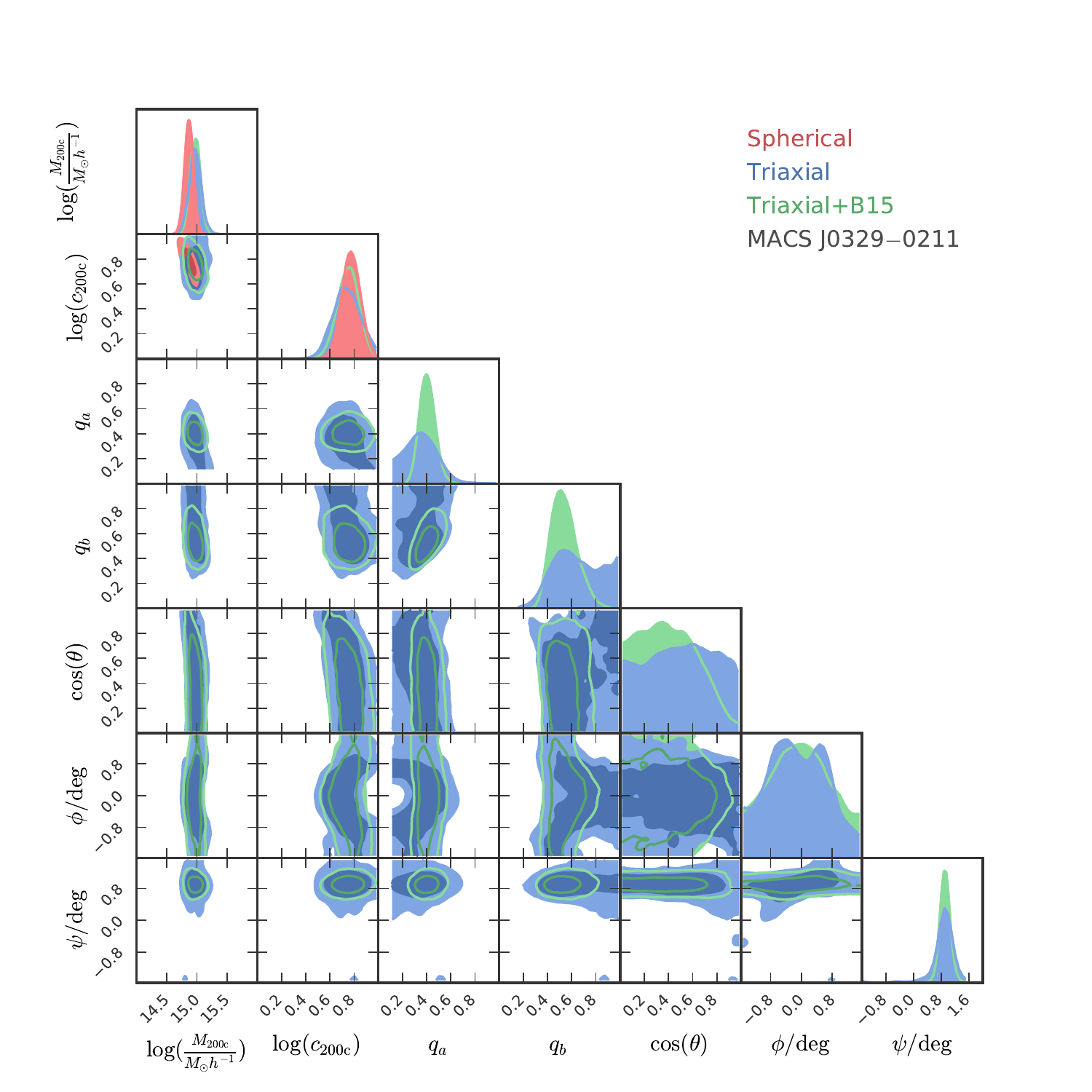}}
\resizebox{!}{0.485\textwidth}{\includegraphics[scale=1.0]{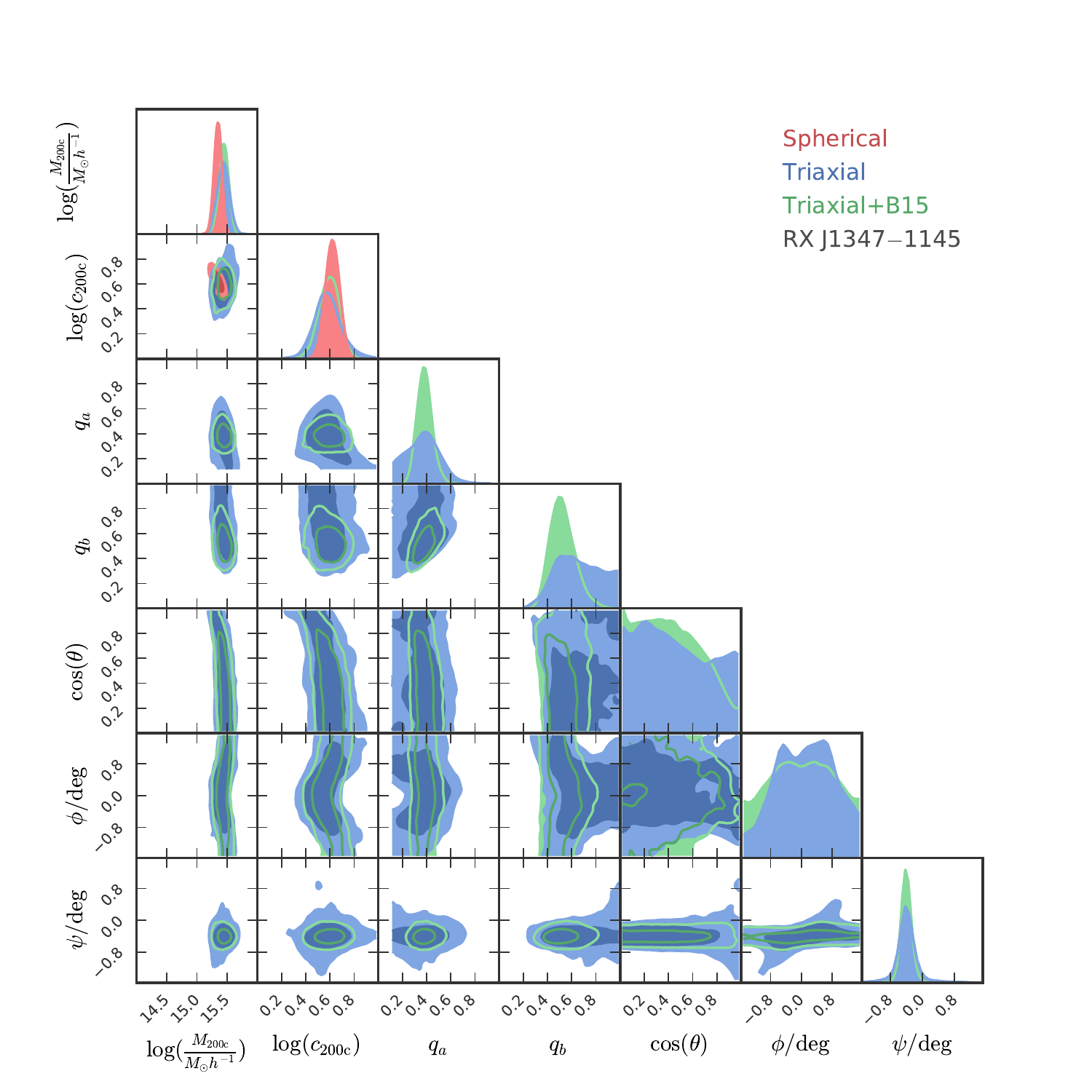}}
\resizebox{!}{0.485\textwidth}{\includegraphics[scale=1.0]{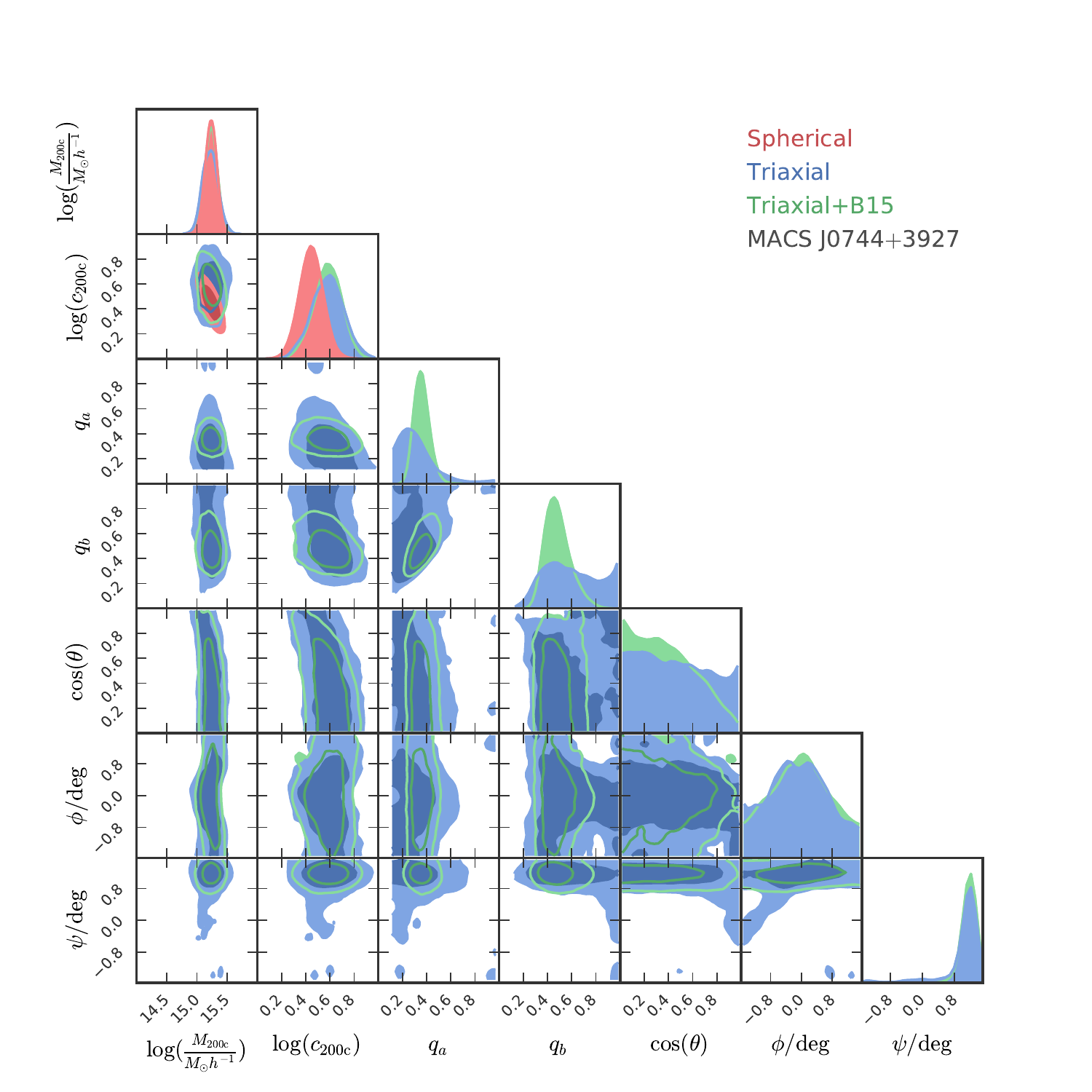}}
\caption{
Continued.
}
\end{figure}
\begin{figure}
\addtocounter{figure}{-1}
\centering
\resizebox{!}{0.485\textwidth}{\includegraphics[scale=1.0]{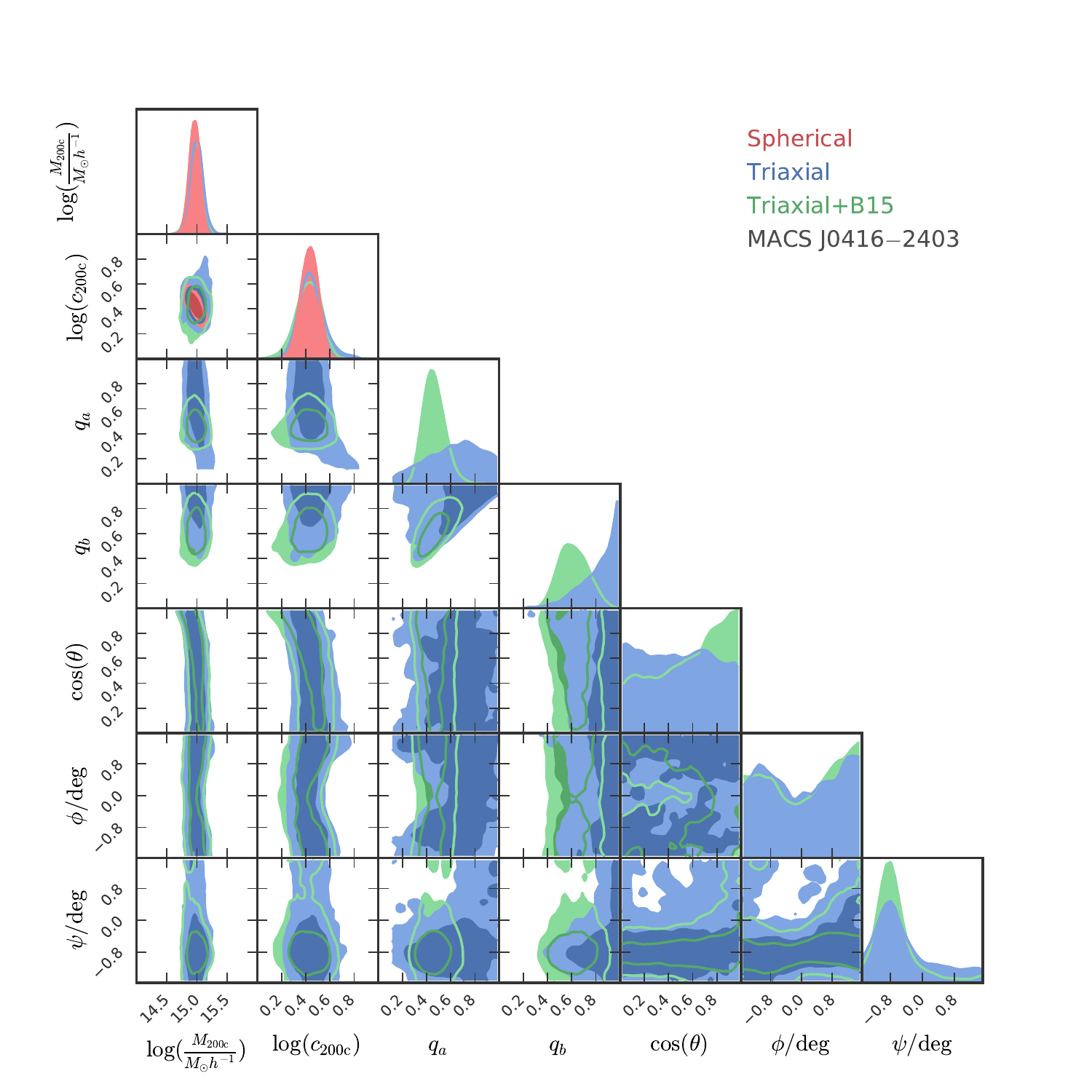}}
\resizebox{!}{0.485\textwidth}{\includegraphics[scale=1.0]{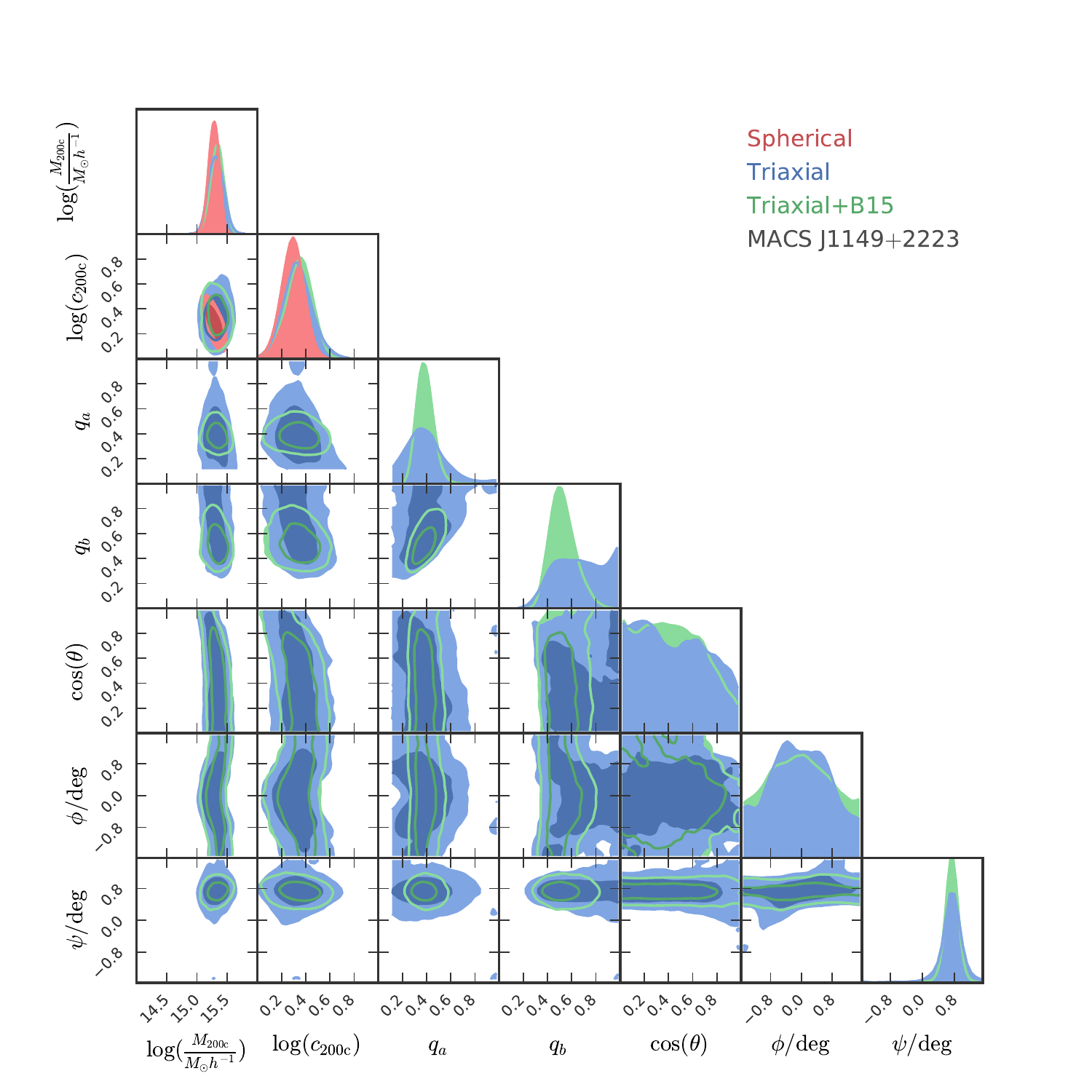}}
\resizebox{!}{0.485\textwidth}{\includegraphics[scale=1.0]{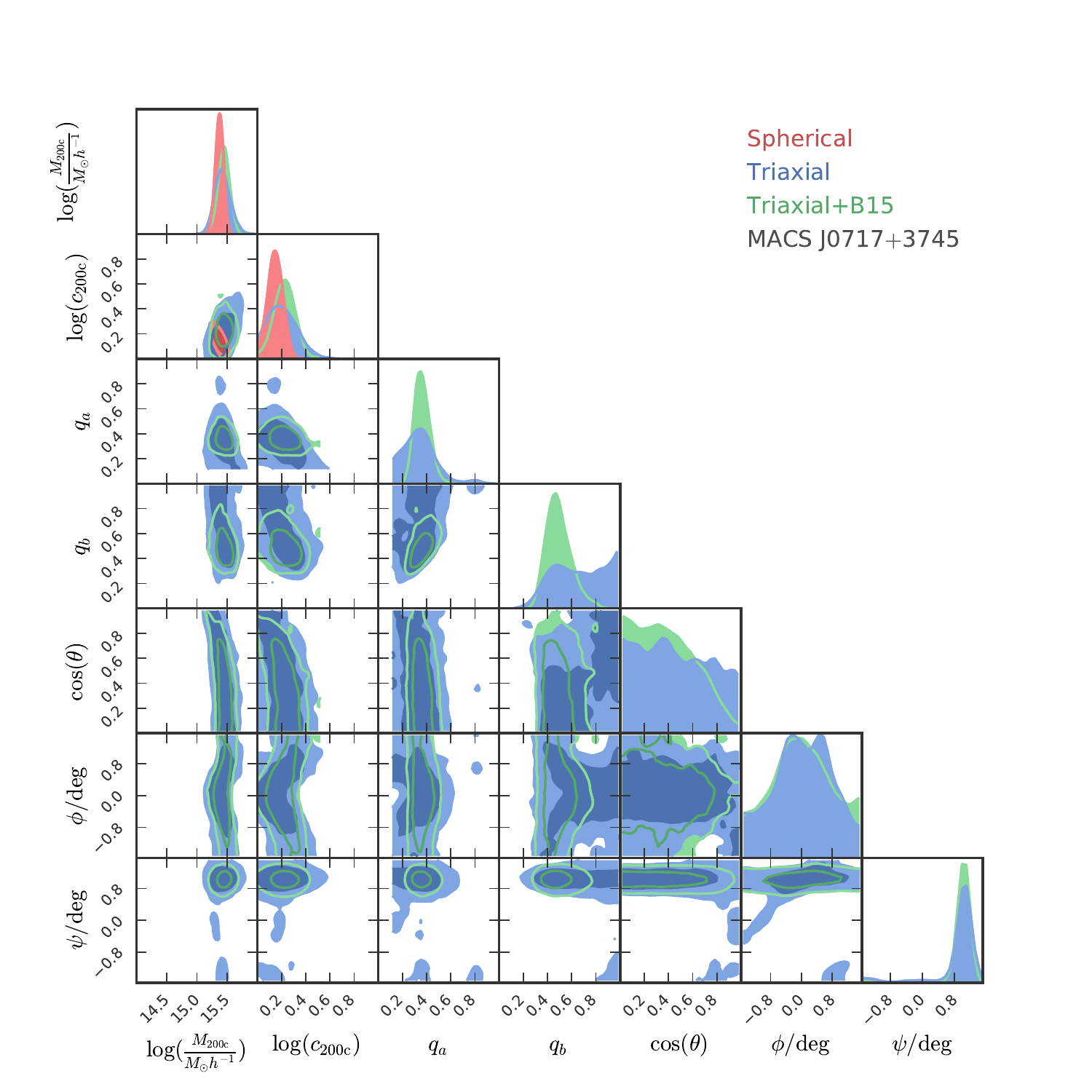}}
\resizebox{!}{0.485\textwidth}{\includegraphics[scale=1.0]{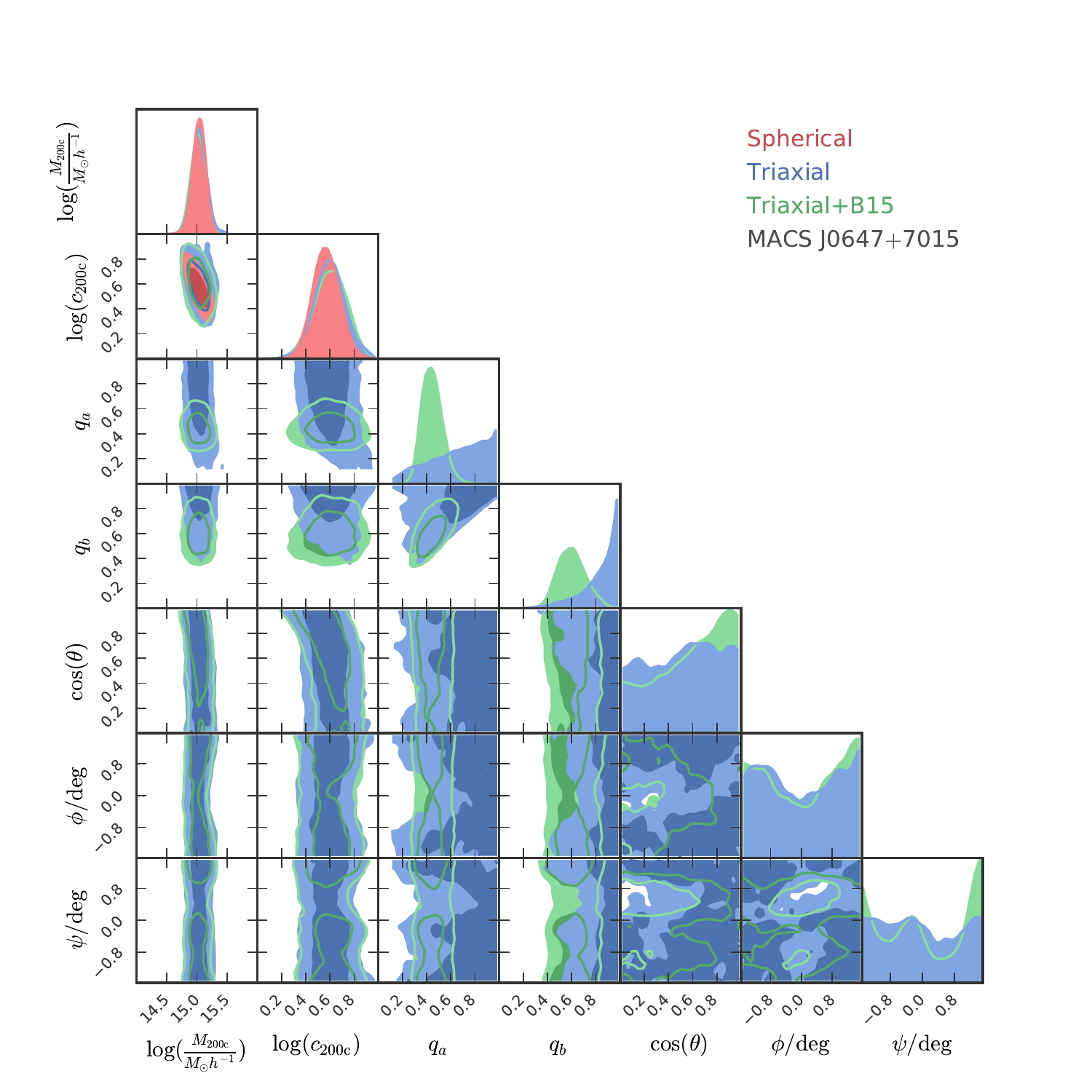}}
\caption{
Continued.
}
 \end{figure}

\end{document}